\documentclass[reqno,12pt]{amsart}\usepackage[]{graphicx}\usepackage[]{xcolor}
\makeatletter
\def\maxwidth{ %
  \ifdim\Gin@nat@width>\linewidth
    \linewidth
  \else
    \Gin@nat@width
  \fi
}
\makeatother

\definecolor{fgcolor}{rgb}{0.345, 0.345, 0.345}

\usepackage{framed}
\makeatletter
 {\par\unskip\endMakeFramed%
 \at@end@of@kframe}
\makeatother

\definecolor{shadecolor}{rgb}{.97, .97, .97}
\definecolor{messagecolor}{rgb}{0, 0, 0}
\definecolor{warningcolor}{rgb}{1, 0, 1}
\definecolor{errorcolor}{rgb}{1, 0, 0}
\usepackage{alltt}

\usepackage{accents,setspace,graphicx,srcltx,enumitem}
\usepackage[round]{natbib}
\usepackage{subfigure}
\usepackage{bbm}
\usepackage{multirow}
\usepackage{amsmath}
\usepackage{amssymb}
\usepackage{amsfonts}
\usepackage{amsaddr}
\usepackage[english]{babel}
 
\usepackage{MnSymbol} 
\usepackage{longtable}
\usepackage{hyperref}
\usepackage[T1]{fontenc}

  \newcommand{\tsigma}{\tilde{\sigma}}
  \newcommand{\cez}{\tilde{W}\gamma}

\newtheorem{theorem}{Theorem}[section]
\newtheorem{proposition}{Proposition}[section]
\newtheorem{assumption}{Assumption}[section]

 \newcounter{assB1}
 \newcounter{assB2}
 \theoremstyle{remark}
 \newtheorem{remark}[]{Remark}

\DeclareMathOperator*{\argmax}{arg\,max}

   \allowdisplaybreaks
  \makeatletter
\def\Let@{\def\\{\notag\math@cr}}
\makeatother

\title{Binary Endogenous Treatment in Stochastic Frontier Models with an Application to Soil Conservation in El Salvador}

\date{\today}
\author{Samuele Centorrino$^\ast$}
\address[S.\ Centorrino]{$(^\ast)$ Corresponding Author.\\
Economics Department, the State University of New York at Stony Brook, USA.\\\textit{Email address}: \texttt{\textup{samuele.centorrino@stonybrook.edu}}.}
\vspace*{-20pt}
\author{Mar\'ia P\'erez Urdiales}
\address[M.\ P\'erez Urdiales]{Inter-American Development Bank, Water and Sanitation Division, USA.\\\textit{Email address}: \texttt{\textup{mariaurdl@iadb.org}}.}
\vspace*{-20pt}
\author{Boris Bravo-Ureta}
\address[B.\ Bravo-Ureta]{Department of Agricultural and Resource Economics, University of Connecticut, USA.\\\textit{Email address}: \texttt{\textup{boris.bravoureta@uconn.edu}}.}
\vspace*{-20pt}
\author{Alan Wall}
\address[A.\ Wall]{Oviedo Efficiency Group and Department of Economics, University of Oviedo, Spain.\\\textit{Email address}: \texttt{\textup{awall@uniovi.es}}.}

\thanks{The authors wish to thank Christopher F. Parmeter, participants in the 11th North American Productivity Workshop (NAPW XI), especially Subal C. Kumbhakar, and seminar participants at the Center of Productivity and Performance of Loughborough University, and the Center for Econometrics and Business Analytics for valuable comments and remarks. The views expressed do not reflect those of the IDB, its Executive Board or its Management.}

\IfFileExists{upquote.sty}{\usepackage{upquote}}{}
\begin{document}

\begin{abstract}
Improving the productivity of the agricultural sector is part of one of the Sustainable Development Goals set by the United Nations. To this end, many international organizations have funded training and technology transfer programs that aim to promote productivity and income growth, fight poverty and enhance food security among smallholder farmers in developing countries. Stochastic production frontier analysis can be a useful tool when evaluating the effectiveness of these programs. However, accounting for treatment endogeneity, often intrinsic to these interventions, only recently has received any attention in the stochastic frontier literature. In this work, we extend the classical maximum likelihood estimation of stochastic production frontier models by allowing both the production frontier and inefficiency to depend on a potentially endogenous binary treatment. We use instrumental variables to define an assignment mechanism for the treatment, and we explicitly model the density of the first and second-stage composite error terms. We provide empirical evidence of the importance of controlling for endogeneity in this setting using farm-level data from a soil conservation program in El Salvador.\\


\end{abstract}

\maketitle

\doublespacing

\newpage 

\section{Introduction} \label{sec:intro}

The global need to increase productivity via technological change and efficiency improvements in the agricultural sector has been recognized in the United Nations 2030 Agenda for Sustainable Development. In particular, the Sustainable Development Goal (SDG) \#2 aims to end hunger and improve agricultural productivity and income among small-scale farmers while promoting resilient agricultural practices and sustainable food production systems. A growing number of governments, development organizations, and agencies are implementing programs targeting this goal. Many of these programs work at the scale of smallholder farms and often include support for the adoption of innovative technologies and practices, as well as funding for technical assistance, agricultural education and training (see \citealp{Bravo2020}, \citealp{Jimi2019} and also \citealp{dejanvry2017}, for a review).

However, participation in these programs is typically  voluntary, which may lead to endogenous selection of participants into treatment. Farmers who choose to participate (i.e., who self-select into the program) may share specific unobserved characteristics that distinguish them from non-participants. For instance, the participants' cultivated land may be more vulnerable to erosion and consequently, they may be less productive than non-participants. If endogenous selection of participants into the treatment is not controlled for, one might conclude that the program is not effective because the agricultural efficiency of those who participate is lower than those who do not.

Stochastic Frontier Analysis (SFA) is a popular method to assess technical efficiency (TE) as a measure of productivity and output shortfall. The production frontier is defined as the maximum quantity of output that can be produced with some input mix, given the technology and the environment. Efficiency (or inefficiency) is measured by the distance of each producer from the frontier, and it is usually modeled using a one-sided unobserved random variable. Similarly, output may be measured with error, or there may be other sources of variation in the outcome not observed by the econometrician that result in a further stochastic component. This \textit{composite} error term is a defining feature of stochastic frontier models. In this framework, endogenous selection has been modeled either by considering the dependence between program participation and the two-sided stochastic component of the error term \citep{Greene2010} or by focusing on the dependence between program participation and the unobserved inefficiency component \citep{Kumbhakar2009}. However, as pointed out by \citet[][Section 6.3.3, p.293]{parmeter2014}, ``a framework where selection is based on [the composite error term] does not exist." We fill this gap in the current literature by developing a framework in which selection is based on the composite error term. Our paper also contributes to the recent literature about tackling endogeneity in stochastic frontier models \citep{amsler2016,amsler2017,centorrinoperez2019}. In all these papers, the endogenous variable is continuously distributed, whereas we study the case in which the endogenous variable is binary. Therefore, in this paper we provide an approach that allows the stochastic frontier model to be used for treatment evaluation.

In our framework, we denote the treatment with a random variable $Z$. The decision of farmer $i$ to participate in the treatment can be described as an index function, $Z^\ast_i$, which depends on observed ($\tilde{W}_i$) and unobserved ($\eta_i$) variables \citep{heckman1985}. If the index $Z^\ast_i$ is greater than a threshold (often normalized to $0$), then the farmer decides to participate in the treatment and $Z_i = 1$; otherwise, she does not and $Z_i = 0$. The decision of farmers to participate in the treatment generates dependence between unobservable individual characteristics (i.e., preferences and/or managerial abilities) and the treatment assignment. Correlation between the unobservable factors, $\eta_i$, and the composite error term in the stochastic frontier framework means that the latter affects the choice of the farmer to undertake the treatment. Hence, the composite error term \textit{confounds} the effect of $Z_i$ on the output and on the mean of the inefficiency term. This endogeneity problem renders standard estimation methods inconsistent. As mentioned above, the challenge is to provide a framework in which the treatment assignment is potentially correlated with both components of the error term.

In the stochastic frontier literature, the issue of endogeneity with respect to the two-sided error component has been previously treated as an issue of \textit{sample selection bias}. The main rationale for this choice is that treated and non-treated units come from two distinct populations. The selection mechanism therefore applies as we only observe the treatment group in one sample and the control group in another sample \citep{Greene2010,Lai2015}. \cite{Bravo2012} considers a framework similar to \citet{Greene2010} to control for selection on unobservables, coupled with a propensity score matching technique to additionally control for selection on observables. 

However, the sample selection framework may not be appropriate to study endogenous selection into treatment for two main reasons. From a methodological perspective, the sample includes observations from both the treatment and the control group coming from the same population (e.g., the population of farmers in a specific region). In this context, there does not seem to be any particular reasons to estimate two different frontiers for a population of farmers based on whether they participate in the treatment or not. From a statistical perspective, the application of the sample selection approach requires splitting the full sample into two subsamples, which effectively reduces the degrees of freedom. Potential differences in output elasticities between treatment and control groups can be appropriately introduced using interaction terms between the treatment and the inputs without splitting the sample, as we explain below.

An alternative approach is proposed by \citet{Kumbhakar2009}. In particular, \citeauthor{Kumbhakar2009}'s (\citeyear{Kumbhakar2009}) paper considers the endogeneity of technology choice (conventional or organic farming) by jointly estimating technology and technology choice using a single-step maximum likelihood method. In their framework, the technology choice directly depends on the inefficiency term, and therefore endogeneity operates through the inefficiency term only. Moreover, they do not consider treatment participation as a potential determinant of inefficiency.

The approaches in both \citet{Greene2010} and \citet{Kumbhakar2009} are based on simulated maximum likelihood, while we provide a closed-form likelihood function, and therefore we do not need to resort to simulations. This has both computational advantages and may improve the finite sample properties of the estimator.

More recently, \citet{chen2020} study a general model with binary endogenous treatment and mediator, that are potentially correlated with the composite error term. Their approach uses a propensity score assumption, which is used to construct moment conditions that are robust to the potential endogeneity of the treatment. However, an important shortcoming is that they do not provide an observation specific estimator of TE.

To the best of our knowledge, there does not exist a maximum likelihood framework in SFA which allows one to control for potential correlation between program participation and both the unobservable idiosyncratic component and the stochastic inefficiency. 

Our contribution to this literature is to provide a model that allows one to control for endogeneity coming from both sources. We allow the treatment to enter the model in a flexible way, so that participation in the program can act as both a determinant of (in)efficiency as well as a facilitating input. This is crucial as it permits testing whether program participation helps farmers produce more efficiently, given the technology; and/or  modifies the technology. Our empirical strategy is to employ instrumental variables (IVs) to construct an auxiliary assignment mechanism for program participation. We then propose a maximum likelihood framework in which we jointly model the density of the first-stage error and the density of the composite error term common to the stochastic production frontier. Under appropriate conditional independence assumptions, we derive the likelihood function in closed form, which allows us to use standard estimation and inference procedures, making the model straightforward to estimate and interpret. We also provide some theoretical results about identification and estimation, with a focus on the parameter capturing dependence between the stochastic inefficiency and the unobservable first-stage error. We show that only the magnitude of the dependence is identified, but not its sign. Moreover, when the true value of this parameter is $0$, the information matrix is singular and the model is only second-order identified. As this limiting case is relevant for practitioners who wish to test the lack of endogeneity with respect to the inefficiency component, we discuss some testing procedures which can be applied in this context, and also informally discuss how to construct tests and confidence intervals that are robust to the lack of first-order identification \citet{rot2000,andrews2001,bottai2003,ekvall2022}.

Our framework is similar in spirit to \citet{Kumbhakar2009}, in that we also use a single-step maximum likelihood method. However, we model the dependence of both components of the error term. Compared to the model proposed by \citet{chen2020}, we impose explicit distributional assumptions on both the inefficiency term and the stochastic component. However, our approach is based on a one-step maximum likelihood estimator and allows one to obtain an estimator of TE for each producer, which is not provided in \citet{chen2020}. The ability to estimate TE for each observation is an essential feature of stochastic frontier models, as it allows comparisons across different observations \citep{farrell1957,jondrow1982}.

We apply the proposed method to a sample of smallholder farms from El Salvador. The data consist of a sample of participants in an environmental program promoting soil conservation practices, as well as a control group of non-participant farmers. In this empirical analysis, standard stochastic frontier estimation does not show any effect of the program, either on the production level or on farmers’ TE. By contrast, our approach reveals that program participation significantly improves TE. These results further highlight the need to control for endogeneity when evaluating such interventions, as this may substantially change the conclusions regarding their effectiveness.

The paper is structured as follows. In Section \ref{sec:model}, we describe the econometric model and our maximum likelihood estimator. Section \ref{sec:montecarlo} contains a finite sample assessment of our method, in which we also discuss the implementation of our estimator. Section \ref{sec:empiricalapplication} contains a description of the sample and outlines our empirical results.

\section{Binary Treatment in Stochastic Production Frontier} \label{sec:model}

\subsection{Model} We consider the following stochastic frontier regression model:
\begin{equation} \label{eq:stmod}
Y = m(X,Z,\beta) + V - U,
\end{equation}
where $Y$ is the logarithm of output; $m(X,Z,\beta)$ is the logarithm of the production frontier, which depends on some unknown vector of parameters $\beta$, some production inputs, $X$, and other \textit{environmental factors}, $Z$; and $\varepsilon = V - U$ is a composite error term. $V$ is a stochastic component with $0$ mean, and $U \geq 0$ is a technical inefficiency term that captures the shortfall of the producer from the frontier. The latter may depend on other observed characteristics of producers (for instance, experience and education) that are often introduced as a scale factor affecting the distribution of $U$ \citep{simar1994,alvarez2006}. Thus, we write $U = U_0 g(Z,\delta)$, where $g(\cdot,\cdot)$ is the \textit{scale function} which is specified by the econometrician and depends on some unknown parameter $\delta$. This function is normalized such that $g(0,\delta) = 1$. For instance, in our empirical illustration, we take $Z$ as the dummy for participation in the program fostering soil-conservation. This is a binary treatment variable that takes value $1$ if the producer participates in the program and $0$ otherwise. This variable can affect both the production frontier and the inefficiency of the producer. To simplify the discussion that follows, we assume that the participation dummy is the only \textit{environmental factor}, so that $Z$ is univariate. This specific model can be easily generalized when $Z$ includes also other exogenous components. 
As $Z$ is binary, we can write the production frontier as
\[
m(X,Z,\beta) = m(X,\beta_0) + Z m(X,\beta_1),
\]
so that the frontier \textit{shifts} from $m(X,\beta_0)$ to $m(X,\beta_0) + m(X,\beta_1)$ as the treatment variable changes from $0$ to $1$. Therefore, our model in \eqref{eq:stmod} becomes
\[
Y = m(X,\beta_0) + Z m(X,\beta_1) + V - U_0 g(Z,\delta).
\]
For instance, in the prevalent case in which the logarithm of the production frontier is linear in parameters (e.g., Cobb-Douglas or translog), this modeling strategy involves the inclusion of the dummy variable for the treatment and the interaction between the treatment dummy and each one of the inputs (or a subset of these regressors) in the production frontier. This specification is in line with \cite{mccloud2008} in that it allows the treatment to affect output through the input coefficients (as a facilitating input that is not necessary for output production), the technological change parameter (generating a frontier shift), and the efficiency term.

Maximum likelihood estimation is a popular approach to obtain estimators of the parameters $(\beta,\delta)$ in a stochastic frontier framework \citep{kumbhakar2003}. Although heavily parametrized, the likelihood specification allows one to identify and estimate the variance of the inefficiency term. This in turn permits the construction of an estimator of TE, which captures the distance of each farmer from the production frontier. 

These maximum likelihood estimators can be based on a variety of assumptions about the distributions of $V$ and $U_0$. However, the most popular model assumes that $V$ follows a normal distribution and that $U_0$ follows a half-normal distribution \citep{aigner1977}. Moreover, one usually assumes that $V$ is independent of $U_0$ and that $(X,Z)$ are fully independent of $(V,U_0)$. 

In our framework, the treatment is not taken to be independent of the joint error term $(V,U_0)$. Volunteering for the treatment can depend both on the inefficiency of the producer and on other preferences that are unobserved to the econometrician. This implies that the treatment is endogenous and, therefore, the stochastic frontier model based on an independence assumption between $Z$ and $(V,U_0)$ would lead to an inconsistent estimator of $(\beta,\delta)$.\footnote{Production inputs can also be correlated with the composite error term \citep{mundlak1961,schmidt1984}. However, we focus here on the endogeneity of the treatment. Constructing an estimator that is also robust to endogeneity in the inputs is possible, although we do not tackle it in this paper \citep[see][]{centorrinoperez2019}.} Our goal is to construct a maximum likelihood estimator that generalizes the normal-half-normal stochastic frontier model when the treatment is allowed to be endogenous.

In econometrics, the use of IVs is a popular method to deal with endogeneity. That is, we assume there exists a vector of instruments, $W$, of dimension $q \geq 1$, which is correlated with $Z$ but independent of $(V,U_0)$. However, $Z$ enters the second-stage equation nonlinearly, so the usual two-stage least-squares approach used in linear IV models would not lead to a consistent estimation of $(\beta,\delta)$ \citep{wooldridge2009,amsler2016}. An alternative approach is based on a \textit{control function} assumption. That is, one can write $Z = \mathbbm{1} \left( \tilde{W} \gamma + \eta \geq 0\right)$, where $\tilde{W} = \left( W, X\right)$, and assume that all the dependence between $Z$ and $(V,U_0)$ is captured by $\eta$. The latter can be considered an omitted variable in the second stage. Thus, once we control for $\eta$, the dependence between $Z$ and $(V,U_0)$ disappears \citep{newey1999,newey2009}. 

In particular, we use a Probit specification to model the treatment assignment (i.e., the first stage equation). Thus, we have that 
\[
P\left( Z = 1 \vert \tilde{W}\right) = 1 - \Phi \left( -\tilde{W} \gamma \right),
\]
where $\Phi \left(\cdot \right)$ is the cdf of a standard normal distribution. The main assumptions of the Probit model are that $\eta \sim N(0,1)$ and that $\tilde{W}$ is independent of $\eta$. The testing procedure developed in \citet{wilde2008} can be used to validate this specification of the treatment assignment mechanism in empirical applications to increase the credibility of the proposed parametric restriction. A maximum likelihood framework further requires specification of the dependence between $\eta$ and $(V,U_0)$. More formally, this is done by modeling the conditional density of $(V,U_0)\vert\eta$. Consistently with the work of \citet{centorrinoperez2019}, we assume that any dependence between $V$ and $U_0$ has to happen through $\eta$. When there is no endogeneity, this assumption is equivalent to the full independence between $V$ and $U_0$ usually imposed in stochastic frontier models.

Our main assumptions are formalized as follows.
\begin{assumption} \label{ass:cindepeta}
\begin{itemize}
\item[(i)] $\tilde{W} \upmodels (V,\eta,U_0)$ and $V \upmodels U_0 \vert \eta$.
\item[(ii)] $Z = \mathbbm{1} \left( \tilde{W} \gamma + \eta \geq 0\right)$, with $\eta \sim N(0,1)$.
\item[(iii)] $V \vert \eta \sim N\left(\rho_V \sigma_V \eta,\sigma^2_V( 1- \rho_V^2) \right)$.
\item[(iv)] $U_0 \vert \eta \sim FN\left( \rho_U \sigma_U \eta,\sigma^2_U(1 - \rho_U^2)\right)$, where $FN$ denotes a folded normal distribution with location parameter $\rho_U \sigma_U \eta$ and scale parameter $\sigma^2_U(1 - \rho_U^2)$.
\end{itemize}
\end{assumption}
The parameters $\rho_V$ and $\rho_U$ capture the dependence between $(V,U_0)$ and $\eta$, respectively. In this respect, Assumptions \ref{ass:cindepeta}(iii)-(iv) seem natural as our model reduces to the special case of a normal-half-normal stochastic frontier specification when $\rho_V = \rho_U = 0$, i.e., when there is no endogeneity. The conditional pdf of $U_0 \vert \eta$ is written as
\begin{equation} \label{eq:densuclosed}
f_{U_0 \vert \eta}(u \vert \eta) = \frac{1}{\sqrt{2\pi (1 - \rho^2_U)\sigma_U^2}}\left\lbrace \exp \left( -\frac{(u - \rho_U \sigma_U \eta)^2}{2(1 - \rho^2_U)\sigma_U^2}\right) + \exp \left( -\frac{(u + \rho_U\sigma_U\eta)^2}{2(1 - \rho^2_U)\sigma_U^2}\right)\right\rbrace.
\end{equation}
\citet{centorrinoperez2019} have shown that this specification of the conditional density of $U_0$ provides a generalization with endogeneity to the normal half-normal stochastic frontier model \citep{aigner1977}. It can be seen that the pdf in Equation \eqref{eq:densuclosed} reduces to the half-normal distribution when $\rho_U = 0$, i.e., when the treatment is assigned independently of the efficiency of the producer. In the following, we refer to $\rho_U$ as a dependence parameter.

Maximum likelihood estimation in stochastic frontier models is usually based on the density of the composite error term $\varepsilon = V - U$. In the case where the treatment is endogenous, our maximum likelihood estimator is based on the joint density of $(\varepsilon,\eta)$, which can be decomposed into the product of the conditional density of $\varepsilon\vert\eta$ and the marginal density of $\eta$ which, in our case, is a standard normal density. 

From \citet{centorrinoperez2019}, the conditional density of $\varepsilon\vert\eta$ is equal to
\begin{align*}
f_{\varepsilon \vert \eta} (\varepsilon \vert \eta) =& \int f_{V\vert\eta} (\varepsilon + u \vert \eta ) \left( g(Z,\delta )\right)^{-1} f_{U_0\vert \eta } (\left( g(Z,\delta )\right)^{-1} u \vert \eta ) du \\
=& \frac{1}{\sqrt{2\pi} \tsigma(Z)} \left\lbrace \Phi \left( \frac{\lambda(Z) \rho_V \sigma_V \eta}{\tsigma(Z)} + \frac{\rho_U\sigma_U(Z)\eta}{\lambda(Z) \tsigma(Z)} -\frac{\lambda(Z) \varepsilon}{\tsigma(Z)}\right) \exp \left( -\frac{(\varepsilon - \rho_V \sigma_V\eta + \rho_U\sigma_U(Z) \eta)^2}{2 \tsigma^2(Z)}\right) \right.\\
& \quad + \left. \Phi \left( \frac{\lambda(Z) \rho_V \sigma_V \eta}{\tsigma(Z)} - \frac{\rho_U\sigma_U(Z)\eta}{\lambda(Z) \tsigma(Z)} -\frac{\lambda(Z) \varepsilon}{\tsigma(Z)}\right) \exp \left( - \frac{(\varepsilon - \rho_V \sigma_V\eta - \rho_U\sigma_U(Z) \eta)^2}{2 \tsigma^2(Z)}\right) \right\rbrace,
\end{align*}
where
\begin{align*}
\sigma^2_U(Z) =& \sigma^2_U \left( g \left( Z,\delta\right) \right)^2,\qquad \tsigma^2_U(Z) = (1 - \rho^2_U) \sigma^2_U(Z), \qquad \tsigma^2_V = (1 - \rho^2_V) \sigma^2_V,\\ \tsigma^2(Z) =& \tsigma^2_U(Z) + \tsigma^2_V, \qquad \lambda(Z) = \frac{\tsigma_U(Z)}{\tsigma_V}.
\end{align*}
\begin{remark}
The fact that $U = g(Z,\delta) U_0$, i.e., the \textit{scaling property} of $U$, is essential to derive the density of $\varepsilon\vert\eta$. Given that $U_0 \upmodels Z \vert \eta$, the scaling properties allows \citet{centorrinoperez2019} to write the density of $U \vert \eta$ as a scaled transformation of the density of $U_0\vert\eta$. 
\end{remark}
Finally, using our assumption that $\eta$ follows a standard normal distribution, the joint density of $(\varepsilon, \eta)$ can be written as
\begin{align*}
f_{\varepsilon,\eta} (\varepsilon ,\eta) =& \frac{1}{2\pi \tsigma(Z)} \left\lbrace \Phi \left( \frac{\lambda(Z) \rho_V \sigma_V \eta}{\tsigma(Z)} + \frac{\rho_U\sigma_U(Z)\eta}{\lambda(Z) \tsigma(Z)} -\frac{\lambda(Z) \varepsilon}{\tsigma(Z)}\right) \exp \left( -\frac{(\varepsilon - \rho_V \sigma_V\eta + \rho_U\sigma_U(Z) \eta)^2}{2 \sigma^2} - \frac{\eta^2}{2}\right) \right.\\
& \quad + \left. \Phi \left( \frac{\lambda(Z) \rho_V \sigma_V \eta}{\tsigma(Z)} - \frac{\rho_U\sigma_U(Z)\eta}{\lambda(Z) \tsigma(Z)} -\frac{\lambda(Z) \varepsilon}{\tsigma(Z)}\right) \exp \left( - \frac{(\varepsilon - \rho_V \sigma_V\eta - \rho_U\sigma_U(Z) \eta)^2}{2 \tsigma^2(Z)}- \frac{\eta^2}{2}\right) \right\rbrace.
\end{align*}

Let $\theta = (\beta^\prime,\delta^\prime,\sigma^2_U,\sigma^2_V,\rho_V,\rho_U,\gamma^\prime)^\prime$ be the vector of parameters of interest. When $Z$ is continuous, at least for identification purposes, we can assume that $\eta$ is observed and thus define the likelihood using the joint density of $\varepsilon$ and $\eta$ obtained above \citep{centorrinoperez2019}. When $Z$ is binary, this is not possible, as $\eta$ cannot be estimated from the data. We thus need to define the joint likelihood differently. 

In similar frameworks (e.g., Probit and Logit models), the observable random variable is discrete, and we usually express the likelihood (conditional on exogenous covariates) as the cdf of a latent error term which follows a known distribution. In our case, we have two observable endogenous variables $(Y,Z)$, and the likelihood is obtained by their density, conditional on the exogenous components, $\tilde{W}$. We aim at rewriting this density in terms of the error components $(\varepsilon,\eta)$. Therefore, as $\eta$ is latent, the likelihood is written with respect to its cdf. In particular, we aim at writing the likelihood as the product between the cdf of $\eta$ conditional on $\varepsilon$ and the pdf of $\varepsilon$. 

To this end, we first consider the following joint probability of the observable endogenous variables. For $Z = 0$, we have 
\begin{equation} \label{eq:likderivation1}
\begin{aligned}
P(Y \leq y ,Z = 0 \vert \tilde{W} = \tilde{w}) =& P( m(x,0,\beta) + \varepsilon \leq y,Z = 0 \vert \tilde{W}= \tilde{w}) \\
=& P(\varepsilon \leq y - m(x,0,\beta) ,\eta \leq -\tilde{w}\gamma ) = F_{\varepsilon,\eta} \left( y -m(x,0,\beta), -\tilde{w}\gamma\right), 
\end{aligned}
\end{equation}
where the second line follows from the assumption of independence between $(\varepsilon,\eta)$ and $\tilde{W}$. A similar derivation holds when $Z = 1$. 

If we take the derivative of the joint probability in equation \eqref{eq:likderivation1} with respect to its first argument, we obtain a function which is a pdf with respect to $\varepsilon$ and a cdf with respect to $\eta$. In particular, we have
\begin{align*}
\partial_1 F_{\varepsilon,\eta} \left( y - m(x,0,\beta), -\tilde{w}\gamma\right) =& \int_{-\infty}^{-\tilde{w}\gamma} f_{\varepsilon \vert \eta } (y - m(x,0,\beta), \eta ) d\eta\\
=&\int_{-\infty}^{-\tilde{w}\gamma} f_{\varepsilon \vert \eta } (y - m(x,0,\beta) \vert \eta ) \phi(\eta) d\eta,
\end{align*}
where the second line follows from Assumption \ref{ass:cindepeta}, and $\phi$ is the pdf of a standard normal distribution.

The likelihood function can thus be obtained as
\begin{equation} \label{eq:fullloglik}
\mathcal{L}(\theta) = \left( \int_{-\tilde{W}\gamma}^\infty f_{\varepsilon \vert \eta } (\varepsilon(\theta) \vert \eta ) \phi(\eta) d\eta \right)^{Z} \left( \int_{-\infty}^{-\tilde{W}\gamma} f_{\varepsilon \vert \eta} (\varepsilon(\theta) \vert \eta ) \phi(\eta) d\eta \right)^{1 - Z},
\end{equation}
where $\theta$ is defined above, and $\varepsilon(\theta) = Y - m(X,Z,\beta)$. 

The integrals appearing in the likelihood function can be solved analytically. In particular, we obtain that the conditional cdf of $\eta \vert \varepsilon$ is a mixture of two conditional skew-normal distributions \citep{azzalini1996,azzalini2013}. For $j = \lbrace 1,2 \rbrace$, we let
\begin{align*}
\Psi_{0,j}(Z,\theta) =&\Phi_2\left(\frac{-\tilde{W} \gamma - \mu_{\eta,j}(Z)\varepsilon(\theta)}{\sigma_{\eta,j}(Z)},\tau_j(Z)\varepsilon(\theta); \rho^\ast(Z)\right),\\
\Psi_{1,j}(Z,\theta) =&\Phi\left( \tau_j(Z)\varepsilon(\theta) \right) - \Phi_2\left(\frac{-\tilde{W} \gamma - \mu_{\eta,j}(Z)\varepsilon(\theta)}{\sigma_{\eta,j}(Z)},\tau_j(Z)\varepsilon(\theta); \rho^\ast(Z)\right),
\end{align*}
where $$\Phi_2(a,b;\rho^\ast) = \int_{-\infty}^a \int_{-\infty}^b \frac{1}{2\pi\sqrt{1 - \rho^{\ast 2}}} \exp \left( -\frac{(\varpi^2_1 - 2\rho^\ast \varpi_1 \varpi_2 + \varpi_2^2)}{2(1 - \rho^{\ast 2})}\right) d\varpi_1 d\varpi_2,$$ \sloppy is the cdf of a bivariate normal random variable with correlation parameter $\rho^\ast$, and $\lbrace \mu_{\eta,j}(Z),\sigma_{\eta,j}(Z),\tau_j(Z),\rho^\ast(Z),\sigma^2_{\varepsilon,j}(Z)\rbrace$ are functions of the parameter $\theta$, whose dependence is suppressed for simplicity. Finally, the likelihood function is 
\begin{align} \label{eq:fullloglik2}
\mathcal{L}(\theta) =& \left( \sum_{j = 1,2} \Psi_{1,j}(Z,\theta) \frac{1}{\sigma_{\varepsilon,j}(Z)} \phi \left(\frac{\varepsilon(\theta)}{\sigma_{\varepsilon,j}(Z)}\right) \right)^Z \left( \sum_{j = 1,2} \Psi_{0,j}(Z,\theta) \frac{1}{\sigma_{\varepsilon,j}(Z)} \phi \left(\frac{ \varepsilon(\theta)}{\sigma_{\varepsilon,j}(Z)}\right) \right)^{1 - Z}.
\end{align}
A detailed derivation is provided in Appendix \ref{sec:appAlikder}. When $\rho_V=\rho_U=0$, 
\begin{align*}
\sigma^2_{\varepsilon,1}(Z) =& \sigma^2_{\varepsilon,2}(Z) = \sigma^2_V + \sigma^2_U(Z)\\
\Psi_{1,1}(Z,\theta) =&  \Psi_{1,2}(Z,\theta) = 1 - \Phi(-\tilde{W} \gamma)\\
\Psi_{0,1}(Z,\theta) =&  \Psi_{0,2}(Z,\theta) = \Phi(-\tilde{W} \gamma),
\end{align*}
and the likelihood reduces to the product of the pdf of a skew-normal distribution (the pdf of $\varepsilon$) and the cdf of a normal distribution (the cdf of $\eta$), which would be the likelihood function if $Z \upmodels (V,U_0)$. This would be the standard approach in SFA \citep{kumbhakar2003}. Also, if $\rho_V= 0$, and we assume that $\sigma^2_U(Z)$ is constant wrt $Z$, we can write $\eta_i = e_i - \rho^2_U U_{0i}/(\sigma^2_V + \sigma^2_U)$, where $e_i \sim N(0,1)$, and our model will collapse to the one proposed by \citet{Kumbhakar2009}.

\subsection{Identification}

Let $\ell(\theta) = \log \mathcal{L}(\theta)$ be the log-likelihood function, and assume that $E \left[ \vert \ell(\theta) \vert \right] < \infty$ for all $\theta \in \Theta$. As we can restrict $\Theta$ to be a compact parameter space, and the likelihood function is continuous in $\theta$, there exists at least one solution to the maximization of the log-likelihood function \citep{gourieroux1990}.

We focus our identification analysis on the parameter $\rho_U$. To this end, we maintain the following assumption.
\begin{assumption} \label{ass:sigmau0pos}
Let $\theta_1 = (\beta^\prime,\delta^\prime,\sigma^2_U,\sigma^2_V,\rho_V,\gamma^\prime)^\prime$. The matrix
\[
E \left[ \nabla_{\theta_{1}\theta^\prime_{1}}^2 \ell(\theta_0) \right]
\]
is negative definite and has full rank. 
\end{assumption}

This assumption imposes that the parameter $\theta_1$ is first-order locally identified \citep{sargan1983}. In particular, we require that the variance of the inefficiency term $\sigma^2_{U,0} > 0$. \citet{lee1986} and \citet{lee1993} have shown that when $\sigma^2_{U,0} = 0$, the stochastic frontier model is not first-order identified. Moreover, in our model, whenever $\sigma^2_U = 0$, $(\delta,\rho_U)$ are not identified. We believe this case is worthy of future investigation, but we rule it out here for simplicity.

\begin{proposition} \label{prop:identif1}
Let Assumptions \ref{ass:cindepeta}-\ref{ass:sigmau0pos} hold, and $\rho_{U,0}$ to be such that
\[
E \left[ \frac{\partial \ell(\theta_{1,0},\rho_{U,0})}{\partial \rho_U} \right] = 0.
\]
We have that
\begin{itemize}
\item[(i)] $E \left[ \frac{\partial \ell(\theta_{1,0},-\rho_{U,0})}{\partial \rho_U}\right] = 0$.
\item[(ii)] $\frac{\partial \ell(\theta_{1},0)}{\partial \rho_U} = 0$, for any $\theta_1$, and 
\[
E \left[ \frac{\partial^2 \ell(\theta_{1,0},0)}{\partial \rho_U^2} \right] =0. 
\]
\end{itemize}
\end{proposition}

This Proposition extends the result shown in \citet{centorrinoperez2019} to the case in which the endogenous variable is binary. Part (i) states that if $\rho_{U,0}$ is a solution of the maximization problem, so is $-\rho_{U,0}$. That is, the sign of $\rho_U$ is not identified. Part (ii) states that $\rho_{U} = 0$ always satisfies the first order conditions of the maximization problem for any value of $\theta_1$. This result implies that the matrix of second derivatives has rank equal to $dim(\theta) - 1$, and the model is not first-order identified at $\rho_U = 0$. A proof of this Proposition is provided in Appendix \ref{pr:prop21}.\footnote{A similar identification problem arises in Zero Inefficiency Stochastic Frontier models, see \citet{kumbhakar2013,rho2015}.}

This identification issue is illustrated in \citet{centorrinoperez2019}, and it follows because the likelihood function is an even function of $\rho_U$ and symmetric about $\rho_U=0$. In our setting, dealing with this identification issue is easier, compared to the framework in \citet{centorrinoperez2019}, as the parameter $\rho_U$ is scalar.

We thus restrict the support of $\rho_U$ to be $[0,1]$ \citep{sundberg1974}. We let $\bar{\Theta}$ to be the parameter's space which embeds the restriction on $\rho_{U}$, and we define
\begin{equation} \label{eq:truepar}
\theta_0 = \argmax_{\theta \in \bar{\Theta}} E \left[ \ell(\theta) \right],
\end{equation}
which exists and is (locally) unique under Assumption \ref{ass:sigmau0pos}.

\subsection{Estimation and Inference} \label{sec:est}

For estimation, we consider an iid sample from the joint distribution of $(Y,X,Z,W)$, which we denote $\lbrace (Y_i,X_i,Z_i,W_i), i = 1,\dots,n \rbrace$, where each observation obeys to the model in \eqref{eq:stmod}.

Let $\ell_n(\theta) = \log\left( \mathcal{L}_n(\theta) \right)$, with
\begin{align} \label{eq:likestimation}
\mathcal{L}_n(\theta) =& \prod_{i = 1}^n \left( \sum_{j = 1,2} \Psi_{1,j}(\theta,Z_i) \frac{1}{\sigma_{\varepsilon,j}(Z_i)} \phi \left(\frac{\varepsilon_i(\theta)}{\sigma_{\varepsilon,j}(Z_i)}\right) \right)^{Z_i}\left( \sum_{j = 1,2} \Psi_{0,j}(\theta,Z_i) \frac{1}{\sigma_{\varepsilon,j}(Z_i)} \phi \left(\frac{\varepsilon_i(\theta)}{\sigma_{\varepsilon,j}(Z_i)}\right) \right)^{1 - Z_i} .
\end{align}
Estimation is straightforward, with the maximum likelihood estimator of the parameter $\theta$ given by 
\[
\hat\theta_n = \argmax_{\theta \in \Theta} \ell_n(\theta),
\]

We analyze our estimator's asymptotic distribution depending on the true value of the parameter $\rho_U$. To simplify our analysis, we make the following high-level assumption.

\begin{assumption} \label{ass:consistencytheta}
$\hat{\theta}_{n} \xrightarrow{p} \theta_{0}$.
\end{assumption}

Under Assumptions \ref{ass:sigmau0pos} and \ref{ass:consistencytheta} and since the log-likelihood function is at least twice continuously differentiable with respect to the parameter $\theta_{0}$, when $\rho_U$ is strictly in the interior of $[0,1]$, standard theory of maximum likelihood estimation applies and we can claim that 
\[
\sqrt{n} \left( \hat{\theta}_{n} - \theta_{0} \right) \xrightarrow{d} N\left(0,\mathcal{I}_{\theta_{0}}^{-1} \right),
\]
where $\mathcal{I}_{\theta_{0}}$ is the Fisher's information matrix. In finite samples, we use a numerical approximation of the first derivative of the likelihood function at $\hat{\theta}_{n}$, and Fisher's information identity to obtain an estimator of the information matrix.

However, the asymptotic distribution and the rate of convergence are non-standard when $\rho_{U,0}$ is equal to $0$. In this case, it follows from the result of Proposition \ref{prop:identif1} that we have a singular Hessian matrix, and one of the parameters of interest is at the boundary of the parameter space. This implies that we do not have the standard $\sqrt{n}$-rate of convergence, and that $\hat{\rho}_U$ is not asymptotically normal \citep{sundberg1974b,andrews1999,rot2000}. However, a reparametrization of the log-likelihood function allows us to obtain some asymptotic results.

Let $\rho_{2,U} = \rho_U^2$. The following theorem gives the asymptotic distribution of our estimator when $\rho_{U,0} = 0$.

% HERE YOU JUST HAVE TO BORROW THE NOTATIONS AND THEOREMS OF ANDREWS (1999) AND SPECIALIZE THEM TO THIS CASE.
\begin{theorem}\label{thm:asnorm}
Let Assumptions \ref{ass:cindepeta}-\ref{ass:sigmau0pos} hold with $\rho_{U,0} = 0$, and $(Z_{\theta_1},Z_{\rho_{2,U}})$ a normal random vector such that $dim(Z_{\theta_1}) = dim(\theta_{1})$, $dim(Z_{\rho_{2,U}}) = 1$, with covariance matrix $\mathcal{I}_1^{-1}$, where $\mathcal{I}_1^{-1}$ is the inverse of
\[
\mathcal{I}_1 = \begin{bmatrix} \mathcal{I}_{\theta_1} & \mathcal{I}^\prime_{\theta_1\rho_{2,U}} \\ \mathcal{I}_{\theta_1\rho_{2,U}} & \mathcal{I}_{\rho_{2,U}} \end{bmatrix}.
\]
Further let $\hat\tau_{\rho_{2,U}} = \max \lbrace Z_{\rho_{2,U}},0\rbrace$.Then
\begin{itemize}
\item[(i)]
\[
\sqrt{n}\begin{pmatrix} \hat{\theta}_1 - \theta_{1,0}\\ \hat\rho_{2,U} \end{pmatrix} \xrightarrow{d} \begin{pmatrix} Z_{\theta_1} - \mathcal{I}_{\theta_1}^{-1} \mathcal{I}^\prime_{\theta_1\rho_{2,U}} \hat\tau_{\rho_{2,U}} \\ \hat\tau_{\rho_{2,U}} \end{pmatrix} 
\]
\item[(ii)] $n^{1/4} \hat\rho_U = O_P(1)$. 
\end{itemize}
\end{theorem}

This theorem reflects existing results in statistics about estimation of a parameter that is not first-order identified \citep{rot2000}. As the order of identification in our model is even, the marginal asymptotic distribution of $\hat\rho_{2,U}$ at $\rho_{U,0} = 0$ is an equal mixture of a point-mass at $0$, and a half-normal distribution \citep{chernoff1954,andrews1999}. The result in Part (ii) is a direct consequence of Part (i). However, it is worth highlighting that our estimator has rates of convergence slower than $\sqrt{n}$, and may not be asymptotically normal, depending on the true value of the parameter $\rho_U$. These results have important implications for conducting inference and constructing confidence intervals for $\rho_U$. 

In particular, an important hypothesis to be tested is whether $Z \upmodels U_0$, i.e., $\rho_U =0$. The trinity of tests is an obvious candidate, but the implementation of these tests is not straightforward because of the non-standard asymptotic properties of the estimator of $\rho_U$. 

\citet{andrews2001} studies the properties of the trinity of tests when some parameters are at the boundary. His theoretical results about the Likelihood Ratio (LR) test can be used following Theorem \ref{thm:asnorm}. Let $\tilde{\theta}_n$ be the estimator of $\theta_0$ under the restriction $\rho_{U,0} = 0$. The LR test statistic can be computed as $LR = 2(\ell_n(\hat\theta_n)-\ell_n(\tilde\theta_n))$. Since $\rho_U = 0$, implies $\rho_{2,U} = 0$, and $\rho_{2,U}$ is a scalar parameter, under the null the LR test has an asymptotic distribution which is an equal mixture between a mass point at $0$ and a $\chi^2$ with one degree of freedom \citep[see][]{self1987,andrews1999,rot2000,andrews2001}. One can then use quantiles from the asymptotic distribution as critical values for the LR test.

Obtaining uniformly valid standard errors and confidence intervals is more cumbersome. Our results suggest that one can first construct a test of the null hypothesis that $\rho_U = 0$. If the null is rejected, we are then able to use standard errors and confidence intervals based on the normal approximation. If the null cannot be rejected, then other methods should be used to obtain confidence intervals. It is unclear whether this approach generates pre-testing bias as in \citet{guggen2010}. This is issue is potentially worthy of future investigation.

\citet{bottai2003} has shown that, when there is a singularity in the Fisher Information matrix, confidence intervals based on inverting the Likelihood Ratio or Wald test fail to have nominal coverage near the point of singularity. However, an appropriately modified version of the score test can be used to obtain confidence intervals with asymptotically uniform nominal coverage. This theoretical result holds, however, only when $\theta$ is a scalar parameter. More recently, \citet{ekvall2022} have provided a generalization for any multivariate parameter $\theta$ and when the rank deficiency of the information matrix is potentially larger than one. This modified version of the score test in a model, such as ours, that is second-order identified, is constructed using the second derivative of the likelihood function at the point of singularity, normalized by the expected value of its square. This methodology could be potentially extended to the model we study in this paper to construct uniformly valid confidence intervals when $\rho_{U,0} = 0$. A formal theoretical result is deferred to future research.

%%%% TECHNICAL EFFICIENCY
\subsection{Technical Efficiency} To complete our framework, we obtain a feasible estimator of technical efficiency, $TE_i = \exp(-U_i)$. Researchers are often interested in obtaining the technical efficiency for each producer. However, in SFA, one can only obtain an estimate of composite error term, $\varepsilon_i$, so that $TE$ is predicted using $E\left[ \exp(-U_i) \vert \varepsilon_i\right]$ \citep{battese1988}. \citet{amsler2017} and \citet{centorrinoperez2019} obtain an estimator of this quantity from the conditional distribution of $U \vert \varepsilon$ and $\eta$. However, as $\eta$ is not observed in our case, we have to follow the standard approach and obtain the estimator of technical inefficiency from the conditional distribution of $U \vert \varepsilon$. The latter distribution can be derived as
\[
f_{U \vert \varepsilon}(u \vert \varepsilon) = \int f_{U \vert \varepsilon,\eta }(u \vert \varepsilon,\eta) f_{\eta \vert \varepsilon}(\eta \vert \varepsilon)d\eta.
\]

Details about the exact computations of this density are given in Appendix \ref{sec:appAlikder}. We let
\begin{align*}
\sigma_{1\star} =& \frac{\tilde\sigma_V \tilde\sigma_U(Z)}{\sigma(Z)} \sqrt{ 1 + \frac{q_1^2(Z) \tsigma^2(Z)}{\tsigma^2(Z) + \rho^2_1(Z)}} \\
\sigma_{2\star} =& \frac{\tilde\sigma_V \tilde\sigma_U(Z)}{\sigma(Z)} \sqrt{ 1 + \frac{q_2^2(Z) \tsigma^2(Z)}{\tsigma^2(Z) + \rho^2_2(Z)}} \\
\mu_{1\star} =& -\frac{\tilde\sigma_V \tilde\sigma_U(Z)}{\sigma(Z)}\varepsilon \left( \frac{\lambda(Z)}{\sigma(Z)} - \frac{q_1(Z) \rho_1(Z)}{\tsigma^2(Z) + \rho^2_1(Z)} \right) \\
\mu_{2\star} =& -\frac{\tilde\sigma_V \tilde\sigma_U(Z)}{\sigma(Z)}\varepsilon \left( \frac{\lambda(Z)}{\sigma(Z)} - \frac{q_2(Z) \rho_2(Z)}{\tsigma^2(Z) + \rho^2_2(Z)} \right),
\end{align*}
where the definition of the other parameters is given in the Appendix, and the dependence on the variable $Z$ has been removed for simplicity. 

We obtain that 
\begin{align}
E \left[ \exp(-U) \vert \varepsilon \right] =& \omega_1(\varepsilon) \exp \left(-\mu_{1\star} + \frac{\sigma^2_{1\star}}{2}\right)\frac{1 - \Phi \left( - \frac{\mu_{1\star}}{\sigma_{1\star}} + \sigma_{1\star}\right)}{\Phi\left( \tau_1(Z) \varepsilon \right)} \nonumber \\
& \quad + \omega_2(\varepsilon) \exp \left(-\mu_{2\star} + \frac{\sigma^2_{2\star}}{2}\right)\frac{1 - \Phi \left( - \frac{\mu_{2\star}}{\sigma_{2\star}} + \sigma_{2\star}\right)}{\Phi\left( \tau_2(Z) \varepsilon \right)} \label{eq:battesecoelli},
\end{align}
where $\omega_l(\varepsilon)$, with $l = 1,2$ are weights, such that $\omega_1(\varepsilon) + \omega_2(\varepsilon) = 1$, whose closed-form expression is given in the Appendix. Finally, the mean technical efficiency \citep{lee1978} can be obtained as
\[
E \left[ \exp(-U) \right] = E \left[E \left[ \exp(-U) \vert \varepsilon \right] \right],
\]
by the law of iterated expectations.

\section{Simulations} \label{sec:montecarlo}

We replicate a similar simulation scheme as in \citet{amsler2017} and \citet{centorrinoperez2019}. We consider the following model
\[
Y_i = \beta_0 + X_{1i} \beta_1 + X_{2i} \beta_2 + X_{1i}Z_{2i}\beta_{1,1} + X_{2i} Z_{2i} \beta_{1,2} + V_i - U_{0i} \exp\left( Z_{1i} \delta_1 + Z_{2i} \delta_2\right),
\]
with $\beta_0 =0$ and $\beta_1 = \beta_2 = \beta_{1,1} = \beta_{1,2} =0.41359$, $\delta_1 = 0$ and $\delta_2 =0$ and where the random variables $(X_{1},X_{2},Z_{1}) \upmodels (V,U_{0})$, and $Z_{2}$ is our endogenous treatment variable. We consider two instruments $(W_{1},W_{2}) \upmodels (V,U_{0})$, such that
\[
Z_{2i} = \mathbbm{1}\left( \gamma_0 + \gamma_1 X_{1i} + \gamma_2 X_{2i} + \gamma_3 Z_{1i} + \gamma_4 W_{1i}+ \gamma_5 W_{2i} + \eta \geq 0\right),
\]
where $\mathbbm{1}\left( \cdot \right)$ is the indicator function, $\gamma_0 = -0.1$, $\gamma_1= \gamma_2 = \gamma_3 = \gamma_5 = 0.31623$, and $\gamma_6 = 1$.

The exogenous variables $(X_1,X_2,Z_1,W_1,R)$ are generated from a joint normal distribution with mean equal to $0$ and covariance matrix with diagonal elements equal to $1$, and off-diagonal elements equal to $0.5$. $W_2 = \mathbbm{1}(R > 0.5)$. $(V,\eta)$ are sampled from a standard bivariate normal distribution with $\rho_V = 0.5$. The stochastic inefficiency term is generated as $U_0 = \sigma_U \vert \rho_U\eta + \sqrt{1 - \rho_U^2} \epsilon \vert$, where $\sigma^2_U = \pi/(\pi-2)$, and $\epsilon \sim N(0,1)$.

We consider three simulation schemes corresponding to alternative values $\rho_U$. In Scheme 1 (S1), $U_0\upmodels \eta$ (i.e. $\rho_U = 0$). In Scheme 2 (S2), $\rho_U = 0.5$. In Scheme 3 (S3), $\rho_U = 0.95$. Sample sizes are fixed to $n =\lbrace 250,500,1000 \rbrace$, and we run $1000$ replications for each scenario.

The implementation of our estimator is rather straightforward. Let $m(X_i,Z_{2i},\beta) = \beta_0 + X_{1i} \beta_1 + X_{2i} \beta_2 + X_{1i}Z_{2i}\beta_{1,1} + X_{2i} Z_{2i} \beta_{1,2}$, the production frontier. The starting point is the log-likelihood function from equation \eqref{eq:likestimation}
\begin{align*}
\ell_n(\theta) =&\sum_{i = 1}^n \left[ Z_{2i} \log \left( \sum_{j = 1,2} \Psi_{1,j}(Z_i,\theta) \frac{1}{\sigma_{\varepsilon,j}(Z_i)} \phi \left(\frac{Y_i - m(X_i,Z_{2i},\beta) }{\sigma_{\varepsilon,j}(Z_i)}\right) \right) \right. \\
&\quad + (1 - Z_{2i})\left. \log\left( \sum_{j = 1,2} \Psi_{0,j}(Z_i,\theta) \frac{1}{\sigma_{\varepsilon,j}(Z_i)} \phi \left(\frac{ Y_i - m(X_i,Z_{2i},\beta)}{\sigma_{\varepsilon,j}(Z_i)}\right) \right) \right],
\end{align*}
which is maximized numerically with respect to the parameter $\theta$. 

An essential step of numerical optimization procedures is to select a starting value. As far as the parameters $(\beta,\delta)$ are concerned, an initial value can be chosen by estimating a stochastic frontier model which does not account for endogeneity. Similarly, an initial value for the parameter $\gamma$ can be obtained by a Probit regression of $Z_{2}$ on the instruments and all the other exogenous variables included in the model. Regarding the parameters $(\rho_V,\rho_U)$, their starting values are taken by uniform draws from the interval $[-1,1]\times [0,1]$. As convergence to a local maximum might be an issue, we draw several random points around the starting value and initialize the search at every one of these points. While computationally expensive, this procedure is more robust to local maxima.

In parametric models with endogeneity, it is also common practice to perform estimation in two steps: first one obtains an estimator of the parameter $\gamma$ from a Probit model, and then one estimates the remaining parameters by holding $\hat{\gamma}$ fixed. While this procedure is computationally more efficient, as it reduces the dimension of the parameters' space, we recommend against it. First of all, there is no guarantee that this procedure yields a numerically equivalent estimator to the maximization of the log-likelihood with respect to the full parameters' space. Moreover, standard errors of the two-step procedure are invalid, and one still needs to obtain the numerical Hessian matrix of the full likelihood to obtain valid standard errors. Our implementation is thus based on the maximization of the log-likelihood with respect to the full parameter vector, $\theta$.

\begin{table}[!h]
\centering
\scriptsize
\begin{tabular}{l | c c c | c c c | c c c} \hline \hline
~ & \multicolumn{3}{c}{$S1$} & \multicolumn{3}{c}{$S2$} & \multicolumn{3}{c}{$S3$}\\ \hline
$n$ & $250$ & $500$ & $1000$ &  $250$ & $500$ & $1000$ & $250$ & $500$ & $1000$\\
% latex table generated in R 4.3.2 by xtable 1.8-4 package
% Thu Dec 21 09:15:28 2023
  \hline
$\beta_0$ & -0.1122 & -0.0553 & -0.0152 & -0.1061 & -0.0514 & -0.0153 & -0.0314 & -0.0216 & -0.0088 \\ 
   & (0.3528) & (0.2188) & (0.1247) & (0.3369) & (0.2028) & (0.1222) & (0.2274) & (0.1561) & (0.1141) \\ 
  $\beta_1$ & 0.0287 & 0.0182 & 0.0114 & 0.0110 & 0.0103 & 0.0044 & -0.0247 & -0.0049 & -0.0011 \\ 
   & (0.1649) & (0.1129) & (0.0787) & (0.1682) & (0.1174) & (0.0807) & (0.1621) & (0.1194) & (0.0764) \\ 
  $\beta_2$ & 0.0343 & 0.0088 & 0.0035 & 0.0231 & -0.0003 & -0.0011 & -0.0160 & -0.0112 & -0.0021 \\ 
   & (0.1694) & (0.1130) & (0.0754) & (0.1745) & (0.1162) & (0.0795) & (0.1678) & (0.1140) & (0.0778) \\ 
  $\beta_{1,1}$ & -0.0259 & -0.0169 & -0.0154 & -0.0102 & -0.0072 & -0.0040 & 0.0206 & 0.0009 & 0.0013 \\ 
   & (0.2169) & (0.1456) & (0.1068) & (0.2195) & (0.1527) & (0.1060) & (0.2026) & (0.1504) & (0.0997) \\ 
  $\beta_{1,2}$ & -0.0381 & -0.0116 & -0.0045 & -0.0284 & 0.0011 & 0.0015 & 0.0066 & 0.0084 & -0.0009 \\ 
   & (0.2109) & (0.1455) & (0.1033) & (0.2143) & (0.1445) & (0.1040) & (0.2076) & (0.1405) & (0.1007) \\ 
  $\delta_{1}$ & -0.0144 & -0.0083 & -0.0021 & -0.0167 & -0.0052 & 0.0002 & 0.0027 & 0.0014 & 0.0018 \\ 
   & (0.2460) & (0.0738) & (0.0402) & (0.2414) & (0.0725) & (0.0383) & (0.0891) & (0.0564) & (0.0366) \\ 
  $\delta_{2}$ & 0.3431 & 0.0078 & 0.0335 & 0.0253 & 0.0397 & 0.0086 & -0.0961 & -0.0434 & -0.0086 \\ 
   & (2.5521) & (1.5419) & (0.4257) & (2.5897) & (0.2663) & (0.1302) & (0.4743) & (0.2317) & (0.1516) \\ 
  $\sigma^2_{U}$ & -0.4698 & -0.2638 & -0.1093 & -0.3479 & -0.1709 & -0.0478 & 0.3203 & 0.1421 & 0.0368 \\ 
   & (1.2665) & (0.8663) & (0.5477) & (1.3266) & (0.9402) & (0.6179) & (1.4489) & (0.9452) & (0.7047) \\ 
  $\sigma^2_{V}$ & 0.1288 & 0.0582 & 0.0191 & 0.1234 & 0.0477 & 0.0103 & -0.0165 & -0.0123 & -0.0003 \\ 
   & (0.4174) & (0.2658) & (0.1545) & (0.4307) & (0.2730) & (0.1644) & (0.3399) & (0.2325) & (0.1773) \\ 
  $\rho_{U,\eta}$ & 0.2290 & 0.0023 & 0.0035 & -0.0198 & -0.0372 & -0.0150 & -0.0041 & 0.0025 & 0.0062 \\ 
   & (0.3119) & (0.2273) & (0.1659) & (0.3167) & (0.2409) & (0.1477) & (0.0901) & (0.0520) & (0.0346) \\ 
  $\rho_{V,\eta}$ & 0.0510 & 0.0279 & 0.0132 & 0.0179 & 0.0117 & 0.0019 & -0.0887 & -0.0461 & -0.0186 \\ 
   & (0.1891) & (0.1268) & (0.0817) & (0.2295) & (0.1478) & (0.0936) & (0.2914) & (0.1875) & (0.1135) \\ 
  $\gamma_{0}$ & 0.0002 & 0.0005 & -0.0007 & -0.0007 & -0.0002 & -0.0010 & -0.0029 & -0.0004 & -0.0017 \\ 
   & (0.1112) & (0.0783) & (0.0550) & (0.1124) & (0.0786) & (0.0547) & (0.1082) & (0.0738) & (0.0496) \\ 
  $\gamma_{1}$ & 0.0064 & 0.0066 & 0.0016 & 0.0060 & 0.0077 & 0.0014 & 0.0063 & 0.0054 & 0.0015 \\ 
   & (0.1389) & (0.0926) & (0.0670) & (0.1392) & (0.0930) & (0.0671) & (0.1210) & (0.0821) & (0.0590) \\ 
  $\gamma_{2}$ & 0.0206 & 0.0043 & 0.0052 & 0.0198 & 0.0041 & 0.0050 & 0.0125 & 0.0045 & 0.0025 \\ 
   & (0.1404) & (0.0916) & (0.0653) & (0.1402) & (0.0917) & (0.0649) & (0.1233) & (0.0788) & (0.0567) \\ 
  $\gamma_{3}$ & 0.0083 & 0.0050 & 0.0062 & 0.0073 & 0.0052 & 0.0054 & 0.0090 & 0.0031 & 0.0037 \\ 
   & (0.1358) & (0.0910) & (0.0652) & (0.1348) & (0.0901) & (0.0645) & (0.1194) & (0.0800) & (0.0555) \\ 
  $\gamma_{4}$ & 0.0138 & 0.0053 & 0.0028 & 0.0158 & 0.0046 & 0.0042 & 0.0152 & 0.0058 & 0.0026 \\ 
   & (0.1306) & (0.0909) & (0.0611) & (0.1307) & (0.0910) & (0.0605) & (0.1161) & (0.0793) & (0.0524) \\ 
  $\gamma_{5}$ & 0.0272 & 0.0126 & 0.0114 & 0.0372 & 0.0173 & 0.0150 & 0.0534 & 0.0225 & 0.0185 \\ 
   & (0.2498) & (0.1789) & (0.1220) & (0.2516) & (0.1773) & (0.1218) & (0.2257) & (0.1489) & (0.1038) \\ 
   \hline

\hline \hline
\end{tabular}
\caption{\textit{Bias and Standard Deviation (in parenthesis) of estimated parameters for all simulation schemes.}}
\label{table:mc1}
\end{table}

We report results of these simulations in Table \ref{table:mc1}. For each parameter, we report the bias and standard deviation (in parenthesis) of the estimator computed over the simulated samples. Our estimator behaves as expected, with a smaller finite-sample bias in S2 and S3. This is due to the slower rate of convergence of $\hat{\rho}_U$ is S1, which may affect the estimation of the other parameters of the model. 

Finally, we report summary statistics for our estimators of technical efficiency using the Battese-Coelli formula provided in Equation \ref{eq:battesecoelli}. To give a reference point to the reader, in all simulation schemes the mean technical efficiency is equal to 
\[
E\left[ \exp(-U) \right] = 0.3847.
\]

\begin{table}[!h]
\centering
\scriptsize
\begin{tabular}{l | c c c | c c c | c c c} 
\hline \hline
~ & \multicolumn{3}{c}{$S1$} & \multicolumn{3}{c}{$S2$} & \multicolumn{3}{c}{$S3$}\\ \hline
$n$ & $250$ & $500$ & $1000$ & $250$ & $500$ & $1000$  & $250$ & $500$ & $1000$ \\
% latex table generated in R 4.3.2 by xtable 1.8-4 package
% Thu Dec 21 09:15:29 2023
  \hline
Min. & 0.000 & 0.001 & 0.001 & 0.000 & 0.001 & 0.000 & 0.000 & 0.000 & 0.000 \\ 
  1st Qu. & 0.285 & 0.268 & 0.257 & 0.283 & 0.266 & 0.255 & 0.256 & 0.254 & 0.251 \\ 
  Median & 0.450 & 0.420 & 0.405 & 0.449 & 0.420 & 0.405 & 0.423 & 0.411 & 0.405 \\ 
  Mean & 0.439 & 0.406 & 0.391 & 0.434 & 0.402 & 0.390 & 0.402 & 0.392 & 0.389 \\ 
  3rd Qu. & 0.579 & 0.545 & 0.530 & 0.576 & 0.544 & 0.530 & 0.551 & 0.536 & 0.531 \\ 
  Max. & 1.000 & 1.000 & 1.000 & 1.000 & 0.990 & 0.833 & 0.993 & 0.855 & 0.831 \\ 
   \hline

\hline \hline 
\end{tabular}
\caption{\textit{Summary measures for the estimator of TE}}
\label{table:techeff}
\end{table}

\section{Soil Conservation in El Salvador} \label{sec:empiricalapplication}

\subsection{Data and Model Specification} We consider data from the \textit{Programa Ambiental de El Salvador} or PAES, an environmental program promoting crop diversification and soil conservation practices. The data set consists of a sample of PAES participants and a control group of nonparticipating farmers.

\begin{figure}[!ht]
\includegraphics[scale=0.3]{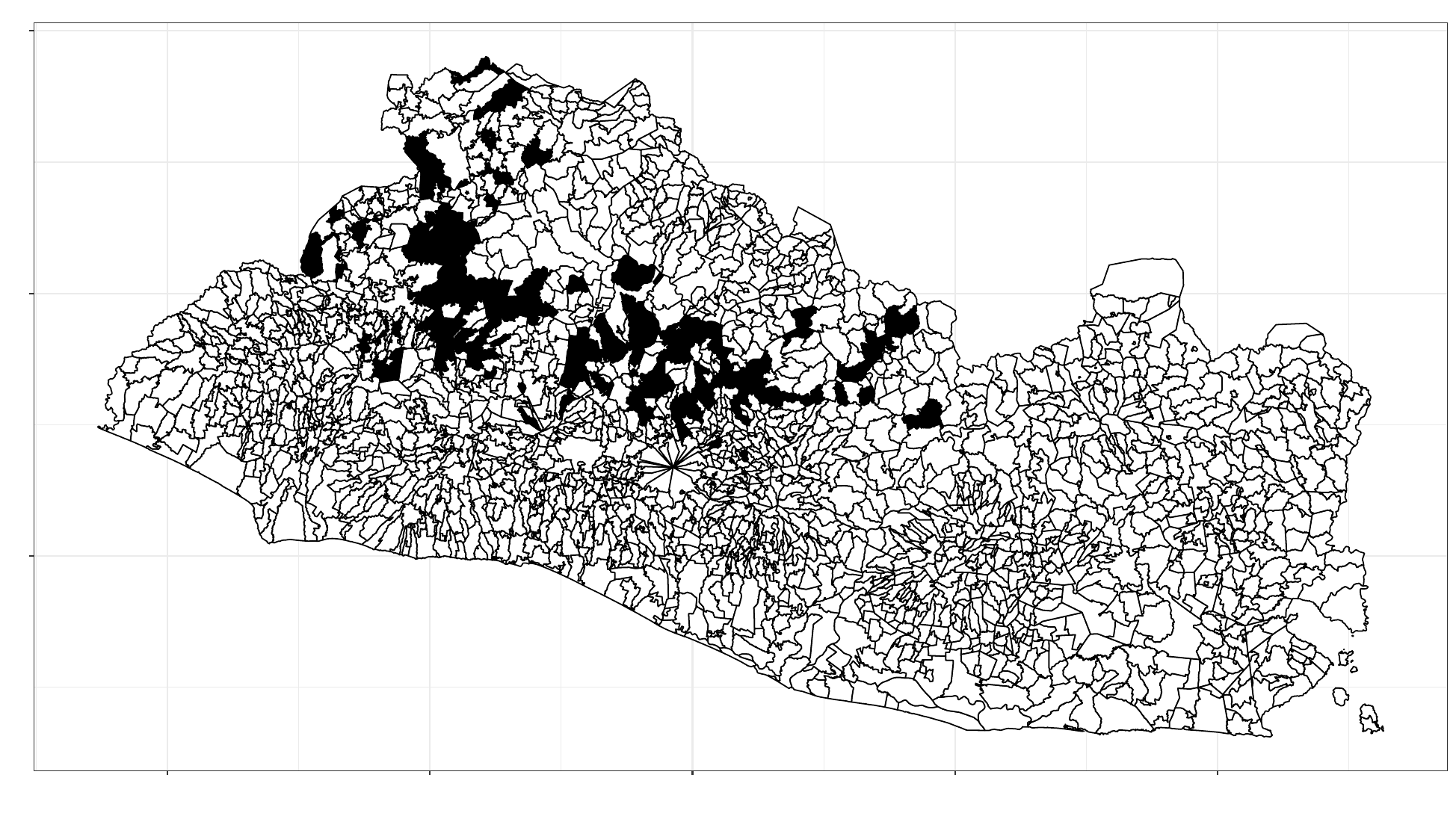}
\caption{El Salvador - Location of cantons}
\label{fig:map}
\end{figure}

The target population of this program was farmers with incomes below the poverty line and producing mostly staple crops, such as corn and beans. The program consisted in promoting soil conservation technologies among participants. The initial fieldwork took place in 2002, and a random sample of participants was re-surveyed in 2005, along with a sample of farmers who never received PAES benefits. Figure \ref{fig:map} shows the cantons (administrative divisions in El Salvador) where participants and non-participants are located (in black). For more details on the program and data collection, see \cite{Bravo2006}.

For this application, we consider the cross-section of farmers surveyed in 2005 and specify the following model:
\begin{equation}
Y_i=\beta_0 + X_i\beta_1 + Z_{2i}\beta_2 + X_i Z_{2i}\beta_{1,2} +V_i-U_{0i}\exp(Z_{1i}\delta_1 + Z_{2i} \delta_2 +Z_{1i} Z_{2i} \delta_{1,2}),
\end{equation}
for $i=1,\dots,n$, with $\beta = (\beta^\prime_1,\beta^\prime_2,\beta^\prime_{1,2})^\prime$, $\delta = (\delta^\prime_1,\delta^\prime_2,\delta^\prime_{1,2})^\prime$, and where $Y_i$ is the total value of \emph{production} (measured in dollars); $X_i$ is a vector of inputs including, \emph{Labor} (number of hired and family workers), \emph{Land} (total area cultivated in manzanas where one manzana=0.7 has), \emph{Fertilizers} (measured in dollars), \emph{Pesticides} (measured in dollars) and \emph{Seeds} (measured in dollars). Output and inputs are expressed in logs. $Z_i = (Z_{1i},Z_{2i})$ is a vector of environmental factors, which is decomposed into two sub-vectors: $Z_{2i}$ can act both as a frontier shifter as well as an inefficiency determinant, while $Z_{1i}$ only contains inefficiency determinants. We consider \emph{Participation} (dummy variable taking value 1 if the farmer is participating in PAES, 0 otherwise) as the frontier shifter, $Z_{2i}$. $Z_{1i}$, instead, includes the log of \emph{Land}; a \emph{Tenure} dummy equal to $1$ if the farmer owns all her cultivated land, and $0$ otherwise; the interaction between the latter two variables; a \emph{Age} dummy equal to $1$ if the farmer is 60 and older and $0$ otherwise; \emph{Education} (measured in years); a \emph{No income} dummy, equal to $1$ if the farmer has no outside sources of income and $0$ otherwise; a \emph{Foot Access} dummy, equal to $1$ that indicates whether farmers have only walking access to their plot; a \emph{Car Access} dummy equal to $1$ that indicates whether farmers also have access by car to their plot; and a risk diversification index, \emph{Risk div}, defined as a continuous variable which compares the relative diversification of each farmer with respect to the average farmer. A higher value of the \emph{Risk div} index implies higher risk diversification. Our indicator ranges over the entire real line.\footnote{We use the Ogive index of risk diversification in \citet{wasylenko1978}. The index is constructed as follows
\begin{equation*}
Risk\;Div\;= -\log \left( \sum_{j=1}^{C_i}\frac{(s_{j} - \bar{s}_j)^{2}}{\bar{s}_j} \right),
\end{equation*}
where $s_{j}$ is the proportion of land devoted by the farmer to crop $j$, $\bar{s}_j$ is the average sample proportion of land devoted to crop $j$, and $C_i$ is the total number of crops cultivated by farmer $i$.} We incorporate first-order interaction terms between $Z_{1}$ and $Z_2$ to account for potential observed treatment heterogeneity and also for potential observable differences between the control and the treatment group, as alluded to in the Introduction. The reason to include the size of cultivated land into the inefficiency determinant is that, given soil erosion, which is common in El Salvador because of the frequent natural disasters, a larger plot requires the implementation of several soil conservation measures which are more costly and time-consuming to implement. The ability of the farmer to effectively implement these measures may also depend on the ownership structure. For this reason, we interact \emph{Land} with \emph{Tenure}.

Finally, $V_i$ is the idiosyncratic error and $U_{0i}$ is the stochastic inefficiency term. Our goal is to estimate the parameters $\lbrace \beta,\delta,\sigma^2_V, \sigma^2_U,\rho_U,\rho_V,\gamma\rbrace$. The total sample size is equal to $n =459$. 

As stated earlier, \emph{Participation} is an endogenous binary treatment variable. Hence, we need to select a vector of IVs to deploy our estimation method. Excluded exogenous variables can be provided by random shocks, and by socio-economic characteristic of the surroundings that can be considered external to the farmer's production choices. Assuming our IVs are valid, the more exogenous variation we can provide, the more efficient our estimator will be. By contrast, and extending the intuition of the linear model \citep[see, e.g.][]{stock2005}, weak instruments usually lead to poor asymptotic approximations and larger standard errors. The type of instrument (discrete vs continuous) should not play a major role. It is usually a good idea, however, to transform a continuous instrument with an irregular distribution, due to, say, mass points, or or very large kurtosis (e.g., variables with multimodal distributions) into a discrete one. Looking at the first-stage likelihood ratio test for the null that all the IVs (or a subset of them) is equal to $0$ is usually a good proxy of their strength, although we are not aware of a formal (or informal) test of IV strength in the context of binary endogenous variables.

El Salvador was hit by a major earthquake in $2001$ (the year prior to the beginning of the PAES program). We use this as an exogenous shock to determine treatment assignment. We construct a binary instrument, \emph{Dist Earthquake}, which is equal to one if the log-distance (measured in kilometers) from the epicenter of the earthquake in 2001 was higher than $3.5$, and $0$ otherwise.\footnote{The density of \emph{Dist Earthquake} is bimodal, with a mode around 1.5 and another one around 3.5. Since the percentage of farmers for whom \emph{Dist Earthquake} is lower than $2.5$ is slightly above $10\%$, we choose 3.5 as our cutoff point.} Our vector of instruments, $W_i$, also includes: \emph{Electricity}, the proportion of families in the farmer's canton with access to electricity; \emph{Bathroom}, the proportion of families in the farmer's canton with access to private bathrooms; \emph{Wage Canton}, the average hourly wage in the canton where the farmer is located; and we also use three dummies for unobserved region characteristics that could have affected the willingness of the farmer to participate in the program. In the following, we let $\tilde{W}_i = (X_i,Z_{1i},W_i)$, to be the vector of included and excluded endogenous variables.

We test the specification of the assignment equation (the regression of the treatment variable on all the other exogenous regressors) as a Probit model using the test in \citet{wilde2008}. We are not able to reject the null hypothesis that the error term follows a normal distribution at any standard significance level, which indicates that our assignment equation is not misspecified (value of the test statistic is $0.042$, with a p-value equal to $0.979$). Moreover, to assess the relevance of the instruments, we construct a likelihood ratio test which compares the unrestricted Probit model, with a Probit model where all coefficients associated with the instruments are set to zero. The null hypothesis of the test is that the instruments do not jointly have a significant effect on participation. The value of the likelihood ratio statistic is $128.76$ which leads to a rejection of the null hypothesis with a p-value equal to $0$. The estimated coefficients for the first-stage equation are given in the Appendix \ref{appBempirics}.

When we ignore selection into treatment, we obtain the MLE assuming that $(U_0,V) \upmodels (X,Z)$ with $V \sim N(0,\sigma^2_V)$ and $U_0 \sim N^{+} (0,\sigma^2_U)$. We remind the reader that the estimator which controls for potential treatment endogeneity would reduce to the former setting when $\rho_V = \rho_U = 0$. 

\begin{table}[!h]
\centering
\scriptsize
\begin{tabular}{l | c l r | c l r } 
\hline \hline
~ & \multicolumn{3}{c}{Exogeneity} & \multicolumn{3}{c}{Endogeneity, $\rho_U=0$} \\ \hline
~ & Estimate & \multicolumn{2}{c}{CI} & Estimate & \multicolumn{2}{c}{CI} \\ \hline
% latex table generated in R 4.3.2 by xtable 1.8-4 package
% Thu Dec 21 09:15:32 2023
  \hline
$\beta_0$ & 3.7411 & [3.0764 & 4.4057] & 3.7137 & [3.0045 & 4.2909] \\ 
  $\beta_{Land}$ & 0.2249 & [0.1157 & 0.3379] & 0.2438 & [0.1444 & 0.3457] \\ 
  $\beta_{Labor}$ & 0.1934 & [0.1191 & 0.2672] & 0.1959 & [0.1328 & 0.2568] \\ 
  $\beta_{Fertilizer}$ & 0.1020 & [0.0263 & 0.1772] & 0.1045 & [0.0412 & 0.1889] \\ 
  $\beta_{Pesticides}$ & 0.0507 & [-0.0111 & 0.1124] & 0.0492 & [0.0046 & 0.1038] \\ 
  $\beta_{Seeds}$ & 0.1332 & [0.0764 & 0.1898] & 0.1362 & [0.0872 & 0.2021] \\ 
  $\beta_{Land \times Z_2}$ & 0.1197 & [-0.0501 & 0.3557] & 0.2127 & [0.0670 & 0.3584] \\ 
  $\beta_{Labor \times Z_2}$ & -0.1370 & [-0.2433 & -0.0304] & -0.1512 & [-0.2396 & -0.0571] \\ 
  $\beta_{Fertilizer \times Z_2}$ & -0.1126 & [-0.2377 & 0.0131] & -0.1054 & [-0.2217 & 0.0017] \\ 
  $\beta_{Pesticides \times Z_2}$ & 0.1266 & [0.0289 & 0.2242] & 0.1389 & [0.0570 & 0.2182] \\ 
  $\beta_{Seeds\times Z_2}$ & 0.0881 & [-0.0126 & 0.1886] & 0.1119 & [0.0306 & 0.1906] \\ 
  $\beta_{Z_2}$ & 0.6653 & [-0.3423 & 1.6773] & 0.2109 & [-0.7217 & 1.1657] \\ 
  $\delta_{Land}$ & 0.0237 & [-0.3282 & 0.4237] & 0.0289 & [-0.3754 & 0.3015] \\ 
  $\delta_{Tenure}$ & -0.4883 & [-1.5927 & 0.0953] & -0.6085 & [-1.4837 & 0.2488] \\ 
  $\delta_{Land \times Tenure}$ & -0.1403 & [-0.6168 & 0.4787] & -0.0917 & [-0.3940 & 0.5131] \\ 
  $\delta_{Age}$ & 0.4653 & [-0.0001 & 1.2973] & 0.4745 & [-0.0023 & 1.0874] \\ 
  $\delta_{Educ}$ & -0.1750 & [-0.4942 & 0.0573] & -0.2197 & [-0.5570 & 0.0074] \\ 
  $\delta_{No income}$ & 0.1079 & [-0.3846 & 0.8403] & 0.1810 & [-0.3027 & 0.8214] \\ 
  $\delta_{Foot access}$ & -1.0207 & [-2.3106 & -0.2441] & -1.1837 & [-2.1660 & -0.6023] \\ 
  $\delta_{Car access}$ & -0.4876 & [-1.2807 & 0.2038] & -0.5814 & [-1.2383 & -0.0847] \\ 
  $\delta_{Risk div}$ & -0.2126 & [-0.5129 & 0.0160] & -0.2203 & [-0.4150 & -0.0189] \\ 
  $\delta_{Land \times Z_2}$ & -0.2187 & [-0.9204 & 1.2145] & 0.2572 & [-0.1642 & 0.6617] \\ 
  $\delta_{Tenure\times Z_2}$ & 0.6209 & [-0.3149 & 1.9683] & 0.5870 & [-0.3339 & 1.7936] \\ 
  $\delta_{Land \times Tenure\times Z_2}$ & 0.3449 & [-0.3869 & 1.0337] & 0.3639 & [-0.2355 & 0.8222] \\ 
  $\delta_{Age\times Z_2}$ & -0.6115 & [-1.5179 & -0.0256] & -0.6974 & [-1.4997 & -0.1376] \\ 
  $\delta_{Educ\times Z_2}$ & 0.2659 & [-0.0097 & 0.6346] & 0.2583 & [-0.0101 & 0.6826] \\ 
  $\delta_{No income\times Z_2}$ & -0.1531 & [-0.9411 & 0.4498] & -0.1997 & [-0.9787 & 0.4204] \\ 
  $\delta_{Foot access\times Z_2}$ & 0.7343 & [-0.3806 & 2.1142] & 1.0512 & [0.4079 & 1.8286] \\ 
  $\delta_{Car access\times Z_2}$ & 0.4337 & [-0.5857 & 1.4887] & 0.9611 & [0.4319 & 1.5592] \\ 
  $\delta_{Risk div\times Z_2}$ & 0.1211 & [-0.1667 & 0.4449] & 0.2663 & [0.0265 & 0.5145] \\ 
  $\delta_{Z_2}$ & -0.7030 & [-5.8932 & 0.8935] & -1.9518 & [-3.3784 & -0.8608] \\ 
  $\rho_{U,\eta}$ & ~ & ~ & ~ & 0.0000 & [0.0000 & 0.0138] \\ 
  $\rho_{V,\eta}$ & ~ & ~ & ~ & 0.3254 & [0.1474 & 0.4922] \\ 
  $\sigma^2_{U}$ & 0.6925 & [0.0564 & 6.4017] & 0.7241 & [0.0997 & 3.5055] \\ 
  $\sigma^2_{V}$ & 0.0978 & [0.0777 & 0.1223] & 0.1133 & [0.0845 & 0.1338] \\ 
   \hline

\hline \hline 
\end{tabular}
\caption{\textit{Estimates of the production frontier with and without accounting for endogeneity.}}
\label{table:elsalres2}
\end{table}

\subsection{Results} Table \ref{table:elsalres2} reports results for our empirical example. We report our point estimates along the $95\%$ confidence intervals. Confidence intervals for model that assumes exogeneity are constructed using a profile-likelihood method \citep{Cox1979}. This is because the sample size is relatively small and the normal approximation implicit in the more standard Wald-type confidence intervals may fail to hold. For the estimator that controls for potential endogeneity, we first perform a test for the absence of dependence between stochastic inefficiency and participation. We find that the null hypothesis of no dependence, i.e., $H_0: \rho_U = 0$, cannot be rejected either using the LR test \`a la \citet{andrews2001} or the modified score test proposed by \citet{bottai2003} and \citet{ekvall2022}. At the $10\%$ level, the critical value of the LR test is equal to $1.642$, and our test statistic is numerically indistinguishable from $0$, while the score test statistic is equal to $0$ for a critical value of $2.706$. This implies that we can exclude any dependence between farmers' stochastic inefficiency and their participation in the program. Hence, we only report results for the endogenous model when $\rho_U$ is restricted to be equal to $0$. When $\rho_U$ is left unrestricted, point estimators of all the other parameters are numerically indistinguishable. Since we assume that $\rho_{U,0} = 0$, the $95\%$ confidence intervals in the endogenous setting are constructed by inverting the score test proposed in \citet{ekvall2022}. 

The first set of columns in Table \ref{table:elsalres2} shows the estimation results assuming exogeneity (parameter estimates and confidence intervals). There is no significant shift in the production frontier for participants. Also, for participants, the output elasticity of \emph{Labor} is lower, and that of \emph{Pesticides} is higher. \emph{Participation} reduces inefficiency (i.e., improves efficiency), although the estimator is not significantly different from zero at the $5\%$ level, according to our profile likelihood confidence intervals. Regarding the other efficiency determinants, we find substantially the same results for both participants and non-participants. However, we note that \emph{Participation} leads to noticeable TE gains for for older farmers. 

The results controlling for endogeneity are reported in the last three columns of Table \ref{table:elsalres2} (parameter estimates and confidence intervals). The qualitative results are broadly consistent with the model assuming exogeneity. Thus, \emph{Participation} does not lead to any significant shift in the production frontier, and for participants, the output elasticity of \emph{Labor} is lower, and that of \emph{Pesticides} is higher. In contrast, in the model controlling for endogeneity, the output elasticity for \emph{Land} and \emph{Seeds} is higher for participants. This is consistent with the view that non-participants cultivate smaller plots, that are less labor intensive and where fertilizers do not help because of poor soil quality, which is adversely affected by erosion. Appropriate use of \emph{Pesticides} and \emph{Seeds} can, however, be beneficial for production. Returns to scale are decreasing for both participants and non-participants, which is consistent with prevailing evidence for developing economies \citep[see, e.g.][]{berry1979}. Also, in the model assuming exogeneity, \emph{Participation} did not have a significant effect on inefficiency, whereas in the model controlling for endogeneity, the negative effect on inefficiency is now significant at the $5\%$ level. In the exogenous model, we obtain that \emph{Participation} would reduce the variance of inefficiency by about $50\%$. However, when \emph{Participation} is considered endogenous, the variance of inefficiency for participants decreases by about $86\%$. Therefore, adopting soil conservation practices brings participating farmers closer to their production frontier. Some other differences between the models concerning inefficiency are also worth highlighting. For example, \emph{Car access} does not affect efficiency in the model assuming exogeneity but reduces inefficiency in the endogeneity model. Similarly, risk diversification (\emph{Riskdiv}) has no effect on efficiency in the model assuming exogeneity but improves efficiency in the endogeneity model. Interestingly, the positive effect of risk diversification on efficiency is found to be lower for participants. Diversification, therefore, seems to be a good strategy for improving TE, though this appears to be more effective for non-participants.

Moreover, the endogenous model also detects a positive correlation between the idiosyncratic component of the error term and \emph{Participation} since $\hat{\rho}_V = 0.325$, which suggests that self-selection into this particular program may be related to a preference component rather than to inefficiency considerations.

\begin{figure}[!h]
\includegraphics[scale=0.6]{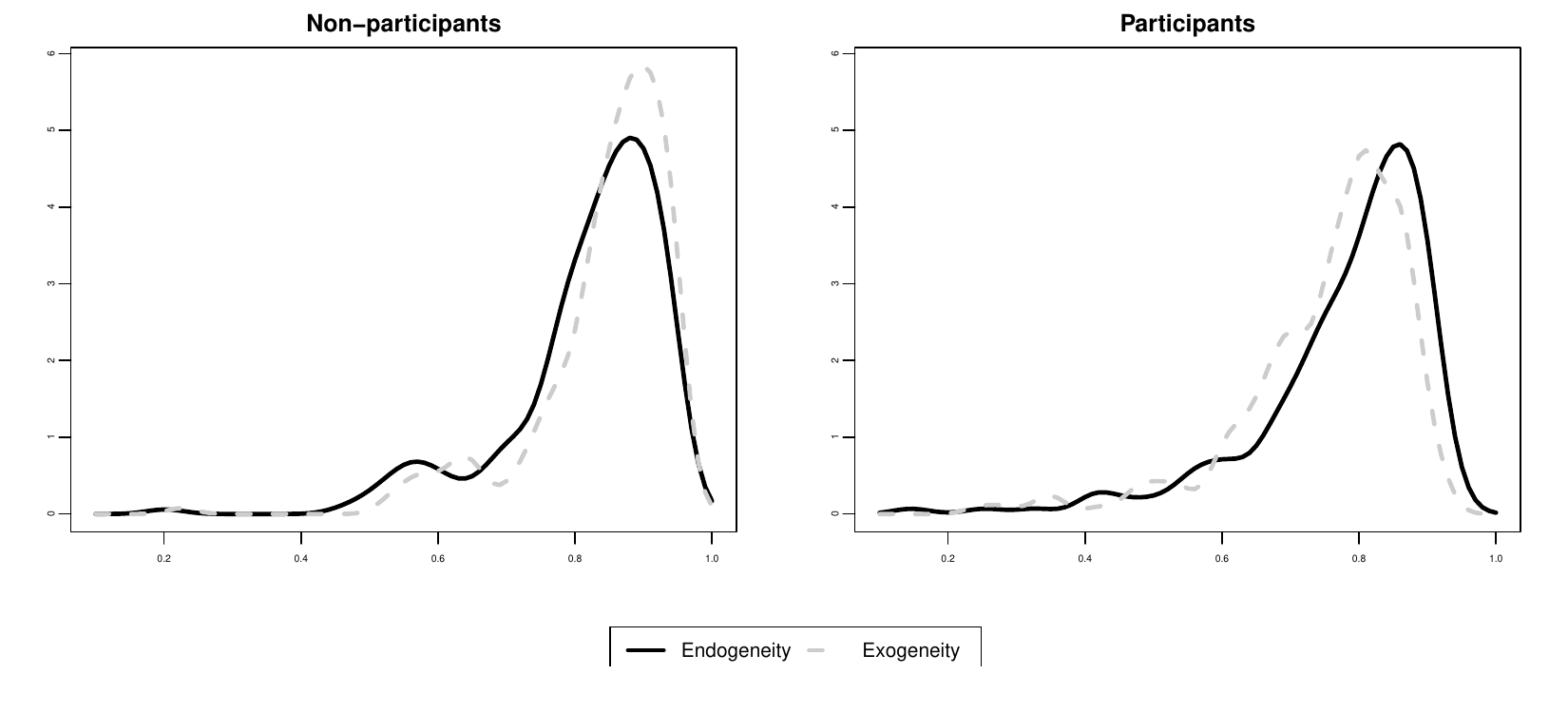}
\caption{Density of estimated technical efficiency.}
\label{fig:fct_elsal_eff_2}
\end{figure}

Finally, we report in Figure \ref{fig:fct_elsal_eff_2} a kernel density estimator of inefficiencies for non-participant farmers (left panel) and participant farmers (right panel) using the models that do and do not control for endogeneity. The dotted gray line is the density of inefficiencies for the model assuming exogeneity, while the solid black line is the density of inefficiencies in the model controlling for endogeneity. The figure clearly illustrates that the model controlling for endogeneity detects larger TE scores in the group of participants and slightly lower ones among non-participants. We test for the difference between the distribution in the TE scores for participants and non-participants using the first-order stochastic dominance test in \citet{linton2005}, which allows for serial dependence in the observations. Complicated and unknown serial dependence can arise in regression residuals, even if the true errors are iid \citep{maddala1988}. \citet{simar2007} argue that this issue is even more relevant for frontier models as changing the value of efficiency close to the frontier would change the relative ranking of all the units in the sample, thus generating dependence between observations.

For participants, our null is that the distribution of efficiency scores predicted by the endogenous model first-order stochastically dominates the distribution of efficiency scores predicted by the exogenous model. That is, for any given value of TE, the probability of being below that value is higher for the exogenous model than for the endogenous one, which would imply that efficiency scores predicted by the endogenous model are consistently higher for participants. Similarly, for non-participants, our null is that the distribution of scores predicted by the exogenous model first-order stochastically dominates the distribution of scores predicted by the endogenous model. The test in \citet{linton2005} is a Kolmogorov-Smirnov-type test. Under the null, there is first-order stochastic dominance. Critical values for the test are obtained by taking subsamples of size $b_n < n$ from $\lbrace (Y_i,X_i,Z_i,W_i),i = 1,\dots,n\rbrace$, and obtaining the estimated parameters and efficiency scores, and test statistics for each subsample \citep{politis1994,linton2005}. The value of the test statistic for participants is $0.205$, and the value of the test statistic for non-participants is $0$. We choose a grid of several subsampling sizes, and we obtain a median critical value of $6.778$ for participants and $4.851$ for non-participants at the $5\%$ level. The null of first-order stochastic dominance cannot be rejected in either case. We can conclude that not controlling for endogeneity in our application would lead to incorrect efficiency scores, with these being overestimated for non-participants and underestimated for participants. 

\section{Conclusions} \label{sec:conclusions}

In the framework of policies aimed at increasing agricultural productivity, Stochastic Production Frontier Models can shed light on the effectiveness of programs targeting this goal. However, controlling for potential endogeneity associated with voluntary program participation is crucial to obtain accurate estimates of the impact of the project. 

In this paper, we propose a method to control for a binary endogenous treatment in Stochastic Production Frontier Models. In particular, we provide a simple maximum likelihood estimator based on distributional assumptions about the first and second-stage error terms. This estimator is in line with a more traditional approach to Stochastic Frontier Estimation, where one is usually interested in estimating the technical efficiency for each producer.

In the empirical application, we estimate Stochastic Production Frontiers for a sample of farmers in El Salvador participating in a soil conservation program and a control group of non-participant farmers. Our results show that the model assuming exogeneity does not detect any significant association between participation in the program and inefficiency. However, when we implement our methodology, we find that participation in the soil conservation program leads to an improvement in the level of technical efficiency. 

The main policy implication of our study is that policymakers wishing to perform program evaluation in a Stochastic Frontier Model context should adequately control for endogeneity issues arising from voluntary program participation. Failure to do so may generate misleading conclusions about the effectiveness of such programs to generate improvements in productive efficiency. This is especially important in guiding future evidence-based policy-making in which the best possible use should be made of scarce resources.

An extension of our maximum likelihood approach to panel data models is a potential avenue for future research \citep{Lai2018,Kutlu2019}. In such a setting, the timing and evolution of the treatment over time, along with the assumptions one imposes on the composite error term, are paramount to understanding how our parametric approach applies, and whether it can allow for richer dependence between treatment and unobservables. 

%%%%%%%%%%%%%%%%%%%%%%%%%%%%%%%%%%%%%%%%%%%%%%%%%%%%%%%%%%%%%%%%%%%%%%%%%%%%%%%%
%%%%%%%%%%%%%%%%%%%%%%%%%%%%%%%%% REFERENCES %%%%%%%%%%%%%%%%%%%%%%%%%%%%%%%%%%%
%%%%%%%%%%%%%%%%%%%%%%%%%%%%%%%%%%%%%%%%%%%%%%%%%%%%%%%%%%%%%%%%%%%%%%%%%%%%%%%%
\bibliography{efficiency}
\bibliographystyle{agsm}

% \clearpage % ensure all floats are processed
% \processdelayedfloats
% \clearpage

\newpage

\appendix

\section{Technical derivations} \label{sec:appA}

\subsection{Derivation of the likelihood function} \label{sec:appAlikder}

We provide here the main steps for the calculation of the conditional distribution of $\eta \vert \varepsilon$. Recall that the joint density of $(\varepsilon, \eta)$ is given by 

\begin{align*}
f_{\varepsilon,\eta} (\varepsilon ,\eta) =& \frac{1}{2\pi \tsigma(Z)} \left\lbrace \Phi \left( \frac{\lambda(Z) \rho_V \sigma_V \eta}{\tsigma(Z)} + \frac{\rho_U\sigma_U(Z)\eta}{\lambda(Z) \tsigma(Z)} -\frac{\lambda(Z) \varepsilon}{\tsigma(Z)}\right) \times \right.\\
& \qquad \exp \left( -\frac{(\varepsilon - \rho_V \sigma_V\eta + \rho_U\sigma_U(Z) \eta)^2}{2 \tsigma^2(Z)} - \frac{\eta^2}{2}\right) \\
& \quad + \Phi \left( \frac{\lambda(Z) \rho_V \sigma_V \eta}{\tsigma(Z)} - \frac{\rho_U\sigma_U(Z)\eta}{\lambda(Z) \tsigma(Z)} -\frac{\lambda(Z) \varepsilon}{\tsigma(Z)}\right)\times \\\
& \qquad \left. \exp \left( - \frac{(\varepsilon - \rho_V \sigma_V\eta - \rho_U\sigma_U(Z) \eta)^2}{2 \tsigma^2(Z)}- \frac{\eta^2}{2}\right) \right\rbrace.
\end{align*}

We analyze the kernel of the two components of this distribution separately. 

\begin{itemize}
\item[(i)] Let $\rho_1(Z) = \rho_V \sigma_V - \rho_U \sigma_U(Z)$. The kernel of the first component is equal to 
\begin{align*}
\frac{1}{\tsigma^2(Z)} & \left(\varepsilon^2 - 2\rho_1(Z) \varepsilon \eta + (\tsigma^2(Z) + \rho^2_1(Z))\eta^2 \right) \\
=&\frac{\tsigma^2(Z) + \rho^2_1(Z)}{2 \tsigma^2(Z)} \left(\eta^2 - 2\frac{\rho_1(Z)\varepsilon \eta}{\tsigma^2(Z) + \rho^2_1(Z)} + \frac{\varepsilon^2}{\tsigma^2(Z) + \rho^2_1(Z)}\right) \\
=& \frac{\tsigma^2(Z) + \rho^2_1(Z)}{2 \tsigma^2(Z)} \left(\eta - \frac{\rho_1(Z)}{\tsigma^2(Z) + \rho^2_1(Z)}\varepsilon \right)^2 + \frac{\varepsilon^2}{2(\tsigma^2(Z) + \rho^2_1(Z))}.
\end{align*}
\item[(ii)] Let $\rho_2(Z) = \rho_V \sigma_V + \rho_U \sigma_U(Z)$. The kernel of the second component can be similarly written as
\begin{align*}
\frac{1}{2 \tsigma(Z)^2} & \left(\varepsilon^2 - 2\rho_2(Z) \varepsilon \eta + (\tsigma^2(Z) + \rho^2_2(Z))\eta^2 \right) \\
=& \frac{\tsigma^2(Z) + \rho^2_2(Z)}{2 \tsigma^2(Z)} \left(\eta - \frac{\rho_2(Z)}{\tsigma^2(Z) + \rho^2_2(Z)}\varepsilon \right)^2 + \frac{\varepsilon^2}{2(\tsigma^2(Z) + \rho^2_2(Z))}.
\end{align*}
\end{itemize}

Therefore, we have that 
\begin{align*}
f_{\varepsilon,\eta} (\varepsilon ,\eta)=& \frac{1}{2\pi \tsigma(Z)} \left\lbrace \Phi \left( \frac{\lambda(Z) \rho_V \sigma_V \eta}{\tsigma(Z)} + \frac{\rho_U\sigma_U(Z)\eta}{\lambda(Z) \tsigma(Z)} -\frac{\lambda(Z) \varepsilon}{\tsigma(Z)}\right) \times \right. \\
& \quad \exp \left( -\frac{\tsigma^2(Z) + \rho^2_1(Z)}{2 \tsigma^2(Z)} \left(\eta - \frac{\rho_1(Z)}{\tsigma^2(Z) + \rho^2_1(Z)}\varepsilon \right)^2 - \frac{\varepsilon^2}{2(\tsigma^2(Z) + \rho^2_1(Z))}\right) \\
& \quad + \Phi \left( \frac{\lambda(Z) \rho_V \sigma_V \eta}{\tsigma(Z)} - \frac{\rho_U\sigma_U(Z)\eta}{\lambda(Z) \tsigma(Z)} -\frac{\lambda(Z) \varepsilon}{\tsigma(Z)}\right) \times \\
& \quad \left. \exp \left( - \frac{\tsigma^2(Z) + \rho^2_2(Z)}{2 \tsigma^2(Z)} \left(\eta - \frac{\rho_2(Z)}{\tsigma^2(Z) + \rho^2_2(Z)}\varepsilon \right)^2 - \frac{\varepsilon^2}{2(\tsigma^2(Z) + \rho^2_2(Z))}\right) \right\rbrace.
\end{align*}
We would like to integrate out the random variable $\eta$. To do so, we rearrange the terms above as follows
\begin{align*}
f_{\varepsilon,\eta} (\varepsilon ,\eta)=& \left\lbrace \Phi \left( q_1 (Z) \eta -\frac{\lambda(Z) \varepsilon}{\tsigma(Z)}\right) \frac{1}{\sqrt{ 2\pi \sigma_{\eta,1}^2(Z)}}\exp \left( -\frac{\left(\eta - \mu_{\eta,1}(Z) \varepsilon \right)^2}{2 \sigma_{\eta,1}^2(Z)} \right) \times \right.\\
& \quad \frac{1}{\sqrt{2\pi \sigma_{\varepsilon,1}^2(Z)}} \exp\left( - \frac{\varepsilon^2}{2\sigma_{\varepsilon,1}^2(Z)}\right) \\
& \quad + \Phi \left( q_2 (Z) \eta -\frac{\lambda(Z) \varepsilon}{\tsigma(Z)}\right) \frac{1}{\sqrt{ 2\pi \sigma_{\eta,2}^2(Z)}}\exp \left( -\frac{\left(\eta - \mu_{\eta,2}(Z)\varepsilon \right)^2}{2 \sigma_{\eta,2}^2(Z)} \right)\times \\
& \quad \left. \frac{1}{\sqrt{2\pi\sigma_{\varepsilon,2}^2(Z)}} \exp\left( -\frac{\varepsilon^2}{2\sigma_{\varepsilon,2}^2(Z)}\right) \right\rbrace \\
=& \frac{1}{2}\left\lbrace \frac{\Phi \left( q_1 (Z) \eta - \frac{\lambda(Z) \varepsilon}{\tsigma(Z)}\right)}{\Phi \left( \tau_1 (Z) \varepsilon \right)} \frac{1}{\sqrt{ 2\pi \sigma_{\eta,1}^2(Z)}}\exp \left( -\frac{\left(\eta - \mu_{\eta,1}(Z)\varepsilon \right)^2}{2 \sigma_{\eta,1}^2(Z)} \right) \times \right. \\
& \quad \frac{2}{\sqrt{2\pi\sigma_{\varepsilon,1}^2(Z)}}\Phi \left(\tau_1 (Z) \varepsilon \right) \exp\left( - \frac{\varepsilon^2}{2\sigma_{\varepsilon,1}^2(Z)}\right) \\
& \quad + \frac{\Phi \left( q_2 (Z) \eta -\frac{\lambda(Z) \varepsilon}{\tsigma(Z)}\right)}{\Phi \left(\tau_2 (Z) \varepsilon \right)} \frac{1}{\sqrt{ 2\pi \sigma_{\eta,2}^2(Z)}}\exp \left( -\frac{\left(\eta - \mu_{\eta,2}(Z)\varepsilon \right)^2}{2 \sigma_{\eta,2}^2(Z)} \right) \times \\
& \quad \left. \frac{2}{\sqrt{2\pi\sigma_{\varepsilon,2}^2(Z)}} \Phi \left(\tau_2 (Z) \varepsilon \right) \exp\left( -\frac{\varepsilon^2}{2\sigma_{\varepsilon,2}^2(Z)}\right) \right\rbrace\\
=& \frac{1}{2}\left\lbrace f_{\varepsilon,\eta,1}\left(\varepsilon,\eta \right) + f_{\varepsilon,\eta,2}\left(\varepsilon,\eta \right)\right\rbrace,
\end{align*}
where

\noindent\begin{minipage}{.5\textwidth}
\begin{align*}
\sigma_{\varepsilon,1}^2(Z) =& \sigma^2_V + \sigma^2_U(Z) - 2\rho_V \sigma_V \rho_U \sigma_U(Z),\\
q_1 (Z) =& \frac{\lambda(Z) \rho_V \sigma_V}{\tsigma(Z)} + \frac{\rho_U\sigma_U(Z)}{\lambda(Z) \tsigma(Z)},\\
\mu_{\eta,1} (Z) =& \frac{\rho_1(Z)}{\sigma_{\varepsilon,1}^2(Z)},\\
\sigma^2_{\eta,1} (Z) =& \frac{\tsigma^2(Z)}{\sigma_{\varepsilon,1}^2(Z)},\\
\tau_1 (Z) =& \frac{\mu_{\eta,1} (Z) q_1 (Z) - \frac{\lambda(Z)}{\tsigma(Z)}}{\sqrt{1 + q^2_1 (Z) \sigma^2_{\eta,1}(Z)}},\\
\end{align*}
\end{minipage}
\begin{minipage}{.5\textwidth}
\begin{align*}
\sigma_{\varepsilon,2}^2(Z) =& \sigma^2_V + \sigma^2_U(Z) + 2\rho_V \sigma_V \rho_U \sigma_U(Z),\\
q_2 (Z) =& \frac{\lambda(Z) \rho_V \sigma_V}{\tsigma(Z)} -\frac{\rho_U\sigma_U(Z)}{\lambda(Z) \tsigma(Z)},\\
\mu_{\eta,2} (Z) =& \frac{\rho_2(Z)}{\sigma_{\varepsilon,2}^2(Z) },\\
\sigma^2_{\eta,2} (Z) =& \frac{\tsigma^2(Z)}{\sigma_{\varepsilon,2}^2(Z) },\\
\tau_2 (Z) =& \frac{ \mu_{\eta,2} (Z) q_2 (Z) - \frac{\lambda(Z)}{\tsigma(Z)}}{\sqrt{1 + q^2_2 (Z)\sigma^2_{\eta,2}(Z)}}.\\
\end{align*}
\end{minipage}
and the normalizing factor $\tau_j(Z)$, with $j = 1,2$ comes from the fact that 
\begin{align*}
\Phi \left( \tau_j(Z)\varepsilon \right) = \int_{-\infty}^\infty \Phi \left( q_j (Z) \eta -\frac{\lambda(Z) \varepsilon}{\tsigma(Z)}\right) \frac{1}{\sqrt{2\pi \sigma_{\eta,j}^2(Z)}}\exp \left( -\frac{\left(\eta - \mu_{\eta,j} (Z)\varepsilon \right)^2}{2\sigma_{\eta,j}^2(Z)} \right)d\eta,
\end{align*}
by Lemma 2.2. in \citet[][p. 26]{azzalini2013}. 

Each component of this density can be interpreted as the pdf of a bivariate skew-normal distribution, properly rearranged into the product of a conditional and a marginal density \citep{azzalini1996}. Therefore, the full joint density is a mixture of two skew normal distribution with equal weights, $0.5$. To obtain the conditional cdf of $\eta\vert\varepsilon$, we need to integrate the above expression appropriately.
The following integral
\begin{align*}
\Psi_{0,j}(Z,\theta) =& \int_{-\infty}^{-\cez} \Phi \left( q_j(Z)\eta -\frac{\lambda(Z) \varepsilon}{\tsigma(Z)}\right)\frac{1}{\sqrt{2\pi \sigma_{\eta,j}^2(Z)}}\exp \left( -\frac{\left(\eta -\mu_{\eta,j} (Z)\varepsilon \right)^2}{2 \sigma_{\eta,j}^2(Z)}\right) d\eta,
\end{align*}
cannot be directly evaluated analytically. However, using the properties of the skew-normal distribution, it can be expressed as the cdf of a bivariate normal distribution. 

Let us define the fictitious random vector $(\eta,\kappa_j)$, for $j = 1,2$, such that the conditional distribution of $(\eta,\kappa_j)\vert\varepsilon$ is a bivariate normal distribution. That is, we have
\[
\begin{matrix} \frac{\eta - \mu_{\eta,j}(Z)\varepsilon}{\sigma_{\eta,j}(Z)} \\ \frac{\kappa_j + q_j (Z) \mu_{\eta,j}(Z)\varepsilon - \frac{\lambda(Z) \varepsilon}{\sigma(Z)}}{\sqrt{1+ q^2_j(Z) \sigma^2_{\eta,j}(Z)}} \varepsilon \end{matrix}\left\vert \varepsilon \right. \sim N \left(\begin{bmatrix} 0 \\ 0 \end{bmatrix}, \begin{bmatrix} 1 & \rho_j^\ast(Z) \\ \rho_j^\ast(Z) & 1\end{bmatrix}\right),
\] 
where, 
\[
\rho^\ast_j(Z) = -\frac{q_j(Z) \sigma_{\eta,j}(Z)}{\sqrt{1 + q^2_j(Z) \sigma^2_{\eta,j}(Z)}}.
\]
Let $\Phi_2(\cdot,\cdot; \rho^\ast(Z))$, be the cdf of a standard bivariate normal distribution with correlation coefficient equal to $\rho^\ast(Z)$. We have that
\begin{align*}
\int_{a_j}^{b_j} & \Phi \left( q_j(Z)\eta -\frac{\lambda(Z) \varepsilon}{\tsigma(Z)}\right)\frac{1}{\sqrt{2\pi \sigma_{\eta,j}^2(Z)}}\exp \left( -\frac{\left(\eta -\mu_{\eta,j} (Z)\varepsilon \right)^2}{2 \sigma_{\eta,j}^2(Z)}\right) d\eta \\
=& \Phi_2\left(\frac{b_j - \mu_{\eta,j}(Z)\varepsilon}{\sigma_{\eta,j}(Z)},\tau_j(Z) \varepsilon; \rho^\ast(Z)\right) - \Phi_2\left(\frac{a_j - \mu_{\eta,j}(Z)\varepsilon}{\sigma_{\eta,j}(Z)},\tau_j(Z) \varepsilon; \rho^\ast(Z)\right).
\end{align*}

This finally implies that 
\begin{align*}
\Psi_{0,j}(Z,\theta) =& \Phi_2\left(\frac{-\tilde{W} \gamma - \mu_{\eta,j}(Z)\varepsilon}{\sigma_{\eta,j}(Z)},\tau_j(Z)\varepsilon; \rho^\ast(Z)\right),
\end{align*}
and
\begin{align*}
\Psi_{1,j}(Z,\theta) =& \Phi\left( \tau_j(Z) \varepsilon \right) - \Phi_2\left(\frac{-\tilde{W} \gamma - \mu_{\eta,j}(Z)\varepsilon}{\sigma_{\eta,j}(Z)},\tau_j(Z)\varepsilon; \rho^\ast(Z)\right).
\end{align*}

Ultimately, these integrals only involve a bivariate normal cumulative distribution function, which is readily available in any standard statistical software. 

Therefore, the likelihood function is given by
\begin{align*}
\mathcal{L}(\theta) =& \left( \sum_{j = 1,2} \Psi_{1,j}(Z,\theta) \frac{1}{\sigma_{\varepsilon,j}(Z)} \phi \left(\frac{\varepsilon}{\sigma_{\varepsilon,j}(Z)}\right) \right)^Z \times \\
& \quad \left( \sum_{j = 1,2} \Psi_{0,j}(Z,\theta) \frac{1}{\sigma_{\varepsilon,j}(Z)} \phi \left(\frac{\varepsilon}{\sigma_{\varepsilon,j}(Z)}\right) \right)^{1 - Z}, 
\end{align*}
where $\phi(\cdot)$ denotes the pdf of a standard normal distribution. 
 
\subsection{Proof of Proposition \ref{prop:identif1}} \label{pr:prop21} By rearranging terms, we can write the likelihood function as follows
\begin{align*}
\mathcal{L}(\theta) =& \left( \Psi_{1,1}(Z,\theta)\sqrt{1 + \frac{4 \rho_V\sigma_V\rho_U \sigma_U(Z)}{\sigma^2_{\varepsilon,1}(Z)}}\exp \left( - \frac{2\varepsilon^2 \rho_V\sigma_V\rho_U \sigma_U(Z)}{\sigma^2_{\varepsilon,1}(Z) \sigma^2_{\varepsilon,2}(Z)}\right) + \Psi_{1,2}(Z,\theta) \right)^Z \times \\
& \quad \left( \Psi_{0,1}(Z,\theta)\sqrt{1 + \frac{4 \rho_V\sigma_V\rho_U \sigma_U(Z)}{\sigma^2_{\varepsilon,1}(Z)}}\exp \left( - \frac{2\varepsilon^2 \rho_V\sigma_V\rho_U \sigma_U(Z)}{\sigma^2_{\varepsilon,1}(Z) \sigma^2_{\varepsilon,2}(Z)}\right) + \Psi_{0,2}(Z,\theta) \right)^{1 - Z} \times \\
& \quad \frac{1}{\sigma_{\varepsilon,2}(Z)} \phi \left(\frac{\varepsilon}{\sigma_{\varepsilon,2}(Z)}\right),
\end{align*}

in a way that the log-likelihood function is given by
\begin{align*}
\ell_n(\theta) =& Z \log \left\lbrace \Psi_{1,1}(Z,\theta)\sqrt{1 + \frac{4 \rho_V\sigma_V\rho_U \sigma_U(Z)}{\sigma^2_{\varepsilon,1}(Z)}}\exp \left( - \frac{2\varepsilon^2 \rho_V\sigma_V\rho_U \sigma_U(Z)}{\sigma^2_{\varepsilon,1}(Z) \sigma^2_{\varepsilon,2}(Z)}\right) + \Psi_{1,2}(Z,\theta) \right\rbrace \\
& \quad + (1 - Z) \log \left\lbrace \Psi_{0,1}(Z,\theta)\sqrt{1 + \frac{4 \rho_V\sigma_V\rho_U \sigma_U(Z)}{\sigma^2_{\varepsilon,1}(Z)}}\exp \left( - \frac{2\varepsilon^2 \rho_V\sigma_V\rho_U \sigma_U(Z)}{\sigma^2_{\varepsilon,1}(Z) \sigma^2_{\varepsilon,2}(Z)}\right) + \Psi_{0,2}(Z,\theta) \right\rbrace \\
& \quad - \frac{1}{2} \log(\sigma^2_{\varepsilon,2}(Z)) - \frac{\varepsilon^2}{2\sigma^2_{\varepsilon,2}(Z)}\\
=&Z \log \left( D_1 \right) + (1 - Z) \log \left( D_0 \right) - \frac{1}{2} \log(\sigma^2_{\varepsilon,2}(Z)) - \frac{\varepsilon^2}{2\sigma^2_{\varepsilon,2}(Z)},
\end{align*}
where the definition of the terms $D_0$ and $D_1$ should be apparent and we have omitted terms that are constant with respect to $\theta$. When $\rho_U = 0$, then $\Psi_{j}(Z,\theta) = \Psi_{j,1}(Z,\theta) = \Psi_{j,2}(Z,\theta)$, $D_j = 2 \Psi_{j}(Z,\theta)$, and 
\begin{align*}
\frac{\partial \Psi_{j,1}(Z,\theta) }{\partial \rho_U} =& - \frac{\partial \Psi_{j,2}(Z,\theta) }{\partial \rho_U},\\
\frac{\partial^2 \Psi_{j,1}(Z,\theta) }{\partial \rho^2_U} = &\frac{\partial^2 \Psi_{j,2}(Z,\theta) }{\partial \rho^2_U}
\end{align*}
for $j = 0,1$. 

To prove Part (i) of the Proposition ($\rho_U$ is only identified up to a sign), we use the fact that the log-likelihood function is an even function of $\rho_U$. That is, $\ell_n(\theta_1,\rho_U) = \ell_n(\theta_1,-\rho_U)$. Therefore, for a given $\theta_1$, the first partial derivative of the log-likelihood wrt $\rho_U$ satisfies
\[
\frac{\partial \ell_n(\theta_1,\rho_U) }{\partial \rho_U} = - \frac{\partial \ell_n(\theta_1,-\rho_U) }{\partial \rho_U}.
\]
At the true parameter value, we therefore have that 
\[
E \left[ \frac{\partial \ell_n(\theta_{1,0},\rho_{U,0}) }{\partial \rho_U} \right] = -E \left[ \frac{\partial \ell_n(\theta_{1,0},-\rho_{U,0}) }{\partial \rho_U} \right] = 0.
\]

To prove Part (ii) (the derivative of the log-likelihood wrt $\rho_U$ is identically equal to $0$ for any value of $\theta_1$), by Rolle's theorem if $\ell_n(\theta_1,\rho_U) = \ell_n(\theta_1,-\rho_U)$, then there must exist $r \in [-\rho_U,\rho_U]$, such that,
\[
\frac{\partial \ell_n(\theta_{1},r) }{\partial \rho_U} = 0.
\]
In particular, since the log-likelihood is continuous and symmetric about $r=0$ as a function of $\rho_U$, using a limiting argument, we must have that $\frac{\partial \ell_n(\theta_{1},0) }{\partial \rho_U} = 0$ for any $\theta_1$. 

We prove the result about the second derivative wrt $\rho_U$ using direct computations. In the following, we let $SN(\xi,\omega^2,\alpha,\tau)$ be an extended skew-normal random variable with location parameter $\xi$, scale parameter $\omega$, slant parameter $\alpha$, and extended parameter $\tau$ \citep[see][p. 36-37]{azzalini2013}. When $\tau=0$, this random variable reduces to a skew-normal. In that case, the last parameter is simply omitted.

To simplify notations let 
\begin{equation} \label{zidefinition}
\zeta = \sqrt{1 + \frac{4 \rho_V\sigma_V\rho_U \sigma_U(Z)}{\sigma^2_{\varepsilon,1}(Z)}} \exp \left( - \frac{2\varepsilon^2 \rho_V\sigma_V\rho_U \sigma_U(Z)}{\sigma^2_{\varepsilon,1}(Z) \sigma^2_{\varepsilon,2}(Z)}\right).
\end{equation}

Thus we have that 
\begin{equation} \label{zifirstderdef}
\begin{aligned}
\frac{\partial \zeta }{\partial \rho_U} =& \frac{1}{\sqrt{1 + \frac{4 \rho_V\sigma_V\rho_U \sigma_U(Z)}{\sigma^2_{\varepsilon,1}(Z)}}} \left( \frac{2\rho_V \sigma_V\sigma_U(Z)}{ \sigma^2_{\varepsilon,1}(Z)} + \frac{4\rho^2_V \sigma^2_V\rho_U\sigma^2_U(Z)}{ \sigma^4_{\varepsilon,1}(Z)}\right) \exp \left( - \frac{2\varepsilon^2 \rho_V\sigma_V\rho_U \sigma_U(Z)}{\sigma^2_{\varepsilon,1}(Z) \sigma^2_{\varepsilon,2}(Z)}\right)\\
& - \sqrt{1 + \frac{4 \rho_V\sigma_V\rho_U \sigma_U(Z)}{\sigma^2_{\varepsilon,1}(Z)}} \left( \frac{2\varepsilon^2 \rho_V\sigma_V\sigma_U(Z)}{\sigma^2_{\varepsilon,1}(Z) \sigma^2_{\varepsilon,2}(Z)} - \frac{2\varepsilon^2 \rho_V\sigma_V\rho_U\sigma_U(Z)}{\sigma^4_{\varepsilon,1}(Z) \sigma^4_{\varepsilon,2}(Z)} \frac{\partial \sigma^2_{\varepsilon,1}(Z) \sigma^2_{\varepsilon,2}(Z)}{\partial \rho_U}\right) \times \\
& \qquad \exp \left( - \frac{2\varepsilon^2 \rho_V\sigma_V\rho_U \sigma_U(Z)}{\sigma^2_{\varepsilon,1}(Z) \sigma^2_{\varepsilon,2}(Z)}\right),
\end{aligned}
\end{equation}
and
\footnotesize
\begin{align*}\label{zisecderdef}
& \frac{\partial^2 \zeta }{\partial \rho^2_U} = -\frac{1}{4 \left( 1 + \frac{4 \rho_V\sigma_V\rho_U \sigma_U(Z)}{\sigma^2_{\varepsilon,1}(Z)}\right)^{\frac{3}{2}}} \left( \frac{4\rho_V \sigma_V\sigma_U(Z)}{ \sigma^2_{\varepsilon,1}(Z)} + \frac{8\rho^2_V \sigma^2_V\rho_U\sigma^2_U(Z)}{ \sigma^4_{\varepsilon,1}(Z)}\right)^2 \exp \left( - \frac{2\varepsilon^2 \rho_V\sigma_V\rho_U \sigma_U(Z)}{\sigma^2_{\varepsilon,1}(Z) \sigma^2_{\varepsilon,2}(Z)}\right)\nonumber\\
& \quad + \frac{1}{2 \sqrt{1 + \frac{4 \rho_V\sigma_V\rho_U \sigma_U(Z)}{\sigma^2_{\varepsilon,1}(Z)}}} \left( \frac{16\rho^2_V \sigma^2_V\sigma^2_U(Z)}{ \sigma^4_{\varepsilon,1}(Z)} + \frac{32\rho^3_V \sigma^3_V\rho_U\sigma^3_U(Z)}{ \sigma^6_{\varepsilon,1}(Z)}\right) \exp \left( - \frac{2\varepsilon^2 \rho_V\sigma_V\rho_U \sigma_U(Z)}{\sigma^2_{\varepsilon,1}(Z) \sigma^2_{\varepsilon,2}(Z)}\right)\nonumber\\
& \quad - \frac{1}{\sqrt{1 + \frac{4 \rho_V\sigma_V\rho_U \sigma_U(Z)}{\sigma^2_{\varepsilon,1}(Z)}}} \left( \frac{4\rho_V \sigma_V\sigma_U(Z)}{ \sigma^2_{\varepsilon,1}(Z)} + \frac{8\rho^2_V \sigma^2_V\rho_U\sigma^2_U(Z)}{ \sigma^4_{\varepsilon,1}(Z)}\right) \times \nonumber\\
&\qquad \left( \frac{2\varepsilon^2 \rho_V\sigma_V\sigma_U(Z)}{\sigma^2_{\varepsilon,1}(Z) \sigma^2_{\varepsilon,2}(Z)} - \frac{2\varepsilon^2 \rho_V\sigma_V\rho_U\sigma_U(Z)}{\sigma^4_{\varepsilon,1}(Z) \sigma^4_{\varepsilon,2}(Z)} \frac{\partial \sigma^2_{\varepsilon,1}(Z) \sigma^2_{\varepsilon,2}(Z)}{\partial \rho_U}\right)\exp \left( - \frac{2\varepsilon^2 \rho_V\sigma_V\rho_U \sigma_U(Z)}{\sigma^2_{\varepsilon,1}(Z) \sigma^2_{\varepsilon,2}(Z)}\right)\\
& \quad - \sqrt{1 + \frac{4 \rho_V\sigma_V\rho_U \sigma_U(Z)}{\sigma^2_{\varepsilon,1}(Z)}} \left( -\frac{4\varepsilon^2 \rho_V\sigma_V\sigma_U(Z)}{\sigma^4_{\varepsilon,1}(Z) \sigma^4_{\varepsilon,2}(Z)} \frac{\partial \sigma^2_{\varepsilon,1}(Z) \sigma^2_{\varepsilon,2}(Z)}{\partial \rho_U} - \frac{4\varepsilon^2 \rho_V\sigma_V\rho_U\sigma_U(Z)}{\sigma^8_{\varepsilon,1}(Z) \sigma^8_{\varepsilon,2}(Z)} \frac{\partial^2 \sigma^2_{\varepsilon,1}(Z) \sigma^2_{\varepsilon,2}(Z)}{\partial \rho^2_U}\right) \times \nonumber\\
& \qquad \exp \left( - \frac{2\varepsilon^2 \rho_V\sigma_V\rho_U \sigma_U(Z)}{\sigma^2_{\varepsilon,1}(Z) \sigma^2_{\varepsilon,2}(Z)}\right)\nonumber\\
& \quad + \sqrt{1 + \frac{4 \rho_V\sigma_V\rho_U \sigma_U(Z)}{\sigma^2_{\varepsilon,1}(Z)}} \left( \frac{2\varepsilon^2 \rho_V\sigma_V\sigma_U(Z)}{\sigma^2_{\varepsilon,1}(Z) \sigma^2_{\varepsilon,2}(Z)} - \frac{2\varepsilon^2 \rho_V\sigma_V\rho_U\sigma_U(Z)}{\sigma^4_{\varepsilon,1}(Z) \sigma^4_{\varepsilon,2}(Z)} \frac{\partial \sigma^2_{\varepsilon,1}(Z) \sigma^2_{\varepsilon,2}(Z)}{\partial \rho_U}\right)^2 \times \nonumber\\
& \qquad \exp \left( - \frac{2\varepsilon^2 \rho_V\sigma_V\rho_U \sigma_U(Z)}{\sigma^2_{\varepsilon,1}(Z) \sigma^2_{\varepsilon,2}(Z)}\right)\nonumber.
\end{align*}
\normalsize
When $\rho_U = 0$, then $\zeta = 1$, and 
\begin{align*}
\left. \frac{\partial \zeta }{\partial \rho_U} \right\vert_{\rho_U = 0} =& \frac{2 \rho_V\sigma_V \sigma_U(Z) }{\sigma^2(Z)} \left( 1 - \frac{\varepsilon^2}{\sigma^2(Z)}\right),\\
\left. \frac{\partial^2 \zeta }{\partial \rho^2_U} \right\vert_{\rho_U = 0} =& \frac{4 \rho^2_V\sigma^2_V \sigma^2_U(Z) }{\sigma^4(Z)} \left( 1 - \frac{\varepsilon^2}{\sigma^2(Z)}\right)^2, 
\end{align*}
where the last expression follows from the fact that 
\begin{align*}
\left. \frac{\partial \sigma^2_{\varepsilon,1}(Z) \sigma^2_{\varepsilon,2}(Z)}{\partial \rho_U}  \right\vert_{\rho_U = 0}=& \left( - 2 \rho_V \sigma_V \sigma_U(Z) \sigma^2_{\varepsilon,2}(Z) +2 \rho_V \sigma_V \sigma_U(Z) \sigma^2_{\varepsilon,1}(Z)\right)_{\rho_U = 0} = 0.
\end{align*}
Therefore, we finally have that 
\[
\left( \left. \frac{\partial \zeta }{\partial \rho_U} \right\vert_{\rho_U = 0} \right)^2 = \left. \frac{\partial^2 \zeta }{\partial \rho^2_U} \right\vert_{\rho_U = 0} .
\]

The first derivative of the log-likelihood can be written as 
\begin{align*}
\frac{\partial \ell_n(\theta)}{\partial \rho_U} =& \frac{Z}{D_1} \left[ \frac{\partial \Psi_{1,1}(Z,\theta) }{\partial \rho_U}\zeta + \Psi_{1,1}(Z,\theta) \frac{\partial  \zeta}{\partial \rho_U}  +\frac{\partial \Psi_{1,2}(Z,\theta) }{\partial \rho_U} \right] \\
& \quad +  \frac{1-Z}{D_0} \left[ \frac{\partial \Psi_{0,1}(Z,\theta) }{\partial \rho_U}\zeta  + \Psi_{0,1}(Z,\theta) \frac{\partial  \zeta}{\partial \rho_U} + \frac{\partial \Psi_{0,2}(Z,\theta) }{\partial \rho_U} \right] \\
&\quad - \frac{\sigma_V \rho_V\sigma_U(Z) }{ \sigma^2_{\varepsilon,2}(Z)} + \frac{\varepsilon^2 \sigma_V \rho_V \sigma_U(Z) }{ \sigma^4_{\varepsilon,2}(Z)}.
\end{align*}

Thus, when $\rho_U = 0$,
\begin{align*}
\left. \frac{\partial \ell_n(\theta)}{\partial \rho_U} \right\vert_{\rho_U = 0}=& \frac{Z}{2\Psi_{1}(Z,\theta)} \left[ \Psi_{1}(Z,\theta) \frac{2 \rho_V\sigma_V \sigma_U(Z) }{\sigma^2(Z)} \left( 1 - \frac{\varepsilon^2}{\sigma^2(Z)}\right) \right]\\
& \quad +  \frac{1-Z}{2\Psi_{0}(Z,\theta)} \left[ \Psi_{0}(Z,\theta) \frac{2 \rho_V\sigma_V \sigma_U(Z) }{\sigma^2(Z)} \left( 1 - \frac{\varepsilon^2}{\sigma^2(Z)}\right) \right] \\
&\quad - \frac{\sigma_V \rho_V\sigma_U(Z) }{ \sigma^2(Z)} + \frac{\varepsilon^2 \sigma_V \rho_V \sigma_U(Z) }{ \sigma^4(Z)}\\
=& \frac{\rho_V\sigma_V \sigma_U(Z) }{\sigma^2(Z)} \left( 1 - \frac{\varepsilon^2}{\sigma^2(Z)}\right) - \frac{\sigma_V \rho_V\sigma_U(Z) }{ \sigma^2(Z)} + \frac{\varepsilon^2 \sigma_V \rho_V \sigma_U(Z) }{ \sigma^4(Z)} = 0,
\end{align*}
which confirms the result shown above.

The second derivative of the log-likelihood is
\begin{align*}
\frac{\partial^2 \ell_n(\theta)}{\partial \rho^2_U} =& -\frac{Z}{D^2_1} \left[ \frac{\partial \Psi_{1,1}(Z,\theta) }{\partial \rho_U}\zeta + \Psi_{1,1}(Z,\theta) \frac{\partial  \zeta}{\partial \rho_U}  + \frac{\partial \Psi_{1,2}(Z,\theta) }{\partial \rho_U} \right]^2 \\
& \quad + \frac{Z}{D_1} \left[ \frac{\partial^2 \Psi_{1,1}(Z,\theta) }{\partial \rho^2_U}\zeta +2 \frac{\partial \Psi_{1,1}(Z,\theta) }{\partial \rho_U} \frac{\partial  \zeta}{\partial \rho_U} + \Psi_{1,1}(Z,\theta) \frac{\partial^2  \zeta}{\partial \rho^2_U}  +\frac{\partial^2 \Psi_{1,2}(Z,\theta) }{\partial \rho^2_U} \right]\\
& \quad - \frac{1-Z}{D^2_0} \left[ \frac{\partial \Psi_{0,1}(Z,\theta) }{\partial \rho_U} + \Psi_{0,1}(Z,\theta) \frac{\partial  \zeta}{\partial \rho_U} + \frac{\partial \Psi_{0,2}(Z,\theta) }{\partial \rho_U} \right]^2 \\
& \quad + \frac{1- Z}{D_0} \left[ \frac{\partial^2 \Psi_{0,1}(Z,\theta) }{\partial \rho^2_U}\zeta +2 \frac{\partial \Psi_{0,1}(Z,\theta) }{\partial \rho_U} \frac{\partial  \zeta}{\partial \rho_U} + \Psi_{0,1}(Z,\theta) \frac{\partial^2  \zeta}{\partial \rho^2_U}  +\frac{\partial^2 \Psi_{0,2}(Z,\theta) }{\partial \rho^2_U} \right]\\
&\quad + \frac{2 \sigma^2_V \rho^2_V\sigma^2_U(Z) }{ \sigma^4_{\varepsilon,2}(Z)} - \frac{4 \varepsilon^2 \sigma^2_V \rho^2_V \sigma^2_U(Z) }{ \sigma^6_{\varepsilon,2}(Z)}.
\end{align*}

Thus, when $\rho_U = 0$,
\begin{align*}
\left. \frac{\partial^2 \ell_n(\theta)}{\partial \rho^2_U}\right\vert_{\rho_U = 0 } =& -\frac{Z}{4\Psi^2_{1}(Z,\theta)} \left[ \Psi_{1}(Z,\theta) \left. \frac{\partial  \zeta}{\partial \rho_U} \right\vert_{\rho_U = 0 } \right]^2 \\
& \quad + \frac{Z}{2\Psi_{1}(Z,\theta)} \left[ 2 \frac{\partial^2 \Psi_{1}(Z,\theta) }{\partial \rho^2_U} +2 \frac{\partial \Psi_{1}(Z,\theta) }{\partial \rho_U} \left. \frac{\partial \zeta}{\partial \rho_U} \right\vert_{\rho_U = 0 } + \Psi_{1}(Z,\theta)\left. \frac{\partial^2  \zeta}{\partial \rho^2_U} \right\vert_{\rho_U = 0 } \right]\\
& \quad - \frac{1-Z}{4\Psi^2_{0}(Z,\theta)} \left[ \Psi_{0}(Z,\theta) \left. \frac{\partial  \zeta}{\partial \rho_U}\right\vert_{\rho_U = 0 } \right]^2 \\
& \quad + \frac{1- Z}{2 \Psi_{0}(Z,\theta)} \left[ 2 \frac{\partial^2 \Psi_{0}(Z,\theta) }{\partial \rho^2_U} +2 \frac{\partial \Psi_{0}(Z,\theta) }{\partial \rho_U} \left. \frac{\partial  \zeta}{\partial \rho_U} \right\vert_{\rho_U = 0 } + \Psi_{0}(Z,\theta) \left. \frac{\partial^2  \zeta}{\partial \rho^2_U} \right\vert_{\rho_U = 0 }  \right]\\
&\quad + \frac{2 \sigma^2_V \rho^2_V\sigma^2_U(Z) }{ \sigma^4(Z)} - \frac{4 \varepsilon^2 \sigma^2_V \rho^2_V \sigma^2_U(Z) }{ \sigma^6(Z)} \\
=& \frac{1}{4} \left( \left. \frac{\partial  \zeta}{\partial \rho_U} \right\vert_{\rho_U = 0 } \right)^2 +\frac{Z}{\Psi_{1}(Z,\theta)}\left[ \frac{\partial^2 \Psi_{1}(Z,\theta) }{\partial \rho^2_U} + \frac{\partial \Psi_{1}(Z,\theta) }{\partial \rho_U} \left. \frac{\partial \zeta}{\partial \rho_U} \right\vert_{\rho_U = 0 } \right]\\
& \quad + \frac{1- Z}{\Psi_{0}(Z,\theta)} \left[\frac{\partial^2 \Psi_{0}(Z,\theta) }{\partial \rho^2_U} + \frac{\partial \Psi_{0}(Z,\theta) }{\partial \rho_U} \left. \frac{\partial  \zeta}{\partial \rho_U} \right\vert_{\rho_U = 0 } \right] \\
&\quad + \frac{2 \sigma^2_V \rho^2_V\sigma^2_U(Z) }{ \sigma^4(Z)} - \frac{4 \varepsilon^2 \sigma^2_V \rho^2_V \sigma^2_U(Z) }{ \sigma^6(Z)}\\
=& \frac{\rho^2_V\sigma^2_V \sigma^2_U(Z) }{\sigma^4(Z)} \left( 1 - \frac{\varepsilon^2}{\sigma^2(Z)}\right)^2 + \frac{2 \sigma^2_V \rho^2_V\sigma^2_U(Z) }{ \sigma^4(Z)} - \frac{4 \varepsilon^2 \sigma^2_V \rho^2_V \sigma^2_U(Z) }{ \sigma^6(Z)}\\
&\quad + \frac{Z}{\Psi_{1}(Z,\theta)}\left[ \frac{\partial^2 \Psi_{1}(Z,\theta) }{\partial \rho^2_U} + \frac{\partial \Psi_{1}(Z,\theta) }{\partial \rho_U}\frac{ 2\rho_V\sigma_V \sigma_U(Z) }{\sigma^2(Z)} \left( 1 - \frac{\varepsilon^2}{\sigma^2(Z)}\right) \right]\\
&\quad + \frac{1- Z}{\Psi_{0}(Z,\theta)} \left[\frac{\partial^2 \Psi_{0}(Z,\theta) }{\partial \rho^2_U} + \frac{\partial \Psi_{0}(Z,\theta) }{\partial \rho_U} \frac{ 2\rho_V\sigma_V \sigma_U(Z) }{\sigma^2(Z)} \left( 1 - \frac{\varepsilon^2}{\sigma^2(Z)}\right) \right]\\
=& \frac{\rho^2_V\sigma^2_V \sigma^2_U(Z) }{\sigma^4(Z)} \left( 3 - \frac{6\varepsilon^2}{\sigma^2(Z)} + \frac{\varepsilon^4 }{ \sigma^4(Z)} \right)\\
&\quad + \frac{Z}{\Psi_{1}(Z,\theta)}\left[ \frac{\partial^2 \Psi_{1}(Z,\theta) }{\partial \rho^2_U} + \frac{\partial \Psi_{1}(Z,\theta) }{\partial \rho_U}\frac{2 \rho_V\sigma_V \sigma_U(Z) }{\sigma^2(Z)} \left( 1 - \frac{\varepsilon^2}{\sigma^2(Z)}\right) \right]\\
&\quad + \frac{1- Z}{\Psi_{0}(Z,\theta)} \left[\frac{\partial^2 \Psi_{0}(Z,\theta) }{\partial \rho^2_U} + \frac{\partial \Psi_{0}(Z,\theta) }{\partial \rho_U} \frac{ 2\rho_V\sigma_V \sigma_U(Z) }{\sigma^2(Z)} \left( 1 - \frac{\varepsilon^2}{\sigma^2(Z)}\right) \right]
\end{align*}

We take the expectation wrt the joint density of $(\varepsilon,Z,\tilde{W})$. Because $(\varepsilon,\eta) \upmodels \tilde{W}$ and $\varepsilon \upmodels Z \vert \eta$, then
\begin{align*}
f_{\varepsilon,Z,\tilde{W}} \left( e,z,\tilde{w} \right) =& f_{\varepsilon \vert Z,\tilde{W}} \left( e \vert z, \tilde{w} \right) P\left( Z=z \vert \tilde{w} \right) f_{\tilde{W}}(\tilde{w}) \\
=& \begin{cases} f_{\varepsilon \vert \eta \leq -,\tilde{W}} \left( e \vert \eta \leq -\tilde{w}\gamma, \tilde{w} \right) \Phi\left( -\tilde{w}\gamma \right)  f_{\tilde{W}}(\tilde{w}) & \\
\qquad = F_{\eta \vert \varepsilon}\left( - \tilde{w} \gamma \vert e \right) f_\varepsilon (e) f_{\tilde{W}}(\tilde{w}) & \text{ if } z = 0 \\
f_{\varepsilon \vert \eta \geq -\tilde{W} \gamma,\tilde{W}} \left( e \vert \eta \geq -\tilde{w}\gamma, \tilde{w} \right) \left( 1 - \Phi\left( -\tilde{w} \gamma \right)\right) f_{\tilde{W}}(\tilde{w}) & \\
\qquad = \left[ 1- F_{\eta \vert \varepsilon}\left( - \tilde{w} \gamma \vert e \right) \right] f_\varepsilon (e) f_{\tilde{W}}(\tilde{w}) & \text{ if } z = 1 \end{cases}
\end{align*}

Therefore, for any function $g(\varepsilon)$,
\[
E \left[ Z g (\varepsilon) \right] = \int_{\tilde{w}} \left( \int_{\varepsilon} \frac{\Psi_{1}(1,\theta)}{\Phi \left( \tau(1) \varepsilon \right)} g(\varepsilon) f_{\varepsilon,1} (\varepsilon) d\varepsilon \right) f_{\tilde{W}}(\tilde{w}) d\tilde{w} = E_{\tilde{W}}\left[ \int_{\varepsilon} \frac{\Psi_{1}(1,\theta)}{\Phi \left( \tau(1) \varepsilon \right)} g(\varepsilon) f_{\varepsilon,1} (\varepsilon) d\varepsilon \right] ,
\]
and 
\[
E \left[ ( 1- Z) g (\varepsilon) \right] = \int_{\tilde{w}} \left( \int_{\varepsilon}\frac{\Psi_{0}(0,\theta)}{\Phi \left( \tau(0) \varepsilon \right)} f_{\varepsilon,0} (\varepsilon) d\varepsilon \right) f_{\tilde{W}}(\tilde{w}) d\tilde{w} = E_{\tilde{W}}\left[ \int_{\varepsilon} \frac{\Psi_{0}(0,\theta)}{\Phi \left( \tau(0) \varepsilon \right)} g(\varepsilon) f_{\varepsilon,0} (\varepsilon)d\varepsilon \right],
\]
with 
\[
f_{\varepsilon,z}(\varepsilon) = \frac{2\Phi \left( \tau(z) \varepsilon \right)}{\sigma(z)}\phi\left( \frac{\varepsilon}{\sigma(z)}\right), \quad z \in \lbrace 0,1 \rbrace.
\]
Therefore,
\begin{align*}
E & \left[ \left. \frac{\partial^2 \ell_n(\theta)}{\partial \rho^2_U}\right\vert_{\rho_U = 0 }\right] =E \left[ \frac{\rho^2_V\sigma^2_V \sigma^2_U(Z) }{\sigma^4(Z)} \left( 3 - \frac{6\varepsilon^2}{\sigma^2(Z)} + \frac{\varepsilon^4 }{ \sigma^4(Z)} \right) \right] \\
&\quad + E\left[ \frac{Z}{\Psi_{1}(Z,\theta)}\left( \frac{\partial^2 \Psi_{1}(Z,\theta) }{\partial \rho^2_U} + \frac{\partial \Psi_{1}(Z,\theta) }{\partial \rho_U}\frac{ 2\rho_V\sigma_V \sigma_U(Z) }{\sigma^2(Z)} \left( 1 - \frac{\varepsilon^2}{\sigma^2(Z)}\right) \right) \right] \\
&\quad + E\left[ \frac{1- Z}{\Psi_{0}(Z,\theta)} \left(\frac{\partial^2 \Psi_{0}(Z,\theta) }{\partial \rho^2_U} + \frac{\partial \Psi_{0}(Z,\theta) }{\partial \rho_U} \frac{ 2\rho_V\sigma_V \sigma_U(Z) }{\sigma^2(Z)} \left( 1 - \frac{\varepsilon^2}{\sigma^2(Z)}\right)\right) \right]\\
=& E \left[ \frac{\rho^2_V\sigma^2_V \sigma^2_U(Z) }{\sigma^4(Z)} \left( 3 - \frac{6E\left[ \varepsilon^2 \vert \eta \right] }{\sigma^2(Z)} + \frac{E\left[ \varepsilon^4 \vert \eta \right] }{ \sigma^4(Z)} \right) \right] \tag{$SD_1$} \\
&\quad + E_{\tilde{W}}\left[ \int_\varepsilon \left( \frac{\partial^2 \Psi_{1}(1,\theta) }{\partial \rho^2_U} + \frac{\partial \Psi_{1}(1,\theta) }{\partial \rho_U}\frac{ 2\rho_V\sigma_V \sigma_U(1) }{\sigma^2(1)} \left( 1 - \frac{\varepsilon^2}{\sigma^2(1)}\right) \right) \frac{2}{\sigma(1)}\phi\left( \frac{\varepsilon}{\sigma(1)}\right) d\varepsilon \right] \tag{$SD_2$}\\
&\quad + E_{\tilde{W}}\left[ \int_\varepsilon \left( \frac{\partial^2 \Psi_{0}(0,\theta) }{\partial \rho^2_U} + \frac{\partial \Psi_{0}(0,\theta) }{\partial \rho_U} \frac{ 2\rho_V\sigma_V \sigma_U(0) }{\sigma^2(0)} \left( 1 - \frac{\varepsilon^2}{\sigma^2(0)}\right) \right) \frac{2}{\sigma(0)}\phi\left( \frac{\varepsilon}{\sigma(0)}\right) d\varepsilon \right]\tag{$SD_3$}.
\end{align*}

%%%%%%%%%%%%%%% FIRST TERM OF THE SECOND DERIVATIVE (SD1)
To treat the term is $SD_1$, when $\rho_U$ is equal to zero, we notice that the density of $\varepsilon$ conditional on $\eta$ is a $SN(\rho_V\sigma_V \eta,\tsigma(z),\lambda(z))$, which implies that
\begin{align*}
E\left[ \varepsilon^2 \vert \eta \right] =& \tsigma^2(Z) + \sigma_V^2 \rho_V^2 \eta^2 - 2 \sqrt{\frac{2}{\pi}}\sigma_U(Z) \sigma_V \rho_V \eta\\
=& \sigma^2(Z) + \sigma_V^2 \rho_V^2 \left( \eta^2 - 1\right) - 2 \sqrt{\frac{2}{\pi}}\sigma_U(Z) \sigma_V \rho_V \eta,
\end{align*}
and 
\begin{align*}
E\left[ \varepsilon^4 \vert \eta \right] =& 3\tsigma^4(Z) - \left( 12\sqrt{\frac{2}{\pi}}\tsigma^2(Z)\sigma_U(Z) - 4\sqrt{\frac{2}{\pi}}\sigma^3_U(Z) \right)\sigma_V \rho_V \eta + 6 \tsigma^2(Z) \sigma_V^2 \rho_V^2 \eta^2 \\
& \quad - 4\sqrt{\frac{2}{\pi}}\sigma_U(Z) \sigma_V^3 \rho_V^3 \eta^3 + \sigma_V^4 \rho_V^4 \eta^4 \\
=&3\sigma^4(Z) + 3 \sigma_V^4 \rho_V^4 - 6 \sigma^2(Z)\sigma_V^2 \rho_V^2 \\
& \quad - \sqrt{\frac{2}{\pi}} \left( 12\tsigma^2_V + 8 \sigma^2_U(Z) \right)\sigma_V \rho_V \eta \sigma_U(z) \\
& \quad + 6 \sigma^2(Z) \sigma_V^2 \rho_V^2 \eta^2 - 6 \sigma_V^4 \rho_V^4 \eta^2 - 4\sqrt{\frac{2}{\pi}}\sigma_U(Z) \sigma_V^3 \rho_V^3 \eta^3 + \sigma_V^4 \rho_V^4 \eta^4\\
=&3\sigma^4(Z) + \rho_V^4 \sigma_V^4\left( 3 - 6 \eta^2 + \eta^4 \right)  + 6 \sigma^2(Z)\sigma_V^2 \rho_V^2 \left( \eta^2 - 1 \right) \\
& \quad - \sqrt{\frac{2}{\pi}} \left( 12\tsigma^2_V + 8 \sigma^2_U(Z) \right)\sigma_V \rho_V \eta \sigma_U(z) \\
& \quad - 4\sqrt{\frac{2}{\pi}}\sigma_U(Z) \sigma_V^3 \rho_V^3 \eta^3 .
\end{align*}
Therefore
\begin{align*}
3 - \frac{6E\left[ \varepsilon^2 \vert \eta \right] }{\sigma^2(Z)} + \frac{E\left[ \varepsilon^4 \vert \eta \right] }{ \sigma^4(Z)} =& 3 - 6 - 6 \frac{\sigma_V^2 \rho_V^2 \left( \eta^2 - 1\right)}{\sigma^2(Z)} + 12 \sqrt{\frac{2}{\pi}} \frac{\sigma_U(Z) \sigma_V \rho_V \eta}{\sigma^2(Z)}\\
& \quad + 3 + \frac{\rho_V^4 \sigma_V^4}{\sigma^4(z)}\left( 3 - 6 \eta^2 + \eta^4 \right)  + 6\frac{\sigma_V^2 \rho_V^2\left( \eta^2 - 1 \right)}{\sigma^2(z)}  \\
& \quad - \sqrt{\frac{2}{\pi}} \left( 12\tsigma^2_V + 8 \sigma^2_U(Z) \right) \frac{\sigma_U(z) \rho_V\sigma_V \eta}{\sigma^4(z)} - 4\sqrt{\frac{2}{\pi}}\frac{\sigma_U(Z) \sigma_V^3 \rho_V^3 \eta^3}{\sigma^4(z)}\\
=& \frac{\sigma_V^4 \rho_V^4}{\sigma^4(Z)} \left[ 3 - 6 \eta^2 + \eta^4 \right] + 12 \sqrt{\frac{2}{\pi}} \frac{\sigma_U(Z) \sigma^3_V \rho^3_V \eta}{\sigma^4(Z)} \\
& \quad + 4 \sqrt{\frac{2}{\pi}} \frac{\sigma^3_U(z) \rho_V\sigma_V \eta}{\sigma^4(z)} - 4\sqrt{\frac{2}{\pi}}\frac{\sigma_U(Z) \sigma_V^3 \rho_V^3 \eta^3}{\sigma^4(z)}.
\end{align*}

When $Z = 0$, the distribution of $\eta \vert \eta \leq -\cez$ is a normal truncated above at $-\tilde{W\gamma}$. When $Z=1$, the distribution of $\eta \vert \eta \geq -\cez$ is instead a normal truncated below at $-\tilde{W\gamma}$. Therefore, when $Z = 0$,
\begin{align*}
E\left[ \eta \vert \eta \leq - \cez\right] =& -\frac{\phi(-\cez)}{\Phi(-\cez)}\\
E\left[ \eta^2 \vert \eta \leq - \cez\right] =& 1 + \cez \frac{\phi(-\cez)}{\Phi(-\cez)}\\
E\left[ \eta^3 \vert \eta \leq - \cez\right] =& -2 \frac{\phi(-\cez)}{\Phi(-\cez)} - (\cez)^2\frac{\phi(-\cez)}{\Phi(-\cez)}\\
E\left[ \eta^4 \vert \eta \leq - \cez\right] =& 3+ 3 \cez \frac{\phi(-\cez)}{\Phi(-\cez)} + (\cez)^3 \frac{\phi(-\cez)}{\Phi(-\cez)},
\end{align*}
and when $Z = 1$
\begin{align*}
E\left[ \eta \vert \eta \geq - \cez\right] =& \frac{\phi(-\cez)}{1 - \Phi(-\cez)}\\
E\left[ \eta^2 \vert \eta \geq - \cez\right] =& 1 - \cez \frac{\phi(-\cez)}{1 - \Phi(-\cez)}\\
E\left[ \eta^3 \vert \eta \geq - \cez\right] =& 2 \frac{\phi(-\cez)}{1 - \Phi(-\cez)} + (\cez)^2\frac{\phi(-\cez)}{1 - \Phi(-\cez)}\\
E\left[ \eta^4 \vert \eta \geq - \cez\right] =& 3- 3 \cez \frac{\phi(-\cez)}{1 - \Phi(-\cez)} - (\cez)^3 \frac{\phi(-\cez)}{1 - \Phi(-\cez)}.
\end{align*}

Thus, we finally have that,
\begin{align*}
SD_1 =& \left( \frac{\rho^6_V\sigma^6_V \sigma^2_U(1) }{\sigma^8(1)} - \frac{\rho^6_V\sigma^6_V \sigma^2_U(0) }{\sigma^8(0)}\right) E_{\tilde{W}}\left[ \left( 3 \cez - (\cez)^3 \right)\phi(-\cez) \right]\\
& \quad - 4\sqrt{\frac{2}{\pi}} \left( \frac{\sigma^3_U(1) \rho_V^5 \sigma_V^5 }{\sigma^8(1)} -\frac{\sigma^3_U(0) \rho_V^5\sigma_V^5 }{\sigma^8(0)} \right) E_{\tilde{W}}\left[ \left( (\cez)^2 - 1 \right) \phi(-\cez) \right] \\
& \quad + 4\sqrt{\frac{2}{\pi}} \left( \frac{\sigma^5_U(1) \rho_V^3 \sigma_V^3 }{\sigma^8(1)} -\frac{\sigma^5_U(0) \rho_V^3\sigma_V^3 }{\sigma^8(0)} \right) E_{\tilde{W}}\left[ \phi(-\cez) \right]
\end{align*}

%%%%%%%%%%%%%%% SECOND TERM OF THE SECOND DERIVATIVE (SD1)
The first derivative of $\Psi_0(z,\theta)$ wrt $\rho_U$ when $\rho_U= 0$ is equal to 
\begin{align*}
\left. \frac{\partial\Psi_{j}(z,\theta)}{\partial \rho_U} \right\vert_{\rho_U = 0} =& \Phi \left( - \lambda(z) \left( \frac{\varepsilon + \cez \rho_V \sigma_V}{\tsigma(z)}\right)\right) \phi \left( - \frac{\sigma(z)}{\tsigma(z)}\left( \cez + \frac{\rho_V\sigma_V}{\sigma^2(z)}\varepsilon\right) \right)\times\\
& \qquad \qquad \left[\frac{\sigma_U(z) \tsigma(z)}{\sigma^3(z)}\varepsilon + \frac{\rho_V \sigma_V \sigma_U(z)}{\sigma(z) \tsigma(z)} \cez \right]\\ 
& \quad + \Phi \left( \frac{\cez - \frac{\rho_V}{\sigma_V}\varepsilon}{\sqrt{1- \rho_V^2}} \right) \phi\left( - \frac{\sigma_U(z)}{\sigma_V \sigma(z)}\varepsilon\right)\frac{\rho_V\sigma_V^2}{\sigma^3(z)}\varepsilon \\
& - \frac{\tsigma^2_V }{\sigma(z)} \frac{1}{\tsigma_V} \phi_{2,\rho^\ast(z) } \left( - \frac{\sigma(z)}{\tsigma(z)}\left( \cez + \frac{\rho_V\sigma_V}{\sigma^2(z)}\varepsilon\right), - \frac{\sigma_U(z)}{\sigma_V \sigma(z)}\varepsilon\right),
\end{align*}
where $\phi_{2,\rho^\ast(z)}$ is the kernel of the standard bivariate normal density with correlation parameter equal to $\rho^\ast(z)$. 

Its second derivative wrt to $\rho_U$ is instead equal to 
\begin{align*}
&\left. \frac{\partial^2 \Psi_0(z,\theta)}{\partial \rho^2_U} \right\vert_{\rho_U = 0} =\Phi \left( - \lambda(z) \left( \frac{\varepsilon + \cez \rho_V \sigma_V}{\tsigma(z)}\right)\right) \phi \left( - \frac{\sigma(z)}{\tsigma(z)}\left( \cez + \frac{\rho_V\sigma_V}{\sigma^2(z)}\varepsilon\right) \right)\times\\
& \qquad \qquad \left\lbrace \frac{\rho_V \sigma_V\sigma^2_U(z)}{\tsigma^3(z)\sigma^5(z)} \left[ 3 \rho_V^4\sigma_V^4 - 5 \rho_V^2 \sigma_V^2 \sigma^2(z) + \sigma^4(z)\right] \varepsilon -\frac{\sigma^2_U(z)}{\tsigma^3(z)\sigma^3(z)}\left[\tsigma^4(z) + \rho^2_V \sigma^2_V \sigma^2_U(z)\right]\cez \right.\\ 
& \qquad \qquad \left. + \frac{\sigma(z)}{\tsigma(z)}\left[ \cez + \frac{\rho_V \sigma_V}{\sigma^2(z)}\varepsilon \right] \left[\frac{\sigma_U(z) \tsigma(z)}{\sigma^3(z)}\varepsilon + \frac{\rho_V \sigma_V \sigma_U(z)}{\sigma(z) \tsigma(z)} \cez \right]^2 \right\rbrace\\ 
& \quad + \Phi \left( \frac{\cez - \frac{\rho_V}{\sigma_V}\varepsilon}{\sqrt{1- \rho_V^2}} \right) \phi\left( - \frac{\sigma_U(z)}{\sigma_V \sigma(z)}\varepsilon\right)\left[ - \frac{\rho_V^2\sigma_U(z)\left( \sigma_U^4(z) + 3 \sigma^2_U(z) \sigma^2_V - \sigma^4_V \right) \varepsilon}{\sigma_V \sigma^5(z)}+ \frac{\rho^2_V\sigma_V^3\sigma_U(z)}{\sigma^7(z)}\varepsilon^3 \right]\\
& + \frac{\tsigma(z)\sigma_V}{\sigma(z)}\frac{1}{\tsigma_V} \phi_{2,\rho^\ast(z) } \left( - \frac{\sigma(z)}{\tsigma(z)}\left( \cez + \frac{\rho_V\sigma_V}{\sigma^2(z)}\varepsilon\right), - \frac{\sigma_U(z)}{\sigma_V \sigma(z)}\varepsilon\right)\times \\
& \quad \left\lbrace \frac{\rho_V(1- \rho_V^2) \sigma_U(z)\sigma^2(z)}{\tsigma^3(z)} + \frac{\rho^3_V(1- \rho_V^2) \sigma^2_V\sigma^3_U(z)}{\tsigma^3(z)\sigma^2(z)}\right. \\
& \quad - \left. \varepsilon^2 \frac{\rho_V (1 - \rho_V^2) \sigma^2_V \sigma_U(z)}{\tsigma(z)\sigma^4(z)} - \cez \varepsilon \frac{(1 - \rho_V^2) \sigma_V \sigma_U(z)}{\tsigma(z)\sigma^2(z)} - (\cez)^2 \frac{\rho_V(1 - \rho_V^2) \sigma^2_V \sigma_U(z)}{\tsigma^3(z)} \right\rbrace,
\end{align*}

Thus
\begin{align*}
& \int_\varepsilon \frac{\partial^2 \Psi_0(z,\theta) }{\partial \rho^2_U} + \frac{\partial \Psi_0(z,\theta) }{\partial \rho_U}\frac{ 2\rho_V\sigma_V \sigma_U(z) }{\sigma^2(z)} \left( 1 - \frac{\varepsilon^2}{\sigma^2(z)}\right) d\varepsilon\\
=& \int_\varepsilon \Phi \left( - \lambda(z) \left( \frac{\varepsilon + \cez \rho_V \sigma_V}{\tsigma(z)}\right)\right) \phi \left( - \frac{\sigma(z)}{\tsigma(z)}\left( \cez + \frac{\rho_V\sigma_V}{\sigma^2(z)}\varepsilon\right) \right) \times\\
& \quad \left\lbrace 2 \left[\frac{\rho_V\sigma_V\sigma^2_U(z) \tsigma(z)}{\sigma^5(z)}\varepsilon + \frac{\rho^2_V \sigma^2_V \sigma^2_U(z)}{\sigma^3(z) \tsigma(z)} \cez \right]\left( 1 - \frac{\varepsilon^2}{\sigma^2(z)}\right) \right.\\
& \quad +\frac{\rho_V \sigma_V\sigma^2_U(z)}{\tsigma^3(z)\sigma^5(z)} \left[ 3 \rho_V^4\sigma_V^4 - 5 \rho_V^2 \sigma_V^2 \sigma^2(z) + \sigma^4(z)\right] \varepsilon -\frac{\sigma^2_U(z)}{\tsigma^3(z)\sigma^3(z)}\left[\tsigma^4(z) + \rho^2_V \sigma^2_V \sigma^2_U(z)\right]\cez \\ 
& \quad + \left. \frac{\sigma(z)}{\tsigma(z)}\left[ \cez + \frac{\rho_V \sigma_V}{\sigma^2(z)}\varepsilon \right] \left[\frac{\sigma_U(z) \tsigma(z)}{\sigma^3(z)}\varepsilon + \frac{\rho_V \sigma_V \sigma_U(z)}{\sigma(z) \tsigma(z)} \cez \right]^2 \right\rbrace \frac{2}{\sigma(z)}\phi\left( \frac{\varepsilon}{\sigma(z)}\right) d\varepsilon \\ 
& \quad + \int_\varepsilon \Phi \left( \frac{\cez - \frac{\rho_V}{\sigma_V}\varepsilon}{\sqrt{1- \rho_V^2}} \right) \phi\left( - \frac{\sigma_U(z)}{\sigma_V \sigma(z)}\varepsilon\right)\left\lbrace 2\frac{\rho^2_V\sigma_V^3\sigma_U(z)}{\sigma^5(z)}\left( \varepsilon - \frac{\varepsilon^3}{\sigma^2(z)}\right) \right.\\
& \quad \left. - \frac{\rho_V^2\sigma_U(z)\left( \sigma_U^4(z) + 3 \sigma^2_U(z) \sigma^2_V - \sigma^4_V \right) \varepsilon}{\sigma_V \sigma^5(z)}+ \frac{\rho^2_V\sigma_V^3\sigma_U(z)}{\sigma^7(z)}\varepsilon^3 \right\rbrace \frac{2}{\sigma(z)} \phi\left( \frac{\varepsilon}{\sigma(z)}\right) d\varepsilon\\
& +\int_\varepsilon \frac{1}{\tsigma_V} \phi_{2,\rho^\ast(z) } \left( - \frac{\sigma(z)}{\tsigma(z)}\left( \cez + \frac{\rho_V\sigma_V}{\sigma^2(z)}\varepsilon\right), - \frac{\sigma_U(z)}{\sigma_V \sigma(z)}\varepsilon\right) \frac{2}{\sigma(z)}\phi\left( \frac{\varepsilon}{\sigma(z)}\right) \\
& \quad \left\lbrace - 2\frac{\rho_V \sigma_V \tsigma^2_V \sigma_U(z) }{\sigma^3 (z)} \left( 1 - \frac{\varepsilon^2}{\sigma^2(z)}\right) + \frac{\rho_V\sigma_V(1- \rho_V^2) \sigma_U(z)\sigma(z)}{\tsigma^2(z)} + \frac{\rho^3_V(1- \rho_V^2) \sigma^4_V\sigma^3_U(z)}{\tsigma(z)\sigma^4(z)}\right. \\
& \quad - \left. \varepsilon^2 \frac{\rho_V (1 - \rho_V^2) \sigma^3_V \sigma_U(z)}{\sigma^5(z)} - \cez \varepsilon \frac{(1 - \rho_V^2) \sigma^2_V \sigma_U(z)}{\sigma^3(z)} - (\cez)^2 \frac{\rho_V(1 - \rho_V^2) \sigma^3_V \sigma_U(z)}{\tsigma^2(z)\sigma(z)} \right\rbrace d\varepsilon\\
% NEW LINE EQUATION
=& \phi(-\cez) \int_\varepsilon \left\lbrace 2 \left[\frac{\rho_V\sigma_V\sigma^2_U(z) \tsigma^2(z)}{\sigma^6(z)}\varepsilon + \frac{\rho^2_V \sigma^2_V \sigma^2_U(z)}{\sigma^4(z)} \cez \right]\left( 1 - \frac{\varepsilon^2}{\sigma^2(z)}\right) \right.\\
& \quad +\frac{\rho_V \sigma_V\sigma^2_U(z)}{\tsigma^2(z)\sigma^6(z)} \left[ 3 \rho_V^4\sigma_V^4 - 5 \rho_V^2 \sigma_V^2 \sigma^2(z) + \sigma^4(z)\right] \varepsilon -\frac{\sigma^2_U(z)}{\tsigma^2(z)\sigma^4(z)}\left[\tsigma^4(z) + \rho^2_V \sigma^2_V \sigma^2_U(z)\right]\cez \\ 
& \quad + \left.\left[ \cez + \frac{\rho_V \sigma_V}{\sigma^2(z)}\varepsilon \right] \left[\frac{\sigma_U(z) \tsigma(z)}{\sigma^3(z)}\varepsilon + \frac{\rho_V \sigma_V \sigma_U(z)}{\sigma(z) \tsigma(z)} \cez \right]^2\right\rbrace \times \\
& \qquad \frac{2}{\tsigma(z)}\Phi \left( - \lambda(z) \left( \frac{\varepsilon + \cez \rho_V \sigma_V}{\tsigma(z)}\right)\right) \phi\left( \frac{\varepsilon + \cez \rho_V \sigma_V}{\tsigma(z)}\right) d\varepsilon \tag{$I$}\\
 -& \Phi(-\cez)\sqrt{\frac{2}{\pi}}\int_\varepsilon \frac{\Phi \left( -\frac{\cez}{\sqrt{1- \rho_V^2}} - \frac{\rho_V}{\sqrt{1- \rho_V^2}}\frac{\varepsilon}{\sigma_V} \right)}{\Phi(-\cez)} \frac{1}{\sigma_V} \phi\left( \frac{\varepsilon}{\sigma_V}\right) \times\\
 & \qquad \left[ \frac{\rho_V^2\sigma_U(z)\left( \sigma_U^4(z) + 3 \sigma^2_U(z) \sigma^2_V - 3\sigma^4_V \right) \varepsilon}{\sigma^6(z)} + \frac{\rho^2_V\sigma_V^4\sigma_U(z)}{\sigma^8(z)}\varepsilon^3\right] d\varepsilon \tag{$II$}\\
 & -\phi(-\cez) \sqrt{\frac{2}{\pi}} \int_\varepsilon \frac{1}{\tsigma_V} \phi\left( \frac{\varepsilon + \cez\rho_V \sigma_V}{\tsigma_V}\right) \\
 & \quad \left\lbrace - 2\frac{\rho_V \sigma_V \tsigma^2_V \sigma_U(z) }{\sigma^4 (z)} \left( 1 - \frac{\varepsilon^2}{\sigma^2(z)}\right) + \frac{\rho_V\sigma_V(1- \rho_V^2) \sigma_U(z)}{\tsigma^2(z)} + \frac{\rho^3_V(1- \rho_V^2) \sigma^4_V\sigma^3_U(z)}{\tsigma(z)\sigma^5(z)}\right. \\
& \quad - \left. \varepsilon^2 \frac{\rho_V (1 - \rho_V^2) \sigma^3_V \sigma_U(z)}{\sigma^6(z)} - \cez \varepsilon \frac{(1 - \rho_V^2) \sigma^2_V \sigma_U(z)}{\sigma^4(z)} - (\cez)^2 \frac{\rho_V(1 - \rho_V^2) \sigma^3_V \sigma_U(z)}{\tsigma^2(z)\sigma^2(z)} \right\rbrace d\varepsilon \tag{$III$}
\end{align*}

We now treat each one each of these terms. Let 
\[
h_{\varepsilon \vert \cez}(\varepsilon \vert \cez) = \Phi \left( -\lambda(z)\left( \frac{\varepsilon + \rho_V \sigma_V \cez}{\tsigma(z)}\right)\right) \frac{2}{\tsigma(z)}\phi\left( \frac{\varepsilon + \cez \rho_V \sigma_V}{\tsigma(z)}\right),
\]
be the density of a $SN(-\cez \rho_V \sigma_V,\tsigma(z),-\lambda(z))$. Therefore,
\begin{align*}
\int_\varepsilon \varepsilon h_{\varepsilon \vert \cez}(\varepsilon \vert \cez) d\varepsilon =& -\cez \rho_V \sigma_V - \sqrt{\frac{2}{\pi}} \sigma_U(z),\\
\int_\varepsilon \varepsilon^2 h_{\varepsilon \vert \cez}(\varepsilon \vert \cez) d\varepsilon =& \tsigma^2(z) + (\cez)^2 \rho^2_V \sigma^2_V + 2 \sqrt{\frac{2}{\pi}} \cez \rho_V \sigma_V \sigma_U(z),\\
\int_\varepsilon \varepsilon^3 h_{\varepsilon \vert \cez}(\varepsilon \vert \cez) d\varepsilon =& -3\tsigma^2(z) \cez \rho_V \sigma_V - 3\sqrt{\frac{2}{\pi}} \tsigma^2(z) \sigma_U(z) - (\cez)^3 \rho^3_V \sigma^3_V\\
& \quad - 3 \sqrt{\frac{2}{\pi}} (\cez)^2 \rho_V^2 \sigma_V^2 \sigma_U(z) + \sqrt{\frac{2}{\pi}} \sigma_U^3(z).
\end{align*}
Thus, we have that
\begin{align*}
I =& \phi(-\cez) \int_\varepsilon \left\lbrace 2 \left[\frac{\rho_V\sigma_V\sigma^2_U(z) \tsigma^2(z)}{\sigma^6(z)}\varepsilon + \frac{\rho^2_V \sigma^2_V \sigma^2_U(z)}{\sigma^4(z)} \cez \right]\left( 1 - \frac{\varepsilon^2}{\sigma^2(z)}\right) \right.\\
& \quad +\frac{\rho_V \sigma_V\sigma^2_U(z)}{\tsigma^2(z)\sigma^6(z)} \left[ 3 \rho_V^4\sigma_V^4 - 5 \rho_V^2 \sigma_V^2 \sigma^2(z) + \sigma^4(z)\right] \varepsilon -\frac{\sigma^2_U(z)}{\tsigma^2(z)\sigma^4(z)}\left[\tsigma^4(z) + \rho^2_V \sigma^2_V \sigma^2_U(z)\right]\cez \\ 
& \quad + \left.\left[ \cez + \frac{\rho_V \sigma_V}{\sigma^2(z)}\varepsilon \right] \left[\frac{\sigma_U(z) \tsigma(z)}{\sigma^3(z)}\varepsilon + \frac{\rho_V \sigma_V \sigma_U(z)}{\sigma(z) \tsigma(z)} \cez \right]^2 \right\rbrace h_{\varepsilon \vert \cez}(\varepsilon \vert \cez) d\varepsilon\\
=&\phi(-\cez)\left\lbrace  -(\cez)^3 \frac{\rho^6_V \sigma^6_V\sigma^2_U(z)}{\sigma^8(z)} + \sqrt{\frac{2}{\pi}}(\cez)^2 \left[ \frac{3 \rho_V^7 \sigma_V^7 \sigma_U^3(z)}{\tsigma^2(z) \sigma^8(z)} - \frac{4 \rho_V^5 \sigma_V^5 \sigma_U^3(z)}{\tsigma^2(z) \sigma^6(z)}\right] \right. \\
& \quad + \cez \left[ 3\frac{\rho^5_V \sigma^5_V \sigma_U^2(z)}{\sigma^8(z)} - 3\frac{\rho^2_V \sigma^2_V \sigma_U^2(z)\tsigma^4(z)}{\sigma^8(z)} + \frac{6}{\pi} \frac{\rho^2_V \sigma^2_V \sigma_U^4(z)\tsigma^2(z)}{\sigma^8(z)}\right]\\ 
& \quad + \left. \sqrt{\frac{2}{\pi}} \left[ - \frac{3\rho^7_V \sigma^7_V \sigma_U^3(z)}{\tsigma^2(z)\sigma^8(z)}+ \frac{4\rho^5_V \sigma^5_V \sigma_U^3(z)}{\tsigma^2(z)\sigma^6(z)} - \frac{\rho_V \sigma_V \sigma_U^5(z)\tsigma^2(z)}{\sigma^8(z)} \right]\right\rbrace.
\end{align*}

Similarly,
\[
h_{\varepsilon}(\varepsilon) = \frac{\Phi \left( -\frac{\cez}{\sqrt{1- \rho_V^2}} - \frac{\rho_V}{\sqrt{1- \rho_V^2}}\frac{\varepsilon}{\sigma_V} \right)}{\Phi(-\cez)} \frac{1}{\sigma_V} \phi\left( \frac{\varepsilon}{\sigma_V}\right),
\]
is the density of an extended skew-normal distribution, $SN(0,\sigma_V,- \frac{\rho_V}{\sqrt{1- \rho_V^2}}, -\cez)$, which implies that
\begin{align*}
\int_\varepsilon \varepsilon h_{\varepsilon}(\varepsilon) =& -\frac{\phi(-\cez)}{\Phi(-\cez)} \rho_V \sigma_V\\
\int_\varepsilon \varepsilon^3 h_{\varepsilon}(\varepsilon) =& -\frac{\phi(-\cez)}{\Phi(-\cez)} \left[ \rho^3_V \sigma^3_V \left( (\cez)^2 - 1 \right) + 3 \rho_V \sigma^3_V\right)
\end{align*}
Thus, 
\begin{align*}
II =& - \Phi(-\cez)\sqrt{\frac{2}{\pi}}\int_\varepsilon \frac{\Phi \left( -\frac{\cez}{\sqrt{1- \rho_V^2}} - \frac{\rho_V}{\sqrt{1- \rho_V^2}}\frac{\varepsilon}{\sigma_V} \right)}{\Phi(-\cez)} \frac{1}{\sigma_V} \phi\left( \frac{\varepsilon}{\sigma_V}\right) \times\\
& \qquad \left[ \frac{\rho_V^2\sigma_U(z)\left( \sigma_U^4(z) + 3 \sigma^2_U(z) \sigma^2_V - 3\sigma^4_V \right) \varepsilon}{\sigma^6(z)} + \frac{\rho^2_V\sigma_V^4\sigma_U(z)}{\sigma^8(z)}\varepsilon^3 \right] d\varepsilon\\
=& \phi(-\cez) \sqrt{\frac{2}{\pi}} \left[ (\cez)^2\frac{\rho_V^5 \sigma^7_V\sigma_U(z)}{\sigma^8(z)} - \frac{\rho_V^5 \sigma^7_V\sigma_U(z)}{\sigma^8(z)} + \frac{\rho_V^3 \sigma_V\sigma^7_U(z)}{\sigma^8(z)} + \frac{4\rho_V^3 \sigma^3_V\sigma^5_U(z)}{\sigma^8(z)} \right].
\end{align*}

Finally, let
\[
h_{\varepsilon \vert \cez}(\varepsilon \vert \cez) = \frac{1}{\tsigma_V}\phi\left( \frac{\varepsilon + \cez \rho_V \sigma_V}{\tsigma_V}\right),
\]
be the density of a $N(-\cez \rho_V \sigma_V,\tsigma_V^2)$. Therefore,
\begin{align*}
\int_\varepsilon \varepsilon h_{\varepsilon \vert \cez}(\varepsilon \vert \cez) d\varepsilon =& -\cez \rho_V \sigma_V\\
\int_\varepsilon \varepsilon^2 h_{\varepsilon \vert \cez}(\varepsilon \vert \cez) d\varepsilon =& \tsigma^2_V + (\cez)^2 \rho^2_V \sigma^2_V,
\end{align*}
and
\begin{align*}
III=&\phi(-\cez)\sqrt{\frac{2}{\pi}} \int_\varepsilon h_{\varepsilon \vert \cez}(\varepsilon \vert \cez) \left\lbrace - 2\frac{\rho_V \sigma_V \tsigma^2_V \sigma_U(z) }{\sigma^4 (z)} \left( 1 - \frac{\varepsilon^2}{\sigma^2(z)}\right) + \frac{\rho_V\sigma_V(1- \rho_V^2) \sigma_U(z)}{\tsigma^2(z)}\right.\\
& \quad + \frac{\rho^3_V(1- \rho_V^2) \sigma^4_V\sigma^3_U(z)}{\tsigma(z)\sigma^5(z)} - \varepsilon^2 \frac{\rho_V (1 - \rho_V^2) \sigma^3_V \sigma_U(z)}{\sigma^6(z)} \\
& \quad \left. - \cez \varepsilon \frac{(1 - \rho_V^2) \sigma^2_V \sigma_U(z)}{\sigma^4(z)} - (\cez)^2 \frac{\rho_V(1 - \rho_V^2) \sigma^3_V \sigma_U(z)}{\tsigma^2(z)\sigma^2(z)}\right\rbrace d\varepsilon\\
=& \phi(-\cez)\sqrt{\frac{2}{\pi}} \left\lbrace -(\cez)^2\frac{\rho_V^5(1 - \rho_V^2) \sigma^7_V\sigma_U(z)}{\tsigma^2(z) \sigma^6(z)} + \frac{\rho_V(1 - \rho_V^2)\sigma_V \sigma^5_U(z)}{\tsigma^2(z) \sigma^4(z)} \right. \\
& \quad \left. + \frac{\rho^5_V(1 - \rho_V^2)\sigma^7_V \sigma_U(z)}{\tsigma^2(z) \sigma^6(z)} - \frac{\rho^3_V(1 - \rho_V^2)\sigma^3_V \sigma^5_U(z)}{\tsigma^2(z) \sigma^6(z)}\right\rbrace\\
\end{align*}

Therefore
\begin{align*}
\int_\varepsilon & \left( \frac{\partial^2 \Psi_0(z,\theta) }{\partial \rho^2_U} + \frac{\partial  \Psi_0(z,\theta) }{\partial \rho_U}\frac{ 2 \rho_V\sigma_V \sigma_U(z) }{\sigma^2(z)} \left( 1 - \frac{\varepsilon^2}{\sigma^2(z)}\right) \right) \frac{2}{\sigma(z)}\phi\left( \frac{\varepsilon}{\sigma(z)}\right) d\varepsilon \\
=& \phi(-\cez)\left\lbrace -(\cez)^3 \frac{\rho^6_V\sigma^6_V\sigma^2_U(z)}{\sigma^8(z)} - (\cez)^2 4\sqrt{\frac{2}{\pi}} \frac{\rho^5_V\sigma^5_V\sigma^3_U(z)}{\sigma^8(z)} \right.\\
& \quad \left. + \cez \frac{3\rho_V^6 \sigma^6_V \sigma^2_U(z)}{\sigma^8(z)} + 4 \sqrt{\frac{2}{\pi}}\frac{\rho_V^5\sigma_V^5\sigma_U^3(z)}{\sigma^8(z)} + 4 \sqrt{\frac{2}{\pi}}\frac{\rho_V^3\sigma_V^3\sigma_U^5(z)}{\sigma^8(z)}\right\rbrace.
\end{align*}

Now, we have
\begin{align*}
\left. \frac{\partial \Phi(\tau_1(z) \varepsilon) }{\partial \rho_U} \right\vert_{\rho_U = 0}=& \frac{\rho_V \sigma^2_V \varepsilon}{\sigma^3(z)}\phi \left( -\frac{\sigma_U(z)\varepsilon}{\sigma_V \sigma(z)}\right) \\
\left. \frac{\partial^2 \Phi(\tau_1(z) \varepsilon) }{\partial \rho^2_U} \right\vert_{\rho_U = 0}=& \left( -\frac{\sigma_U(Z) \rho_V^2\left( \sigma_U^4(Z) + 3\sigma^2_U(Z)\sigma_V^2 - \sigma^4_V\right) \varepsilon}{\sigma_V \sigma^5(z)} + \frac{\rho^2_V \sigma_V^4\sigma_U(Z)\varepsilon^3}{\sigma_V \sigma^7(Z)}\right) \phi \left( -\frac{\sigma_U(z)\varepsilon}{\sigma_V \sigma(z)}\right),
\end{align*}
so that 
\begin{align*}
 E_{\tilde{W}}& \left[ \int_\varepsilon \left( -\frac{\sigma_U(Z) \rho_V^2\left( \sigma_U^4(Z) + 3\sigma^2_U(Z)\sigma_V^2 - \sigma^4_V\right) \varepsilon}{\sigma_V \sigma^5(Z)} + \frac{\rho^2_V \sigma_V^4\sigma_U(Z)\varepsilon^3}{\sigma_V \sigma^7(Z)} \right. \right.\\
& \qquad \left. \left. + \frac{2\rho^2_V \sigma^4_V \sigma_U(Z) \varepsilon}{\sigma_V \sigma^5(Z)} - \frac{2\rho^2_V \sigma^4_V\sigma_U(Z) \varepsilon^3}{\sigma_V \sigma^7(Z)} \right) \phi \left( -\frac{\sigma_U(Z)\varepsilon}{\sigma_V \sigma(Z)} \right) \frac{2}{\sigma(Z)}\phi\left( \frac{\varepsilon}{\sigma(Z)}\right) d\varepsilon \right] \\
=& -\sqrt{\frac{2}{\pi}}E_{\tilde{W}}\left[ \frac{\sigma_U(Z) \rho_V^2\left( \sigma_U^4(Z) + 3\sigma^2_U(Z)\sigma_V^2 - 3\sigma^4_V\right)}{\sigma^6(Z)} \int_\varepsilon  \frac{\varepsilon}{\sigma_V} \phi \left( \frac{\varepsilon}{\sigma_V} \right) d\varepsilon \right] \\
& \quad - \sqrt{\frac{2}{\pi}}E_{\tilde{W}}\left[ \frac{\rho^2_V \sigma^4_V\sigma_U(Z)}{\sigma^8(Z)} \int_\varepsilon  \frac{\varepsilon^3}{\sigma_V} \phi \left( \frac{\varepsilon}{\sigma_V} \right) d\varepsilon \right] = 0,
\end{align*}
where the final result follows because
\begin{align*}
\int_\varepsilon & \frac{\varepsilon}{\sigma_V} \phi \left( \frac{\varepsilon}{\sigma_V} \right) d\varepsilon = 0,\\
\int_\varepsilon & \frac{\varepsilon^3}{\sigma_V} \phi \left( \frac{\varepsilon}{\sigma_V} \right) d\varepsilon = 0,
\end{align*}
by the properties of the pdf of a standard normal distribution. Thus, we can finally write
\begin{align*}
SD_2 =& -\frac{\rho^6_V\sigma^6_V\sigma^2_U(1)}{\sigma^8(1)} E_{\tilde{W}} \left[ \left( 3 \cez - (\cez)^3\right) \phi(-\cez) \right] \\
& \quad + 4\sqrt{\frac{2}{\pi}} \frac{\rho^5_V\sigma^5_V\sigma^3_U(1)}{\sigma^8(1)} E_{\tilde{W}} \left[ \left( (\cez)^2 - 1 \right) \phi(-\cez) \right] \\
& \quad - 4 \sqrt{\frac{2}{\pi}}\frac{\rho_V^3\sigma_V^3\sigma_U^5(1)}{\sigma^8(1)} E_{\tilde{W}} \left[ \phi(-\cez) \right]\\
SD_3 =& \frac{\rho^6_V\sigma^6_V\sigma^2_U(0)}{\sigma^8(0)} E_{\tilde{W}} \left[ \left( 3 \cez - (\cez)^3\right) \phi(-\cez) \right] \\
& \quad - 4\sqrt{\frac{2}{\pi}} \frac{\rho^5_V\sigma^5_V\sigma^3_U(0)}{\sigma^8(0)} E_{\tilde{W}} \left[ \left( (\cez)^2 - 1 \right) \phi(-\cez) \right] \\
& \quad + 4 \sqrt{\frac{2}{\pi}}\frac{\rho_V^3\sigma_V^3\sigma_U^5(0)}{\sigma^8(0)} E_{\tilde{W}} \left[ \phi(-\cez) \right],
\end{align*}
which implies that
\begin{align*}
SD_2 + SD_3=&  -\left( \frac{\rho^6_V\sigma^6_V\sigma^2_U(1)}{\sigma^8(1)} - \frac{\rho^6_V\sigma^6_V\sigma^2_U(0)}{\sigma^8(0)} \right) E_{\tilde{W}} \left[ \left( 3 \cez - (\cez)^3\right) \phi(-\cez) \right]\\
& \quad + 4\sqrt{\frac{2}{\pi}} \left( \frac{\rho^5_V\sigma^5_V\sigma^3_U(1)}{\sigma^8(1)} - \frac{\rho^5_V\sigma^5_V\sigma^3_U(0)}{\sigma^8(0)} \right) E_{\tilde{W}} \left[ \left( (\cez)^2 - 1 \right) \phi(-\cez) \right]\\
& \quad - 4 \sqrt{\frac{2}{\pi}}\left( \frac{\rho_V^3\sigma_V^3\sigma_U^5(1)}{\sigma^8(1)} - \frac{\rho_V^3\sigma_V^3\sigma_U^5(0)}{\sigma^8(0)} \right) E_{\tilde{W}} \left[ \phi(-\cez) \right], 
\end{align*}

and the final result follows from $ SD_1 + SD_2 + SD_3 = 0$. This concludes the proof. 

\subsection{Conditional density of $U\vert\varepsilon$} 
From \citet{centorrinoperez2019}, we have that
\begin{align*}
f_{U\vert\varepsilon,\eta} (u \vert \varepsilon ,\eta ) =& \frac{1}{\sqrt{2\pi} \frac{\tilde\sigma_V \tilde\sigma_U(Z)}{\tsigma(Z)}}\left\lbrace \frac{f_{\varepsilon,\eta,1}\left(\varepsilon,\eta \right)}{f_{\varepsilon,\eta}\left(\varepsilon,\eta \right)}\left[ \Phi\left( q_1(Z)\eta -\frac{\lambda(Z) }{\tsigma(Z)}\varepsilon \right)\right]^{-1} \exp \left( -\frac{\left(u - \frac{\tilde\sigma_V \tilde\sigma_U(Z)}{\tsigma(Z)} \left( q_1(Z)\eta -\frac{\lambda(Z)}{\tsigma(Z)}\varepsilon \right) \right)^2}{2\frac{\tilde\sigma^2_V \tilde\sigma^2_U(Z)}{\tsigma^2(Z)}}\right) \right. \\
&\quad \left. + \frac{f_{\varepsilon,\eta,2}\left(\varepsilon,\eta \right)}{f_{\varepsilon,\eta}\left(\varepsilon,\eta \right)} \left[ \Phi\left( q_2(Z)\eta -\frac{\lambda(Z)}{\tsigma(Z)}\varepsilon \right)\right]^{-1} \exp \left( -\frac{\left(u - \frac{\tilde\sigma_V \tilde\sigma_U(Z)}{\tsigma(Z)} \left( q_2(Z)\eta -\frac{\lambda(Z)}{\tsigma(Z)}\varepsilon \right) \right)^2}{2\frac{\tilde\sigma^2_V \tilde\sigma^2_U(Z)}{\tsigma^2(Z)}}\right) \right\rbrace \\
=& \frac{f_{\varepsilon,\eta,1}\left(\varepsilon,\eta \right)}{2 f_{\varepsilon,\eta}\left(\varepsilon,\eta \right)} f_{U\vert\varepsilon,\eta,1} (u \vert \varepsilon ,\eta ) + \frac{f_{\varepsilon,\eta,2}\left(\varepsilon,\eta \right)}{2 f_{\varepsilon,\eta}\left(\varepsilon,\eta \right)} f_{U\vert\varepsilon,\eta,2} (u \vert \varepsilon ,\eta ).
\end{align*}

Let
\begin{align*}
f_{\varepsilon,1}\left( \varepsilon \right) =& \frac{2}{\sqrt{2\pi(\tsigma^2(Z) + \rho^2_1(Z))}} \Phi \left(\tau_1 (Z) \varepsilon \right) \exp\left( -\frac{\varepsilon^2}{2(\tsigma^2(Z) + \rho^2_1(Z))}\right)\\
f_{\varepsilon,2}\left( \varepsilon \right) =& \frac{2}{\sqrt{2\pi(\tsigma^2(Z) + \rho^2_2(Z))}} \Phi \left(\tau_2 (Z) \varepsilon \right) \exp\left( -\frac{\varepsilon^2}{2(\tsigma^2(Z) + \rho^2_2(Z))}\right),
\end{align*}
which implies
\[
f_{\varepsilon} (\varepsilon) = f_{\varepsilon,1}\left( \varepsilon \right) + f_{\varepsilon,2}\left( \varepsilon \right).
\]

Using these notations, we can rewrite
\begin{align} \label{eq:etaconddens}
f_{\eta \vert \varepsilon} (\eta \vert \varepsilon) =& \frac{f_{1}\left( \varepsilon \right)}{f_{\varepsilon} (\varepsilon) }\frac{\Phi \left( q_1 (Z) \eta - \frac{\lambda(Z) \varepsilon}{\tsigma(Z)}\right)}{\Phi \left( \tau_1 (Z) \varepsilon \right)} \times \nonumber \\
& \quad \sqrt{\frac{\tsigma^2(Z) + \rho^2_1(Z)}{2\pi \tsigma^2(Z)}}\exp \left( -\frac{\tsigma^2(Z) + \rho^2_1(Z)}{2 \tsigma^2(Z)} \left(\eta - \frac{\rho_1(Z)}{\tsigma^2(Z) + \rho^2_1(Z)}\varepsilon \right)^2 \right)\nonumber \\
& \quad + \frac{f_{2}\left( \varepsilon \right)}{f_{\varepsilon} (\varepsilon) } \frac{\Phi \left( q_2 (Z) \eta -\frac{\lambda(Z) \varepsilon}{\tsigma(Z)}\right)}{\Phi \left(\tau_2 (Z) \varepsilon \right)} \times \\ 
& \quad \sqrt{\frac{\tsigma^2(Z) + \rho^2_2(Z)}{2\pi \tsigma^2(Z)}} \exp \left( - \frac{\tsigma^2(Z) + \rho^2_2(Z)}{2 \tsigma^2(Z)} \left(\eta -\frac{\rho_2(Z)}{\tsigma^2(Z) + \rho^2_2(Z)}\varepsilon \right)^2 \right) \nonumber\\
=& \frac{f_{1}\left( \varepsilon \right)}{f_{\varepsilon} (\varepsilon) } f_{\eta \vert \varepsilon, 1} (\eta \vert \varepsilon) + \frac{f_{2}\left( \varepsilon \right)}{f_{\varepsilon} (\varepsilon) } f_{\eta \vert \varepsilon, 2} (\eta \vert \varepsilon). \nonumber
\end{align}

We multiply the previous equation by the conditional density of $U\vert(\varepsilon,\eta)$ to get
\[
f_{U,\eta\vert \varepsilon} (u ,\eta \vert \varepsilon) = f_{U\vert\varepsilon,\eta} (u \vert \varepsilon ,\eta ) f_{\eta \vert \varepsilon} (\eta \vert \varepsilon).
\]

These computations give
\begin{align*}
f_{U,\eta \vert\varepsilon} (u , \eta \vert \varepsilon ) =& \frac{1}{2\pi \frac{\tilde\sigma_V \tilde\sigma_U(Z)}{\tsigma(Z)}}\left\lbrace \frac{f_{1}\left( \varepsilon \right)}{f_{\varepsilon} (\varepsilon) } \left[ \Phi \left( \tau_1 (Z) \varepsilon \right) \right]^{-1} \exp \left( -\frac{\left(u - \frac{\tilde\sigma_V \tilde\sigma_U(Z)}{\tsigma(Z)} \left( q_1(Z)\eta -\frac{\lambda(Z)}{\tsigma(Z)}\varepsilon \right) \right)^2}{2\frac{\tilde\sigma^2_V \tilde\sigma^2_U(Z)}{\tsigma^2(Z)}}\right) \right. \times \\
&\quad \sqrt{\frac{\tsigma^2(Z) + \rho^2_1(Z)}{\tsigma^2(Z)}}\exp \left( -\frac{\tsigma^2(Z) + \rho^2_1(Z)}{2 \tsigma^2(Z)} \left(\eta - \frac{\rho_1(Z)}{\tsigma^2(Z) + \rho^2_1(Z)}\varepsilon \right)^2 \right) \\
&\quad + \frac{f_{2}\left( \varepsilon \right)}{f_{\varepsilon} (\varepsilon) } \left[ \Phi \left( \tau_2 (Z) \varepsilon \right) \right]^{-1} \exp \left( -\frac{\left(u - \frac{\tilde\sigma_V \tilde\sigma_U(Z)}{\tsigma(Z)} \left( q_2(Z)\eta -\frac{\lambda(Z)}{\tsigma(Z)}\varepsilon \right) \right)^2}{2\frac{\tilde\sigma^2_V \tilde\sigma^2_U(Z)}{\tsigma^2(Z)}}\right) \\
&\quad \left.\sqrt{\frac{\tsigma^2(Z) + \rho^2_2(Z)}{\tsigma^2(Z)}}\exp \left( -\frac{\tsigma^2(Z) + \rho^2_2(Z)}{2 \tsigma^2(Z)} \left(\eta - \frac{\rho_2(Z)}{\tsigma^2(Z) + \rho^2_2(Z)}\varepsilon \right)^2 \right) \right\rbrace.
\end{align*}

Now, we let
\begin{align*}
\Sigma_{U\eta, 1 \vert \varepsilon} =& \begin{bmatrix} \frac{\tilde\sigma^2_V \tilde\sigma^2_U(Z)}{\tsigma^2(Z)} \left( 1 + \frac{q^2_1(Z)\tsigma^2(Z)}{\tsigma^2(Z) + \rho^2_1(Z)} \right) & \frac{\tilde\sigma_V \tilde\sigma_U(Z)}{\tsigma(Z)} \frac{q_1(Z)\tsigma^2(Z)}{\tsigma^2(Z) + \rho^2_1(Z)} \\ \frac{\tilde\sigma_V \tilde\sigma_U(Z)}{\tsigma(Z)} \frac{q_1(Z)\tsigma^2(Z)}{\tsigma^2(Z) + \rho^2_1(Z)} & \frac{\tsigma^2(Z)}{\tsigma^2(Z) + \rho^2_1(Z)} \end{bmatrix}\\
\Sigma_{U\eta, 2 \vert \varepsilon} =& \begin{bmatrix} \frac{\tilde\sigma^2_V \tilde\sigma^2_U(Z)}{\tsigma^2(Z)} \left( 1 + \frac{q^2_2(Z)\tsigma^2(Z)}{\tsigma^2(Z) + \rho^2_2(Z)} \right) & \frac{\tilde\sigma_V \tilde\sigma_U(Z)}{\tsigma(Z)} \frac{q_2(Z)\tsigma^2(Z)}{\tsigma^2(Z) + \rho^2_2(Z)} \\ \frac{\tilde\sigma_V \tilde\sigma_U(Z)}{\tsigma(Z)} \frac{q_2(Z)\tsigma^2(Z)}{\tsigma^2(Z) + \rho^2_2(Z)} & \frac{\tsigma^2(Z)}{\tsigma^2(Z) + \rho^2_2(Z)} \end{bmatrix},
\end{align*}
in a way that the expression above can be rewritten as

\begin{align*}
f_{U,\eta \vert\varepsilon}  & (u , \eta \vert \varepsilon ) =\\
& \frac{1}{2\pi \frac{\tilde\sigma_V \tilde\sigma_U(Z)}{\tsigma(Z)}} \left\lbrace \frac{f_{1}\left( \varepsilon \right)}{f_{\varepsilon} (\varepsilon) } \left[ \Phi \left( \tau_1 (Z) \varepsilon \right) \right]^{-1} \sqrt{\frac{\tsigma^2(Z) + \rho^2_1(Z)}{\tsigma^2(Z)}} \right. \times \\
&\quad \exp \left( -\frac{1}{2} \begin{pmatrix} u - \frac{\tilde\sigma_V \tilde\sigma_U(Z)}{\tsigma(Z)} \left( \frac{q_1(Z) \rho_1(Z)}{\tsigma^2(Z) + \rho^2_1(Z)} -\frac{\lambda(Z)}{\tsigma(Z)} \right) \varepsilon \\ \eta - \frac{\rho_1(Z)}{\tsigma^2(Z) + \rho^2_1(Z)}\varepsilon \end{pmatrix}^\prime \Sigma^{-1}_{U\eta, 1 \vert \varepsilon} \begin{pmatrix} u - \frac{\tilde\sigma_V \tilde\sigma_U(Z)}{\tsigma(Z)} \left( \frac{q_1(Z) \rho_1(Z)}{\tsigma^2(Z) + \rho^2_1(Z)} -\frac{\lambda(Z)}{\tsigma(Z)} \right) \varepsilon \\ \eta - \frac{\rho_1(Z)}{\tsigma^2(Z) + \rho^2_1(Z)}\varepsilon \end{pmatrix} \right) \\
&\quad + \frac{f_{2}\left( \varepsilon \right)}{f_{\varepsilon} (\varepsilon) } \left[ \Phi \left( \tau_2 (Z) \varepsilon \right) \right]^{-1} \sqrt{\frac{\tsigma^2(Z) + \rho^2_2(Z)}{\tsigma^2(Z)}} \times \\
&\quad \left. \exp \left( -\frac{1}{2} \begin{pmatrix} u - \frac{\tilde\sigma_V \tilde\sigma_U(Z)}{\tsigma(Z)} \left( \frac{q_2(Z) \rho_2(Z)}{\tsigma^2(Z) + \rho^2_2(Z)} -\frac{\lambda(Z)}{\tsigma(Z)} \right) \varepsilon \\ \eta - \frac{\rho_2(Z)}{\tsigma^2(Z) + \rho^2_2(Z)}\varepsilon \end{pmatrix}^\prime \Sigma^{-1}_{U\eta, 2 \vert \varepsilon} \begin{pmatrix} u - \frac{\tilde\sigma_V \tilde\sigma_U(Z)}{\tsigma(Z)} \left( \frac{q_2(Z) \rho_2(Z)}{\tsigma^2(Z) + \rho^2_2(Z)} -\frac{\lambda(Z)}{\tsigma(Z)} \right) \varepsilon \\ \eta - \frac{\rho_2(Z)}{\tsigma^2(Z) + \rho^2_2(Z)}\varepsilon \end{pmatrix}\right) \right\rbrace.
\end{align*}

By integrating this joint conditional density with respect to $\eta$, we finally obtain
\begin{equation} \label{eq:cdensueps}
\begin{aligned}
f_{U\vert \varepsilon} \left( u \vert \varepsilon \right) =& \frac{f_{\varepsilon,1}\left( \varepsilon\right)}{f_{\varepsilon}\left( \varepsilon\right)}\frac{\left[ \Phi\left( \tau_1(Z) \varepsilon \right) \right]^{-1}}{\sqrt{2\pi} \sigma_{1\star}} \exp \left( -\frac{\left(u - \mu_{1\star} \right)^2}{2\sigma^2_{1\star}}\right) \\
& \quad + \frac{f_{\varepsilon,2}\left( \varepsilon\right)}{f_{\varepsilon}\left( \varepsilon\right)} \frac{\left[ \Phi\left( \tau_2(Z) \varepsilon \right) \right]^{-1}}{\sqrt{2\pi} \sigma_{2\star}} \exp \left( -\frac{\left(u - \mu_{2\star} \right)^2}{2\sigma^2_{2\star}}\right),
\end{aligned}
\end{equation}
which is a mixture of two half-normal densities, with weights given by 
\begin{align*}
\omega_1(\varepsilon) = \frac{f_{\varepsilon,1}\left( \varepsilon\right)}{f_{\varepsilon}\left( \varepsilon\right)}, \qquad \omega_2(\varepsilon) = \frac{f_{\varepsilon,2}\left( \varepsilon\right)}{f_{\varepsilon}\left( \varepsilon\right)}.
\end{align*}

Hence
\begin{align*}
E \left[ \exp(-U) \vert \varepsilon \right] =& \left\lbrace \omega_1(\varepsilon)\left[ \Phi\left( \tau_1(Z) \varepsilon \right) \right]^{-1} \int_{0}^\infty \frac{1}{\sqrt{2\pi} \sigma_{1\star}} \exp \left( -u - \frac{\left(u - \mu_{1\star} \right)^2}{2\sigma^2_{1\star}}\right) du \right. \\
&\quad \left. + \omega_2(\varepsilon)\left[ \Phi\left( \tau_2(Z) \varepsilon \right) \right]^{-1} \int_{0}^\infty \frac{1}{\sqrt{2\pi} \sigma_{2\star}} \exp \left( -u - \frac{\left(u - \mu_{2\star} \right)^2}{2\sigma^2_{2\star}}\right) du \right\rbrace.
\end{align*}

The final expression in \eqref{eq:battesecoelli} follows by the properties of the cdf of the univariate normal distribution.

\newpage 
\section{Additional simulation results} \label{appBsimulation}

We compare the bias and standard deviation of our proposed MLE approach (CPUW) with the classical stochastic frontier model that ignores the endogeneity issue (EX), and with the estimator in \citet{Kumbhakar2009} which imposes $\rho_V = 0$ (KTS). Results are reported in Table \ref{tab:appsimcomp} below.

%\begin{table}[!h]
\begin{center}
\tiny
\begingroup
\renewcommand\arraystretch{0.58}
\begin{longtable}{l | c c c | c c c | c c c} \hline \hline
~ & \multicolumn{3}{c}{$n = 250$} & \multicolumn{3}{c}{$n = 500$} & \multicolumn{3}{c}{$n = 1000$}\\ \hline
~ & CPUW & EX & KTS & CPUW & EX & KTS & CPUW & EX & KTS \\ \hline \hline
\multicolumn{10}{l}{\textsc{Scheme 1}} \\ \hline
% latex table generated in R 4.3.2 by xtable 1.8-4 package
% Thu Dec 21 09:15:33 2023
  \hline
$\beta_0$ & -0.1122 & -0.0142 & -0.3498 & -0.0553 & -0.0141 & -0.3386 & -0.0152 & -0.0122 & -0.2864 \\ 
   & (0.3528) & (0.0272) & (0.5023) & (0.2188) & (0.0188) & (0.4647) & (0.1247) & (0.0133) & (0.4239) \\ 
  $\beta_1$ & 0.0287 & -0.0084 & -0.0186 & 0.0182 & -0.0101 & -0.0222 & 0.0114 & -0.0103 & -0.0202 \\ 
   & (0.1649) & (0.1505) & (0.1656) & (0.1129) & (0.1077) & (0.1136) & (0.0787) & (0.0749) & (0.0811) \\ 
  $\beta_2$ & 0.0343 & -0.0065 & -0.0152 & 0.0088 & -0.0172 & -0.0306 & 0.0035 & -0.0195 & -0.0295 \\ 
   & (0.1694) & (0.1563) & (0.1671) & (0.1130) & (0.1072) & (0.1148) & (0.0754) & (0.0731) & (0.0793) \\ 
  $\beta_{1,1}$ & -0.0259 & -0.0451 & -0.0545 & -0.0169 & -0.0436 & -0.0471 & -0.0154 & -0.0489 & -0.0518 \\ 
   & (0.2169) & (0.2095) & (0.2203) & (0.1456) & (0.1443) & (0.1489) & (0.1068) & (0.1042) & (0.1095) \\ 
  $\beta_{1,2}$ & -0.0381 & -0.0541 & -0.0650 & -0.0116 & -0.0417 & -0.0437 & -0.0045 & -0.0373 & -0.0409 \\ 
   & (0.2109) & (0.2051) & (0.2132) & (0.1455) & (0.1440) & (0.1502) & (0.1033) & (0.1024) & (0.1069) \\ 
  $\delta_{1}$ & -0.0144 & 0.0711 & 0.0866 & -0.0083 & 0.0653 & 0.0702 & -0.0021 & 0.0681 & 0.0709 \\ 
   & (0.2460) & (0.0777) & (0.2366) & (0.0738) & (0.0523) & (0.1185) & (0.0402) & (0.0361) & (0.0745) \\ 
  $\delta_{2}$ & 0.3431 & -0.3667 & -4.9447 & 0.0078 & -0.3623 & -4.3551 & 0.0335 & -0.3660 & -4.1441 \\ 
   & (2.5521) & (0.1626) & (8.2058) & (1.5419) & (0.1091) & (7.5421) & (0.4257) & (0.0799) & (7.4282) \\ 
  $\sigma^2_{U}$ & -0.4698 & 0.9661 & -0.1173 & -0.2638 & 1.0132 & -0.0454 & -0.1093 & 1.0202 & 0.0938 \\ 
   & (1.2665) & (0.6630) & (1.6361) & (0.8663) & (0.4690) & (1.4951) & (0.5477) & (0.3313) & (1.3269) \\ 
  $\sigma^2_{V}$ & 0.1288 & -0.0221 & 0.2178 & 0.0582 & -0.0128 & 0.2343 & 0.0191 & 0.0014 & 0.2220 \\ 
   & (0.4174) & (0.1692) & (0.4300) & (0.2658) & (0.1196) & (0.3812) & (0.1545) & (0.0881) & (0.3455) \\ 
  $\rho_{U,\eta}$ & 0.2290 & ~ & 0.3840 & 0.0023 & ~ & 0.3336 & 0.0035 & ~ & 0.3101 \\ 
   & (0.3119) & ~ & (0.4101) & (0.2273) & ~ & (0.3639) & (0.1659) & ~ & (0.3361) \\ 
  $\rho_{V,\eta}$ & 0.0510 & ~ & ~ & 0.0279 & ~ & ~ & 0.0132 & ~ & ~ \\ 
   & (0.1891) & ~ & ~ & (0.1268) & ~ & ~ & (0.0817) & ~ & ~ \\ 
  $\gamma_{0}$ & 0.0002 & ~ & 0.0071 & 0.0005 & ~ & 0.0042 & -0.0007 & ~ & 0.0024 \\ 
   & (0.1112) & ~ & (0.1158) & (0.0783) & ~ & (0.0795) & (0.0550) & ~ & (0.0560) \\ 
  $\gamma_{1}$ & 0.0064 & ~ & 0.0042 & 0.0066 & ~ & 0.0063 & 0.0016 & ~ & 0.0017 \\ 
   & (0.1389) & ~ & (0.1404) & (0.0926) & ~ & (0.0937) & (0.0670) & ~ & (0.0677) \\ 
  $\gamma_{2}$ & 0.0206 & ~ & 0.0199 & 0.0043 & ~ & 0.0032 & 0.0052 & ~ & 0.0047 \\ 
   & (0.1404) & ~ & (0.1426) & (0.0916) & ~ & (0.0934) & (0.0653) & ~ & (0.0666) \\ 
  $\gamma_{3}$ & 0.0083 & ~ & 0.0081 & 0.0050 & ~ & 0.0038 & 0.0062 & ~ & 0.0046 \\ 
   & (0.1358) & ~ & (0.1394) & (0.0910) & ~ & (0.0927) & (0.0652) & ~ & (0.0663) \\ 
  $\gamma_{4}$ & 0.0138 & ~ & 0.0168 & 0.0053 & ~ & 0.0075 & 0.0028 & ~ & 0.0038 \\ 
   & (0.1306) & ~ & (0.1331) & (0.0909) & ~ & (0.0953) & (0.0611) & ~ & (0.0632) \\ 
  $\gamma_{5}$ & 0.0272 & ~ & 0.0396 & 0.0126 & ~ & 0.0226 & 0.0114 & ~ & 0.0166 \\ 
   & (0.2498) & ~ & (0.2560) & (0.1789) & ~ & (0.1824) & (0.1220) & ~ & (0.1259) \\ 
   \hline

\hline \hline
\multicolumn{10}{l}{\textsc{Scheme 2}} \\ \hline
% latex table generated in R 4.3.2 by xtable 1.8-4 package
% Thu Dec 21 09:15:33 2023
  \hline
$\beta_0$ & -0.1061 & -0.0108 & -0.2020 & -0.0514 & -0.0107 & -0.1982 & -0.0153 & -0.0084 & -0.1209 \\ 
   & (0.3369) & (0.0264) & (0.4361) & (0.2028) & (0.0194) & (0.3968) & (0.1222) & (0.0136) & (0.3114) \\ 
  $\beta_1$ & 0.0110 & -0.0462 & -0.0282 & 0.0103 & -0.0417 & -0.0269 & 0.0044 & -0.0450 & -0.0250 \\ 
   & (0.1682) & (0.1512) & (0.1668) & (0.1174) & (0.1092) & (0.1178) & (0.0807) & (0.0749) & (0.0804) \\ 
  $\beta_2$ & 0.0231 & -0.0376 & -0.0201 & -0.0003 & -0.0494 & -0.0347 & -0.0011 & -0.0513 & -0.0314 \\ 
   & (0.1745) & (0.1578) & (0.1709) & (0.1162) & (0.1083) & (0.1141) & (0.0795) & (0.0742) & (0.0801) \\ 
  $\beta_{1,1}$ & -0.0102 & 0.0120 & -0.0349 & -0.0072 & 0.0078 & -0.0323 & -0.0040 & 0.0097 & -0.0337 \\ 
   & (0.2195) & (0.2050) & (0.2202) & (0.1527) & (0.1461) & (0.1555) & (0.1060) & (0.1021) & (0.1078) \\ 
  $\beta_{1,2}$ & -0.0284 & -0.0041 & -0.0486 & 0.0011 & 0.0121 & -0.0278 & 0.0015 & 0.0154 & -0.0276 \\ 
   & (0.2143) & (0.2054) & (0.2178) & (0.1445) & (0.1409) & (0.1457) & (0.1040) & (0.0999) & (0.1066) \\ 
  $\delta_{1}$ & -0.0167 & 0.0718 & 0.0752 & -0.0052 & 0.0672 & 0.0650 & 0.0002 & 0.0703 & 0.0678 \\ 
   & (0.2414) & (0.0778) & (0.1454) & (0.0725) & (0.0524) & (0.0965) & (0.0383) & (0.0358) & (0.0561) \\ 
  $\delta_{2}$ & 0.0253 & -0.3984 & -2.3954 & 0.0397 & -0.3962 & -1.9677 & 0.0086 & -0.3989 & -1.3042 \\ 
   & (2.5897) & (0.1556) & (5.6336) & (0.2663) & (0.1087) & (4.8683) & (0.1302) & (0.0776) & (3.8200) \\ 
  $\sigma^2_{U}$ & -0.3479 & 1.2520 & 0.4775 & -0.1709 & 1.3060 & 0.5358 & -0.0478 & 1.3209 & 0.7687 \\ 
   & (1.3266) & (0.7172) & (1.5219) & (0.9402) & (0.5166) & (1.3642) & (0.6179) & (0.3661) & (1.0605) \\ 
  $\sigma^2_{V}$ & 0.1234 & -0.0909 & 0.0673 & 0.0477 & -0.0845 & 0.0829 & 0.0103 & -0.0729 & 0.0437 \\ 
   & (0.4307) & (0.1665) & (0.3953) & (0.2730) & (0.1166) & (0.3434) & (0.1644) & (0.0853) & (0.2712) \\ 
  $\rho_{U,\eta}$ & -0.0198 & ~ & 0.0548 & -0.0372 & ~ & 0.0548 & -0.0150 & ~ & 0.0446 \\ 
   & (0.3167) & ~ & (0.3651) & (0.2409) & ~ & (0.2931) & (0.1477) & ~ & (0.2144) \\ 
  $\rho_{V,\eta}$ & 0.0179 & ~ & ~ & 0.0117 & ~ & ~ & 0.0019 & ~ & ~ \\ 
   & (0.2295) & ~ & ~ & (0.1478) & ~ & ~ & (0.0936) & ~ & ~ \\ 
  $\gamma_{0}$ & -0.0007 & ~ & 0.0126 & -0.0002 & ~ & 0.0126 & -0.0010 & ~ & 0.0119 \\ 
   & (0.1124) & ~ & (0.1184) & (0.0786) & ~ & (0.0805) & (0.0547) & ~ & (0.0570) \\ 
  $\gamma_{1}$ & 0.0060 & ~ & 0.0072 & 0.0077 & ~ & 0.0072 & 0.0014 & ~ & 0.0019 \\ 
   & (0.1392) & ~ & (0.1389) & (0.0930) & ~ & (0.0930) & (0.0671) & ~ & (0.0671) \\ 
  $\gamma_{2}$ & 0.0198 & ~ & 0.0033 & 0.0041 & ~ & 0.0033 & 0.0050 & ~ & 0.0048 \\ 
   & (0.1402) & ~ & (0.1414) & (0.0917) & ~ & (0.0928) & (0.0649) & ~ & (0.0658) \\ 
  $\gamma_{3}$ & 0.0073 & ~ & 0.0018 & 0.0052 & ~ & 0.0018 & 0.0054 & ~ & 0.0024 \\ 
   & (0.1348) & ~ & (0.1377) & (0.0901) & ~ & (0.0923) & (0.0645) & ~ & (0.0654) \\ 
  $\gamma_{4}$ & 0.0158 & ~ & 0.0067 & 0.0046 & ~ & 0.0067 & 0.0042 & ~ & 0.0037 \\ 
   & (0.1307) & ~ & (0.1327) & (0.0910) & ~ & (0.0945) & (0.0605) & ~ & (0.0623) \\ 
  $\gamma_{5}$ & 0.0372 & ~ & 0.0283 & 0.0173 & ~ & 0.0283 & 0.0150 & ~ & 0.0239 \\ 
   & (0.2516) & ~ & (0.2579) & (0.1773) & ~ & (0.1798) & (0.1218) & ~ & (0.1251) \\ 
   \hline

\hline \hline
\multicolumn{10}{l}{\textsc{Scheme 3}} \\ \hline
% latex table generated in R 4.3.2 by xtable 1.8-4 package
% Thu Dec 21 09:15:33 2023
  \hline
$\beta_0$ & -0.0314 & -0.0074 & 0.0395 & -0.0216 & -0.0061 & 0.0538 & -0.0088 & -0.0037 & 0.0747 \\ 
   & (0.2274) & (0.0273) & (0.2108) & (0.1561) & (0.0205) & (0.1486) & (0.1141) & (0.0142) & (0.1013) \\ 
  $\beta_1$ & -0.0247 & -0.1491 & -0.0627 & -0.0049 & -0.1420 & -0.0423 & -0.0011 & -0.1447 & -0.0396 \\ 
   & (0.1621) & (0.1506) & (0.1562) & (0.1194) & (0.1150) & (0.1157) & (0.0764) & (0.0739) & (0.0747) \\ 
  $\beta_2$ & -0.0160 & -0.1420 & -0.0532 & -0.0112 & -0.1459 & -0.0473 & -0.0021 & -0.1477 & -0.0412 \\ 
   & (0.1678) & (0.1606) & (0.1629) & (0.1140) & (0.1121) & (0.1113) & (0.0778) & (0.0752) & (0.0768) \\ 
  $\beta_{1,1}$ & 0.0206 & 0.1794 & 0.0170 & 0.0009 & 0.1686 & -0.0113 & 0.0013 & 0.1754 & -0.0139 \\ 
   & (0.2026) & (0.1926) & (0.2017) & (0.1504) & (0.1444) & (0.1499) & (0.0997) & (0.0981) & (0.1005) \\ 
  $\beta_{1,2}$ & 0.0066 & 0.1629 & 0.0015 & 0.0084 & 0.1754 & -0.0055 & -0.0009 & 0.1758 & -0.0152 \\ 
   & (0.2076) & (0.2018) & (0.2089) & (0.1405) & (0.1383) & (0.1419) & (0.1007) & (0.0969) & (0.1032) \\ 
  $\delta_{1}$ & 0.0027 & 0.0700 & 0.0489 & 0.0014 & 0.0697 & 0.0488 & 0.0018 & 0.0718 & 0.0500 \\ 
   & (0.0891) & (0.0716) & (0.0723) & (0.0564) & (0.0489) & (0.0474) & (0.0366) & (0.0347) & (0.0324) \\ 
  $\delta_{2}$ & -0.0961 & -0.4802 & -0.4692 & -0.0434 & -0.4752 & -0.4473 & -0.0086 & -0.4729 & -0.4316 \\ 
   & (0.4743) & (0.1370) & (0.2135) & (0.2317) & (0.0914) & (0.1094) & (0.1516) & (0.0676) & (0.0748) \\ 
  $\sigma^2_{U}$ & 0.3203 & 2.1138 & 1.8042 & 0.1421 & 2.1568 & 1.8418 & 0.0368 & 2.1510 & 1.8773 \\ 
   & (1.4489) & (0.8585) & (1.0776) & (0.9452) & (0.5858) & (0.7884) & (0.7047) & (0.4296) & (0.5622) \\ 
  $\sigma^2_{V}$ & -0.0165 & -0.3017 & -0.2732 & -0.0123 & -0.2940 & -0.2578 & -0.0003 & -0.2833 & -0.2565 \\ 
   & (0.3399) & (0.1422) & (0.2112) & (0.2325) & (0.0983) & (0.1495) & (0.1773) & (0.0710) & (0.1097) \\ 
  $\rho_{U,\eta}$ & -0.0041 & ~ & -0.0035 & 0.0025 & ~ & -0.0035 & 0.0062 & ~ & -0.0037 \\ 
   & (0.0901) & ~ & (0.1106) & (0.0520) & ~ & (0.0592) & (0.0346) & ~ & (0.0416) \\ 
  $\rho_{V,\eta}$ & -0.0887 & ~ & ~ & -0.0461 & ~ & ~ & -0.0186 & ~ & ~ \\ 
   & (0.2914) & ~ & ~ & (0.1875) & ~ & ~ & (0.1135) & ~ & ~ \\ 
  $\gamma_{0}$ & -0.0029 & ~ & 0.0327 & -0.0004 & ~ & 0.0327 & -0.0017 & ~ & 0.0324 \\ 
   & (0.1082) & ~ & (0.1057) & (0.0738) & ~ & (0.0709) & (0.0496) & ~ & (0.0493) \\ 
  $\gamma_{1}$ & 0.0063 & ~ & 0.0031 & 0.0054 & ~ & 0.0031 & 0.0015 & ~ & 0.0004 \\ 
   & (0.1210) & ~ & (0.1213) & (0.0821) & ~ & (0.0813) & (0.0590) & ~ & (0.0594) \\ 
  $\gamma_{2}$ & 0.0125 & ~ & 0.0030 & 0.0045 & ~ & 0.0030 & 0.0025 & ~ & 0.0012 \\ 
   & (0.1233) & ~ & (0.1227) & (0.0788) & ~ & (0.0796) & (0.0567) & ~ & (0.0570) \\ 
  $\gamma_{3}$ & 0.0090 & ~ & -0.0031 & 0.0031 & ~ & -0.0031 & 0.0037 & ~ & -0.0038 \\ 
   & (0.1194) & ~ & (0.1202) & (0.0800) & ~ & (0.0810) & (0.0555) & ~ & (0.0561) \\ 
  $\gamma_{4}$ & 0.0152 & ~ & 0.0029 & 0.0058 & ~ & 0.0029 & 0.0026 & ~ & 0.0004 \\ 
   & (0.1161) & ~ & (0.1170) & (0.0793) & ~ & (0.0805) & (0.0524) & ~ & (0.0532) \\ 
  $\gamma_{5}$ & 0.0534 & ~ & 0.0348 & 0.0225 & ~ & 0.0348 & 0.0185 & ~ & 0.0330 \\ 
   & (0.2257) & ~ & (0.2298) & (0.1489) & ~ & (0.1552) & (0.1038) & ~ & (0.1070) \\ 
   \hline

\hline \hline
\caption[]{\textit{Bias and Standard Deviation (in parenthesis) of estimated parameters for all simulation schemes and across different estimation techniques.}\label{tab:appsimcomp}}\\
\end{longtable}
\endgroup
\end{center}

In general, the EX estimator performs very poorly for the estimation of the inefficiency parameter in all settings. In particular, $\sigma^2_U$ is estimated to be much larger than its true value, and $\delta_2$ to be smaller. This bias does not decrease with the sample size, and it becomes larger as we introduce endogeneity coming from the dependence between $U_0$ and $\eta$. All other parameters, including $\sigma^2_V$, are estimated more precisely. In particular, their bias decreases with the sample size, but it becomes larges as $\rho_U$ increases. For small sample sizes ($n=250$), it appears that the efficiency frontier is estimated more precisely with EX than with CPUW. However, this advantage quickly disappears as we increase the sample size. 

The KTS estimator has comparable performance to the EX estimator (as it effectively ignores the dependence between $V$ and $\eta$), and it overestimate $\sigma^2_U$ and underestimate $\delta_2$. However, it can pin down precisely the value of $\rho_U$, because of the conditional independence between $(V,U_0) \vert \eta$, so that imposing $\rho_V = 0$ does not affect its ability to precisely estimate $\rho_U$. Although, when $\rho_U = 0$, KTS largely overestimate $\rho_U$ compared to CPUW.

\newpage 

\section{Additional material for empirical application} \label{appBempirics}

In this section, we provide some additional information about the empirical application. 

Table \ref{tab:descrstat} contains descriptive statistics from the main variables used in the analysis. The variables are divided by category for convenience of the reader. Table \ref{tab.fs.probit} contains the MLE of the assignment equation with their $95\%$ confidence intervals.

% latex table generated in R 4.3.2 by xtable 1.8-4 package
% Thu Dec 21 09:15:33 2023
\begin{table}[ht]
\centering
\begin{tabular}{ccccc}
  \hline
\hline
 & Mean & St.Dev. & Min & Max \\ 
  \hline
\hline
Output & 1198.647 & 964.018 & 26.000 & 6816.000 \\ 
   \hline
\multicolumn{1}{l}{\textit{Inputs}} &  &  &  &  \\ 
   \hline
Land & 1.732 & 1.821 & 0.250 & 28.000 \\ 
  Labor & 75.505 & 51.461 & 7.000 & 484.000 \\ 
  Fertilizers & 209.037 & 159.201 & 9.000 & 1945.500 \\ 
  Pesticides & 111.615 & 112.140 & 2.000 & 1116.600 \\ 
  Seeds & 62.546 & 56.921 & 1.320 & 750.000 \\ 
   \hline
\multicolumn{1}{l}{\textit{Environmental variables}} &  &  &  &  \\ 
   \hline
Tenure & 0.658 & 0.475 & 0.000 & 1.000 \\ 
  Age & 0.257 & 0.438 & 0.000 & 1.000 \\ 
  Education & 7.022 & 3.391 & 0.000 & 16.000 \\ 
  No Income & 0.673 & 0.470 & 0.000 & 1.000 \\ 
  Foot access & 0.488 & 0.500 & 0.000 & 1.000 \\ 
  Car access & 0.466 & 0.499 & 0.000 & 1.000 \\ 
  Risk div & 1.199 & 1.758 & -5.828 & 5.513 \\ 
  Participation & 0.468 & 0.500 & 0.000 & 1.000 \\ 
   \hline
\multicolumn{1}{l}{\textit{Instruments}} &  &  &  &  \\ 
   \hline
Dist Earthquake & 0.756 & 0.430 & 0.000 & 1.000 \\ 
  Wage canton & 4.060 & 0.987 & 0.000 & 6.000 \\ 
  Prop of families with electricity & 0.820 & 0.232 & 0.000 & 1.000 \\ 
  Prop of families with bathroom & 0.776 & 0.258 & 0.000 & 1.000 \\ 
   \hline
\hline
\end{tabular}
\caption{Descriptive Statistics} 
\label{tab:descrstat}
\end{table}

\begin{table}[!h]
\centering
\begin{tabular}{l | c l r} 
\hline \hline
~ & Estimate & \multicolumn{2}{c}{CI} \\ \hline
% latex table generated in R 4.3.2 by xtable 1.8-4 package
% Thu Dec 21 09:15:33 2023
  \hline
$\gamma_0$ & -2.3912 & [-4.2259 & -0.7052] \\ 
  $\gamma_{Land}$ & 0.2533 & [-0.0217 & 0.5675] \\ 
  $\gamma_{Labor}$ & -0.0076 & [-0.1971 & 0.1926] \\ 
  $\gamma_{Fertilizer}$ & 0.0474 & [-0.1482 & 0.2324] \\ 
  $\gamma_{Pesticides}$ & 0.0379 & [-0.1399 & 0.2061] \\ 
  $\gamma_{Seeds}$ & 0.1117 & [-0.0428 & 0.2890] \\ 
  $\gamma_{Tenure}$ & 0.2303 & [-0.1764 & 0.6603] \\ 
  $\gamma_{Tenure \times Land}$ & -0.2096 & [-0.5034 & 0.0411] \\ 
  $\gamma_{Age}$ & 0.3106 & [0.0203 & 0.6091] \\ 
  $\gamma_{Educ}$ & 0.2021 & [0.0768 & 0.3345] \\ 
  $\gamma_{NoIncome}$ & 0.1490 & [-0.1199 & 0.4328] \\ 
  $\gamma_{Foot Access}$ & 0.2833 & [-0.3062 & 0.8727] \\ 
  $\gamma_{Car Access}$ & 0.0490 & [-0.5310 & 0.6455] \\ 
  $\gamma_{Risk Div}$ & -0.0221 & [-0.1467 & 0.1099] \\ 
  $\gamma_{DistEarthquake}$ & -0.8975 & [-1.2014 & -0.6175] \\ 
  $\gamma_{Wage}$ & 0.0637 & [-0.0659 & 0.2036] \\ 
  $\gamma_{FamElect}$ & 0.7387 & [0.1938 & 1.4231] \\ 
  $\gamma_{FamBath}$ & -0.4484 & [-0.9240 & 0.0658] \\ 
  $\gamma_{Reg2}$ & 0.7886 & [0.4113 & 1.1983] \\ 
  $\gamma_{Reg3}$ & -0.1212 & [-0.4182 & 0.1838] \\ 
  $\gamma_{Reg4}$ & 1.7186 & [1.3361 & 2.1848] \\ 
   \hline

\hline \hline 
\end{tabular}
\caption{\textit{Estimation of the first-stage equation.}}
\label{tab.fs.probit}
\end{table}

\end{document}